\documentclass[a4paper,fleqn,usenatbib]{mnras}

\usepackage{ae,aecompl}
\usepackage{graphicx}	
\usepackage{amsmath}	
\usepackage{amssymb}	
\usepackage{mathtools}
\usepackage{hyperref}
\usepackage{rotating}
\usepackage{caption}

\usepackage{tikz}
\newcommand{\dittotikz}{%
    \tikz{
        \draw [line width=0.12ex] (-0.2ex,0) -- +(0,0.8ex)
            (0.2ex,0) -- +(0,0.8ex);
        \draw [line width=0.08ex] (-0.6ex,0.4ex) -- +(-1.5em,0)
            (0.6ex,0.4ex) -- +(1.5em,0);
    }%
}

\usepackage{newtxtext,newtxmath}

\title[DEVILS Multi-wavelength Photometry]{Deep Extragalactic VIsible Legacy Survey (DEVILS): Consistent Multi-wavelength Photometry for the DEVILS Regions (COSMOS, XMMLSS \& ECDFS)}
\author[L. J. M. Davies et. al.]{L. J. M. Davies$^{1}$\thanks{E-mail:
 luke.j.davies@uwa.edu.au}, J. E. Thorne$^{1}$,  A. S. G. Robotham$^{1,2}$, S. Bellstedt$^{1}$, S. P. Driver$^{1}$,  N. J. Adams$^{3}$, \newauthor M. Bilicki$^{4}$, R. A. A. Bowler $^{3}$, M. Bravo$^{1}$, L. Cortese$^{1,2}$, C. Foster$^{2,5}$, M. W. Grootes$^{6}$, B. H{\"a}u{\ss}ler$^{7}$,  \newauthor  A. Hashemizadeh$^{1}$,  B. W. Holwerda$^{8}$, P. Hurley$^{9}$, M. J. Jarvis$^{3,10}$, C. Lidman$^{11,12}$,  N. Maddox$^{13}$,   \newauthor  M. Meyer$^{1,2}$,   M. Paolillo$^{14,15,16}$, S. Phillipps$^{17}$,  M. Radovich$^{18}$,  M. Siudek$^{19,20}$, M. Vaccari$^{10,21}$,   \newauthor  R. A. Windhorst$^{22}$ \\
$^{1}$ ICRAR, The University of Western Australia, 35 Stirling Highway, Crawley, WA 6009, Australia\\
$^{2}$ ARC Centre of Excellence for All Sky Astrophysics in 3 Dimensions (ASTRO 3D)\\
$^{3}$ Sub-department of Astrophysics, University of Oxford, Denys Wilkinson Building, Keble Road, Oxford OX13RH, UK\\
$^{4}$ Center for Theoretical Physics, Polish Academy of Sciences, al. Lotnik{\'o}w 32/46, 02-668 Warsaw, Poland \\
$^{5}$ Sydney Institute for Astronomy, School of Physics, A28, The University of Sydney, NSW, 2006, Australia\\
$^{6}$ Netherlands eScience Center, Science Park 140, 1098 XG Amsterdam, The Netherlands \\
$^{7}$ European Southern Observatory, Alonso de Cordova 3107, Vitacura, Santiago, Chile \\
$^{8}$ Department of Physics and Astronomy, 102 Natural Science Building, University of Louisville, Louisville KY 40292, USA\\
$^{9}$ Astronomy Centre, Department of Physics \& Astronomy, University of Sussex, Brighton, BN1 9QH, UK\\
$^{10}$ Department of Physics and Astronomy, University of the Western Cape, Robert Sobukwe Road, Bellville 7535, South Africa\\
$^{11}$ Research School of Astronomy and Astrophysics, The Australian National University, ACT 2601, Australia\\
$^{12}$ Centre for Gravitational Astrophysics, College of Science, The Australian National University, ACT 2601, Australia\\
$^{13}$ Faculty of Physics, Ludwig-Maximilians-Universit\"at, Scheinerstr. 1, 81679 Munich, Germany\\
$^{14}$ INAF - Osservatorio Astronomico di Capodimonte, Salita Moiariello 16, I-80131, Napoli, Italy\\
$^{15}$ Dipartimento di Fisica, Universit{\~a}  di Napoli "Federico II," via Cinthia 9, I-80126 Napoli, Italy\\
$^{16}$ INFN - Sezione di Napoli, via Cinthia 9, I-80126 Napoli, Italy\\
$^{17}$ H. H. Wills Physics Laboratory, University of Bristol, Tyndall Avenue, Bristol, BS8 1TL, UK \\
$^{18}$ INAF - Osservatorio Astronomico di Padova, vicolo dell'Osservatorio 5, I-35122 Padova, Italy \\
$^{19}$ Institut de F\'{\i}sica d'Altes Energies (IFAE), The Barcelona Institute of Science and Technology, 08193 Bellaterra (Barcelona), Spain\\
$^{20}$ National Centre for Nuclear Research, ul. Pasteura 7, 02-093, Warsaw, Poland\\
$^{21}$ INAF - Istituto di Radioastronomia, via Gobetti 101, 40129 Bologna, Italy\\
$^{22}$ School of Earth \& Space Exploration, Arizona State University, P.O. Box 871404, Tempe, AZ 85287-1404, USA
}

\date{Accepted XXX. Received YYY; in original form ZZZ}

\pubyear{2020}

\begin{document}
\label{firstpage}
\pagerange{\pageref{firstpage}--\pageref{lastpage}}
\maketitle

\begin{abstract}

The Deep Extragalactic VIsible Legacy Survey (DEVILS) is an ongoing high-completeness, deep spectroscopic survey of $\sim$60,000 galaxies to Y$<$21.2\,mag, over $\sim$6\,deg$^2$ in three well-studied deep extragalactic fields: D10 (COSMOS), D02 (XMM-LSS) and D03 (ECDFS). Numerous DEVILS projects all require consistent, uniformly-derived and state-of-the-art photometric data with which to measure galaxy properties. Existing photometric catalogues in these regions either use varied photometric measurement techniques for different facilities/wavelengths leading to inconsistencies, older imaging data and/or rely on source detection and photometry techniques with known problems. Here we use the \textsc{ProFound} image analysis package and state-of-the-art imaging datasets (including Subaru-HSC, VST-VOICE, VISTA-VIDEO and UltraVISTA-DR4) to derive matched-source photometry in 22 bands from the FUV to 500$\mu$m. This photometry is found to be consistent, or better,  in colour-analysis to previous approaches using fixed-size apertures (which are specifically tuned to derive colours), but produces superior total source photometry, essential for the derivation of stellar masses, star-formation rates, star-formation histories, etc. Our photometric catalogue is described in detail and, after internal DEVILS team projects, will be publicly released for use by the broader scientific community.

\end{abstract}


\section{Introduction}
\label{sec:into}

The advent of large area, high-completeness, spectroscopic surveys of the local Universe, coupled with deep panchromatic multi-wavelength imaging, and resulting robustly derived galaxy and environmental properties, have been revolutionary over the past two decades. Surveys such as the Sloan Digital Sky Survey \citep[SDSS,][]{York00, Abazajian09} and the Galaxy and Mass Assembly \citep[GAMA,][]{Driver11,Hopkins13,Liske15,Driver16, Baldry18} have compiled extensive databases of galaxy locations, distances, environments and physical properties leading to a complete and comprehensive analysis of `typical' galaxies ($i.e.$ $\gtrsim$0.1M$^{*}$, one tenth the characteristic mass in the stellar mass function, and above) in the relatively local Universe. These surveys have fundamentally changed our understanding of the distribution of stellar mass \citep[$e.g.$][]{Baldry12, Wright17}, star-formation \citep[$e.g.$][]{Kauffmann03, Brinchmann04, Davies16b, Davies17, Davies19b}, gas and stellar phase metallicity \citep[$e.g.$][]{Tremonti04, Lara-lopez13, Bellstedt21}, large-scale structure and group environments \citep[$e.g.$][]{Tegmark04,Yang07, Robotham11, Blake13, Alpaslan14}, pairs of galaxies \citep[$e.g.$][]{Patton13, Robotham14} and the impact of these environments on galaxy properties \cite[$e.g.$][]{Blanton05, Ellison08, Peng10, Patton11, Robotham13, Alpaslan16, Davies15b, Davies16a, Davies19a}. 

However, this approach of successfully combining extensive panchromatic multi-wavelength imaging with high completeness spectroscopy to consistently measure galaxy and environmental properties has largely been limited to the relatively local Universe. While this can tell us a wealth of information regarding the end-point of the galaxy evolution process to date, it tells us little about the astrophysical process occurring over the preceding billions of years of Universal evolution. While numerous surveys have aimed at probing the evolution of galaxies outside of the local Universe ($z>0.3$), they can not be well-matched to the local studies of SDSS and GAMA due to sometimes complex selections functions, incompleteness and/or varying analysis techniques.  Until recently, the state of the art survey which probed this epoch \citep[zCOSMOS-bright,][]{Lilly07} was encumbered by its very small area ($\sim$1deg$^{2}$), sparse sampling, and complex footprint (due to slit-mask spectroscopy). This ultimately leads to low completeness \citep[only $\sim50\%$ to $i<22$, see][]{Davies15a}. Other surveys at this epoch have focused on the sparse sampling of colour-selected populations over large volumes, such as the VIMOS Public Extragalactic Redshift Survey \citep[VIPERS,][]{Guzzo14,Scodeggio18} and DEEP2/3 \citep{Cooper12, Newman13}. The designs of these surveys, while matched to their specific science goals, are not tuned to provide a comparable sample to more local studies and are inadequate for studying the evolution of galaxy groups, mergers, sub-Mpc structure and the influence of environment on galaxy evolution - a key requirement in understanding the evolutionary processes which are shaping galaxies.    

The new Deep Extragalactic VIsible Legacy Survey \citep[DEVILS,][]{Davies18} is a high completeness, deep spectroscopic survey aimed at providing a direct counterpart to GAMA but at a much earlier epoch ($0.3<z<1$). Unlike previous samples DEVILS does not apply any colour selection, does not sparsely sample the population ($i.e.$ is flux-limited only), and will use identical source selection and analysis techniques to samples in the local Universe. This will allow similar studies to GAMA and SDSS, but at a much earlier point in cosmic time. Importantly, within DEVILS we will apply identical data processing and analysis techniques to GAMA to minimise any potential errors/inconsistencies in the approach applied, $i.e.$ it is difficult to compare measurements from surveys at low and high redshift, which use different selection methods, approaches and measurement techniques leading to potential errors when probing galaxy evolution processes. 

The foundation of many derived astrophysical properties of galaxies are robust observed images/photometric measurements coupled with redshifts. With just these data products in hand, an extensive array of measurements can be made, such as stellar masses \citep{Kauffmann03, Baldry10, Taylor15}, star-formation rates \citep{Brinchmann04, Davies16b} and star-formation histories \citep[$e.g.$][Thorne at al 2021]{Kauffmann03, Bellstedt20}, metallicities \citep[$.e.g.$][Thorne et al in prep.]{Tremonti04, Bellstedt21}, morphological classifications \citep[$e.g.$][]{Moffett16, Hashemizadeh21} and structural component analyses \citep[$e.g.$][]{Lange15}. The DEVILS spectroscopic observations are currently being undertaken at the Anglo-Australian Telescope (AAT) \citep[see][ and below]{Davies18}. However, the next stage of producing these derived properties for our sample is to acquire robust photometric measurements of galaxies in the DEVILS regions.

Importantly for direct comparisons to $z\sim0$, this photometry must be derived using an almost identical method to that used in GAMA. Recently, \cite{Bellstedt20}, produced new multi-wavelength photometric measurements for the GAMA sample using the \textsc{ProFound} \citep{Robotham18} source detection and photometry code. In this work we apply the same process (albeit with small changes where appropriate) to the deep photometric data available in the DEVILS regions. While there are existing photometric catalogues in these fields \citep[most notably in the COSMOS region from COSMOS2015 and G10/COSMOS,][]{Laigle16,Andrews17}, these catalogues typically used imaging data that has since been superseded by more recent higher quality data, and disparate measurement techniques at different wavelengths, which are then matched together in catalogue space. In addition, these samples were compiled with pre-Gaia astrometry, which can now be improved upon using the updated astrometric accuracy Gaia provides. This can lead to inconsistencies/errors from $e.g.$ errors in position matching and/or blending of sources. In addition, no consistent photometry currently exists across all of the DEVILS regions (COSMOS, XMMLSS, ECDFS). With this in mind, in this work we derive new photometric measurements in the DEVILS regions using the most up-to-date imaging available, in a consistent way from FUV to FIR wavelengths and in a largely identical manner to GAMA to allow direct comparisons to $z\sim0$.  

This photometry is already being used in a number of projects within the DEVILS survey \citep[$e.g.$][Thorne et al in prep., Hashemizadeh et al in prep.]{Thorne21, Hashemizadeh21, Koushan21} and will form the basis for many subsequent studies. In this paper we fully document the source finding, photometric derivation, masking, star-galaxy separation and catalogue compilation of the photometry used in these science projects. Following this we also utilise our new photometry catalogues to produce and tabulate deep multi-wavelength number counts for broader use. Once our catalogues have been used for internal team projects, they will be made publicly available to the wider community via the DEVILS data release 1 (Davies et al in prep.).       

All magnitudes reported here are in the AB system and when necessary we assume a cosmology with H$_{0}$ = 70\,km\,s$^{-1}$\,Mpc$^{1}$, $\Omega_{m}$ = 0.3, and $\Omega_{\Lambda}$= 0.7.

\section{The Deep Extragalactic VIsible Legacy Survey}

Briefly, DEVILS is an ongoing spectroscopic and multi-wavelength survey being undertaken with the Anglo-Australian Telescope (AAT). The survey aims to build a high completeness ($>$85\%) sample of $\sim$60,000 galaxies to Y$<$21.2\,mag in three well-studied deep extragalactic fields: D10 (COSMOS), D02 (ECDFS) and D03 (XMM-LSS). This sample will provide the first high spectroscopic completeness flux-limited sample at intermediate redshifts, allowing for the robust parametrisation of group and pair environments in the distant Universe. In addition to the spectroscopic survey, the DEVILS team will also compile an extensive database of multi-wavelength-derived properties for sources in the DEVILS spectroscopic sample, such as spectral energy distribution (SED) fits, star-formation histories, stellar masses, star-formation rates and metallicities (see Thorne et al in prep.), and morphologies and structural parameters (see Hashemizadeh et al, in prep.) - with the work presented here forming the basis of this database. The spectroscopic component of the survey is currently $\sim$70\% complete, and is due to finish spectroscopic observations in late 2021. The science goals of the project are varied, from the environmental impact on galaxy evolution at intermediate redshift, to the evolution of the halo mass function over the last $\sim$7\,billion years. For full details of the survey science goals, survey design, target selection and spectroscopic observations see \cite{Davies18}. In this paper we describe our new multi-wavelength photometric catalogue, which will be used in the core DEVILS science projects.

\subsection{The DEVILS Regions}

The DEVILS regions were specifically selected to cover areas with extensive existing and upcoming imaging campaigns to facilitate broad range of science topics. 

The D10 field represents a sub-region of the Cosmic Evolution Survey field \citep[COSMOS,][]{Scoville07}, covering 1.5deg$^{2}$ of the UltraVISTA \citep{McCracken12} region and centred at R.A.=150.04$^{\circ}$, Dec=2.22$^{\circ}$. This region is covered by an extensive array of imaging and spectroscopic data ranging from x-ray to low-frequency radio continuum observations (see Table \ref{tab:D10photom} for data in the FUV-FIR reguime). The DEVILS spectroscopic target catalogue in this region is selected from the UltraVISTA DR4 Y-band observations and includes all extragalactic sources to Y$<$21.2\,mag \citep[see][]{{Davies18}}. We do not discuss in detail the DEVILS spectroscopic campaign further in this work.

The D02 region covers the central 3\,deg$^{2}$ of the XMM Large-Scale-Structure field (XMMLSS) centred at R.A.=35.53$^{\circ}$, Dec=-4.70$^{\circ}$. This region is also covered by extensive imaging data, which is largely consistent in facility/filter coverage to D10. The main and important difference to the imaging data in D10, is that the NIR imaging comes from the Vista Deep Extragalactic Survey \citep[VIDEO,][]{Jarvis13}, which is consistent in facility/filters to UltraVISTA but shallower in depth (see Table \ref{tab:D10photom}). 

Finally, the D03 region covers the central 1.5\,deg$^{2}$ of the Extended Chandra Deep Field South (ECDFS) centred at R.A.=53.15$^{\circ}$, Dec=-28.0$^{\circ}$. This region  contains similar photometric datasets to D02 in the UV and NIR-FIR, but differs in the optical (described below and see Table \ref{tab:D10photom}). The DEVILS spectroscopic target catalogues in D02 and D03 are selected from the VIDEO Y-band and also includes all extragalactic sources to Y$<$21.2\,mag \citep[see][]{{Davies18}}.

\section{Imaging Datasets}
\label{sec:D10}

\begin{figure*}
\begin{center}
\includegraphics[scale=0.7]{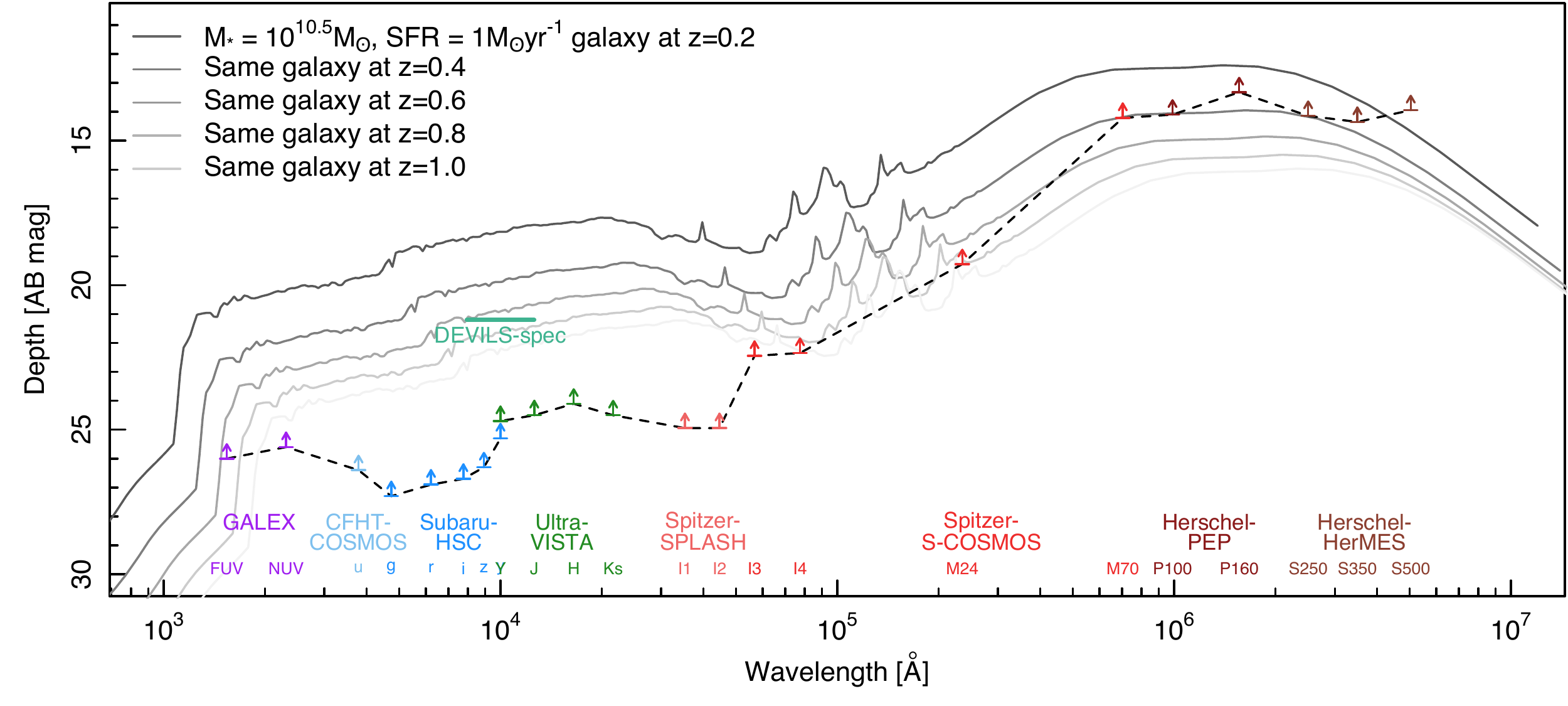}
\caption{The survey-quoted nominal 5$\sigma$ point source photometric depth of multi-wavelength data in the D10 region displaying an example of the typical depth of data covering the DEVILS regions (arrows joined by dashed line). The grey curves show a M$_{*}$=10$^{10.5}$\,M$_{\odot}$ and SFR= 1\,M$_{\odot}$\,yr$^{-1}$ galaxy at $z$=0.2, 0.4, 0.6, 0.8 and 1.0, generated using \textsc{ProSpect}. The depth of imaging in the D10 region means that a typical galaxy will be detected in $GALEX$-NUV to  $Spitzer$-IRAC out to $z\sim1$. The imaging data extends to much greater depth than the DEVILS spectroscopic campaign.  }
\label{fig:covarage}
\end{center}
\end{figure*}

\begin{figure*}
\begin{center}
\includegraphics[scale=0.38]{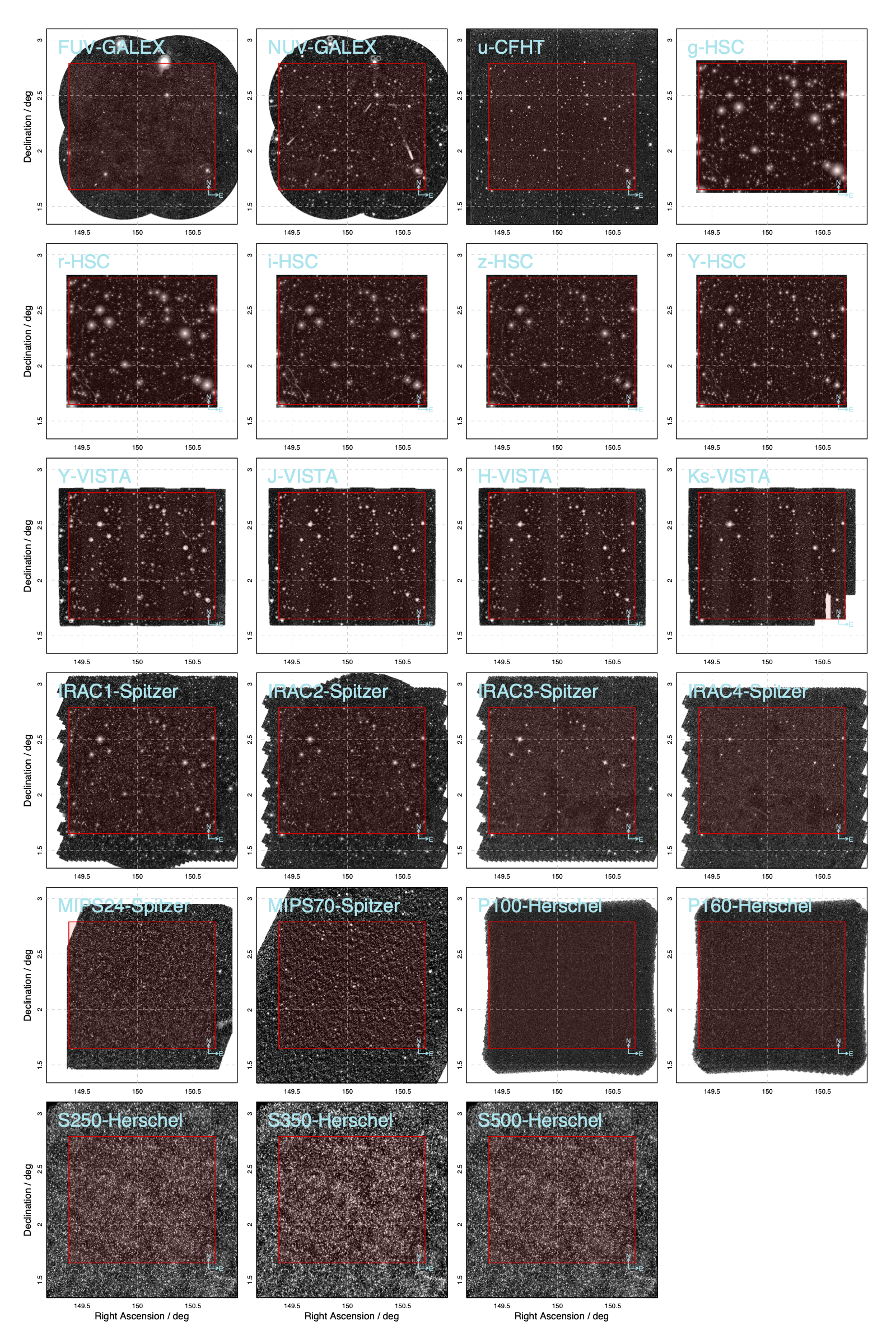}
\caption{Multi-wavelength data covering the D10 region. The D10 spectroscopic region is displayed as the red box. Note that the HSC data are only downloaded to cover the D10 region (with a small buffer). The raw HSC data covers a larger area.}
\label{fig:images}
\end{center}
\end{figure*}

\begin{figure*}
\begin{center}
\includegraphics[scale=0.5]{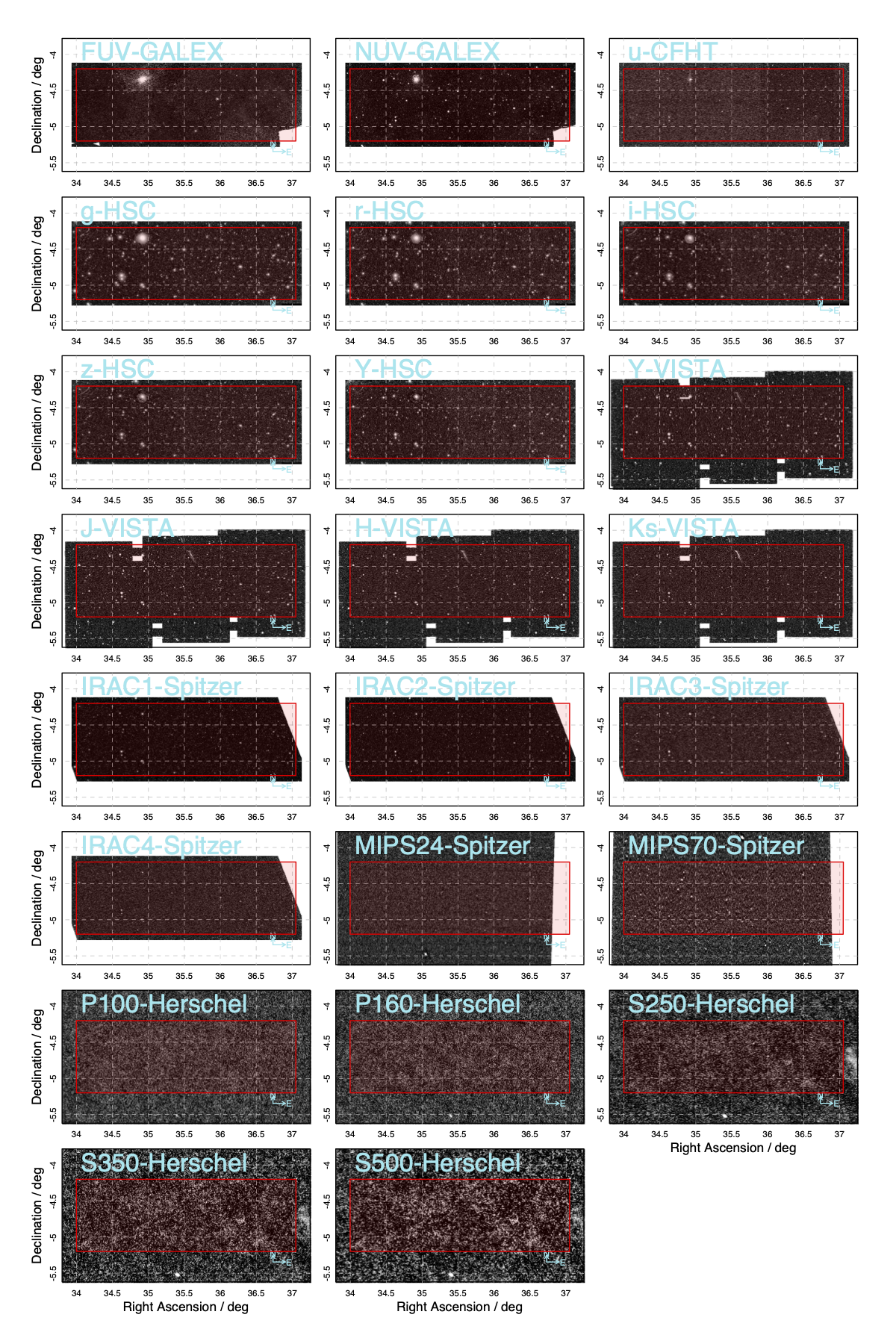}
\caption{Multi-wavelength data covering the D02 region. The D02 spectroscopic region is displayed as the red box. Data from GALEX, CFHT, HSC and $Spitzer$-IRAC is combined using \textsc{SWARP} to only cover the D02 region, but the original data does extend to larger areas. }
\label{fig:imagesD02}
\end{center}
\end{figure*}

\begin{figure*}
\begin{center}
\includegraphics[scale=0.38]{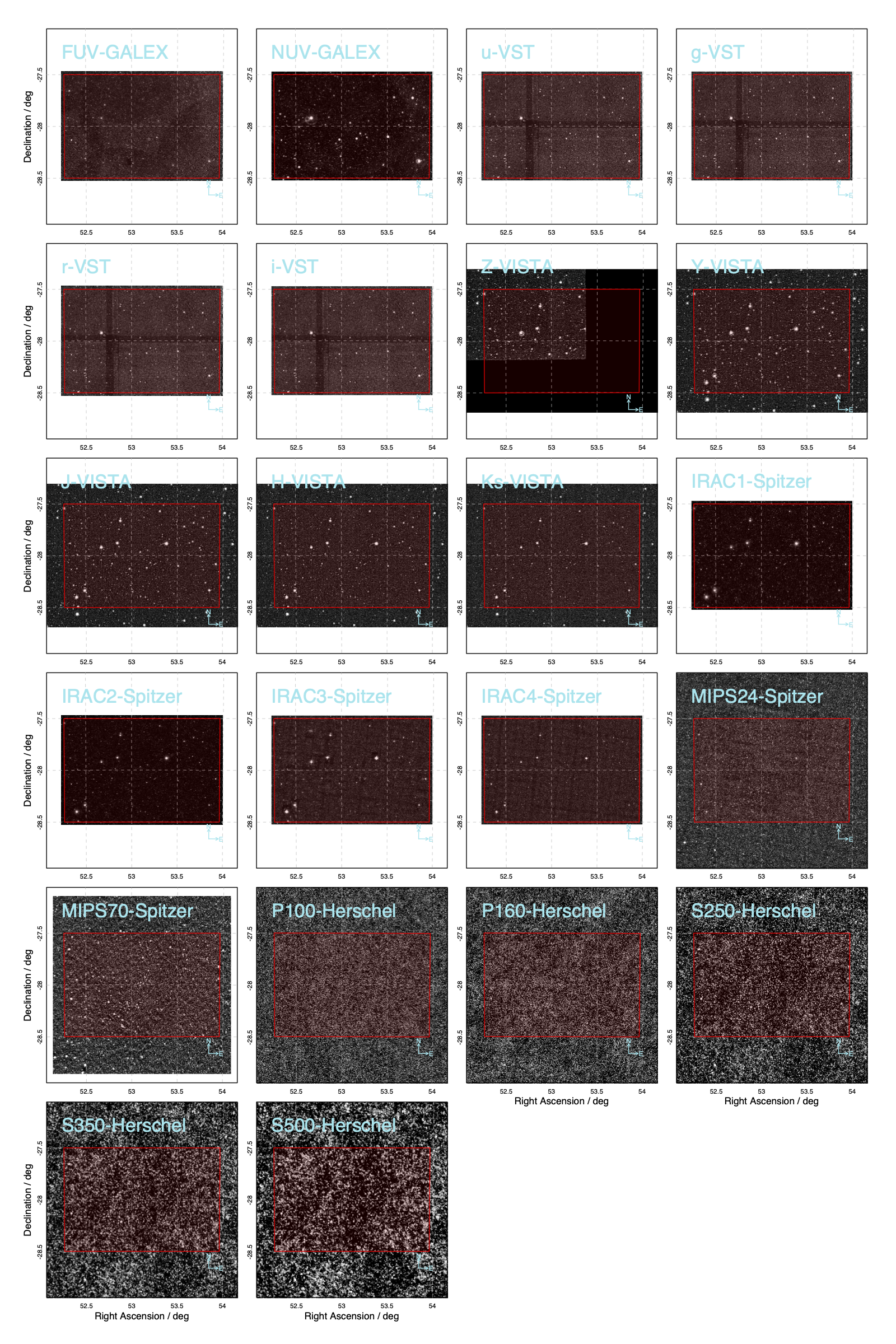}
\caption{Multi-wavelength data covering the D03 region. The D03 spectroscopic region is displayed as the red box. Note that data from GALEX, VST and $Spitzer$-IRAC is combined using \textsc{SWARP} to only cover the D03 region. The original data does extend to larger areas. Note that Z-band VISTA data does not cover the full region, but is the only available deep Z-band imaging the D03 field.}
\label{fig:imagesD03}
\end{center}
\end{figure*}

In this section, we briefly describe the various imaging datasets in each of the DEVILS regions, and the surveys in which they were undertaken. The coverage of these data in each of these fields are displayed in Figures \ref{fig:images}, \ref{fig:imagesD02}, \ref{fig:imagesD03} and a summary of all data is given in Table \ref{tab:D10photom}. In order to display the depth of the various imaging datasets in comparison to the typical galaxies we will parameterise with DEVILS,  Figure \ref{fig:covarage} shows the point source depth of the D10 imaging in-comparison to the spectral energy distribution of a typical galaxy within the DEVILS spectroscopic sample (note D02/D03 are roughly comparable in depth, but see Table \ref{tab:D10photom} for differences). In the following subsections, unless otherwise stated, when combining data using \textsc{SWARP} \citep{Bertin02} we use the following settings: i) we set the centre of each field to the centre of the respective DEVILS region, ii) we set the pixel scale to that of the VISTA imaging used in our detection image (see Section \ref{sec:multi} for more details), iii) we set the image size to cover the full DEVILS region $\pm$ a 0.05\,deg buffer, iv) we do not use a weight map or mask in the combination, v) we apply a median combining process where frames overlap in sky position (this is only true for the GALEX and CFHT - u data in D02 and D03, all other data does not require combination), and vi) we do not apply a background subtraction at the SWARP combination stage as \textsc{ProFound} performs its own background subtraction when extracting photometric measurements.

\subsection{Ultraviolet}

\subsubsection{GALEX (D10/D02/D03)}

The COSMOS region was observed using the Galaxy Evolution Explorer (GALEX) space telescope as part of the Deep Imaging Survey \citep{Zamojski07}.  The telescope was equipped with a 0.5\,m mirror, a circular field of view with 1.2\,deg diameter, a 1.5\,$^{\prime\prime}$  pixel$^{-1}$ detector and two passbands in the far and near-ultraviolet (FUV and NUV, respectively). Observations in the D10/COSMOS region consisted of four pointings in each band with exposure times of 45,000\,s in FUV with a point spread function (PSF) full width at half-maximum (FWHM) of 5.4\,$^{\prime\prime}$ , and 50,000\,s in NUV with a PSF FWHM of 5.6\,$^{\prime\prime}$ . This data reaches a 5$\sigma$ point source limiting depth of 26.0 and 25.6\,mag (AB) in the FUV and NUV channels, respectively (see Table \ref{tab:D10photom} and Figure \ref{fig:covarage}).  In this work we combine the four GALEX pointings in each band into a single mosaic covering the full D10 region using \textsc{SWARP}. In this process background subtraction is turned off due to the low photon statistics in the GALEX imaging resulting in poor background subtraction (see top left panels of Figure \ref{fig:images}).  

D02 and D03 have also been observed using the Galaxy Evolution Explorer (GALEX) space telescope as part of various imaging programs undertaken by different survey teams. To maximise the depth and coverage of the data available, we identify all GALEX pointings in the MAST database that overlap with the D02 and D03 spectroscopic regions. The images are then combined using \textsc{SWARP}, with background subtraction turned off (see top left panels of Figure \ref{fig:imagesD02} and \ref{fig:imagesD03}). While this process results in slightly varying depth across the image, our resultant photometry uses a local sky background estimate to determine photometric errors, and thus takes the varying depth into account. For reference, the 5$\sigma$ point-source depth is measured to vary from $\sim23.5-25.5$\,mag across the region. 

\subsection{Optical}

\subsubsection{CFHT - u (D10/D02)}

The $u$-band images in D10 and D02 were obtained as part of the Canada-France-Hawaii Telescope Legacy Survey (CFHT-LS\footnote{\url{http://www.cfht.hawaii.edu/Science/CFHLS/}}). In D10, the CFHT-LS deep survey covers the central 1\,deg$^{2}$ of the field, but is also supplemented by independent CFHT observations outlined in \cite{Capak07} to cover the full 1.5\,deg$^{2}$.  To maximise the depth of these data and to have coverage over the full D10 region, we combine the CFHT-LS and \cite{Capak07} data into a single image using \textsc{SWARP}. Both data sets were obtained using Mega-Prime camera on CFHT with a 1\,deg$^{2}$ field of view with a native resolution of 0.18\,$^{\prime\prime}$ pixel$^{-1}$.  For CFHT-LS the worst case seeing was 0.9\,$^{\prime\prime}$  and the 5$\sigma$ point source limiting magnitude is 26.6\,mag. For the \cite{Capak07} data the worst case seeing was also 0.9\,$^{\prime\prime}$  and the 5$\sigma$ point source limiting magnitude is 26.4\,mag. The D02 region is covered by a combination of the CFHT-LS deep 1 region (D1, 5$\sigma$ point source depth $\sim$26.25\,mag) and wide 1 region (W1, 5$\sigma$ point source depth $\sim$25.3\,mag). To maximise the depth and coverage, we obtain all D1 and W1 data from the CFHT archive and combine using SWARP.

\subsubsection{Subaru-HSC - g,r,i,z (D10/D02)}

For our optical, $griz$, imaging in the D10 and D02 region we use the new second public data release \citep[PDR2][]{Aihara19} of the Subaru Telescope's Hyper Suprime-Cam Subaru Strategic Program \citep[HSC-SSP, see][]{Aihara18a,Aihara18b}. HSC is an optical imaging camera on the 8.2m mirror Subaru Telescope, covering a 1.5\,deg diameter field of view to 0.168\,$^{\prime\prime}$  pixel$^{-1}$ resolution. As part of the Deep+Ultra-Deep (DUD) SSP program, the HSC team have released 31\,deg$^2$ of DUD imaging including the extended-COSMOS snd XMMLSS regions. Data reaches a 5$\sigma$ point source limiting magnitude of 27.3, 26.9, 26.7, 26.3, 25.3\,mag in $grizy$, respectively (quoted by the HSC-SSP team). To obtain these data, we independently download overlapping 0.25 $\times$ 0.25\,deg regions in all bands from the HSC data access cutout server\footnote{\url{https://hsc-release.mtk.nao.ac.jp/das\_cutout/pdr2/}} and combined using \textsc{SWARP}.  

\subsubsection{VST-VOICE - u,g,r,i (D03)}
Both CFHT-LS and HSC-SSP programs do not cover the D03 region due to its southern nature. As such, for our optical, $ugri$ imaging we use the first data release of the VST Optical Imaging of the CDFS and ES1 Fields \citep[VOICE,][]{Vaccari16} Survey.  The VOICE data covers 4\,deg$^{2}$ of the ECDFS region with four pointings reaching a 5$\sigma$ point source limiting magnitude of 25.3, 26.4, 26.1, 25.2\,mag in $ugri$, respectively.  The survey is undertaken on the 2.6m VLT Survey Telescope (VST) using OmegaCAM with a  0.21\,$^{\prime\prime}$ /pixel resolution.  In addition, the $ugri$ imaging is also supplemented by partial coverage of the field in the VIDEO Z-band (see below).

\subsection{Near Infrared}

\subsubsection{UltraVISTA-DR4 (D10)}

The near-IR UltraVISTA survey \citep{McCracken12} is a recently completed survey on the Visible and Infrared Survey Telescope for Astronomy (VISTA) using the VISTA Infrared Camera (VIRCAM). VIRCAM covers a 1.65\,deg diameter field of view with 0.34\,$^{\prime\prime}$  pixel$^{-1}$ native resolution. The UltraVISTA team observe a single pointing centred on the COSMOS region, and are resampled to a 0.15\,$^{\prime\prime}$  pixel$^{-1}$ resolution.  Here we use the fourth UltraVISTA public data release (DR4\footnote{\url{http://ultravista.org/release4/}}), which comprises of stacked Y, J, H, Ks images using data taken between 2009 and 2016, with a total exposure time of $>$11\,h in each band/pixel. Typical FWHM seeing of the stacked data is $\sim$0.75\,$^{\prime\prime}$  and the 5$\sigma$ point source limiting magnitudes are 24.7, 24.5, 24.1, 24.5\,mag, in the deep regions and 25.8, 25.6, 25.2, 24.9\,mag, in the ultra-deep region \citep[see][for details]{McCracken12} in Y, J , H and Ks respectively. Data are downloaded from the ESO public archive. 

\subsubsection{VIDEO (D02/D03)}

Complementary to the UltraVISTA imaging in D10, the VISTA Deep Extragalactic Observations \citep[VIDEO][]{Jarvis13} Survey has covered 12\,deg$^{2}$ of three deep extragalactic fields using VISTA-VIRCAM in Z, Y, J, H and Ks bands - completely covering the D02 and D03 regions in YJHKs. Unlike UltraVISTA, VIDEO data are resampled to a 0.2\,$^{\prime\prime}$  pixel$^{-1}$ resolution. For our catalogues we use the VIDEO team-internal 2019-11-08 stacked images, reaching full survey depth. Data with $>0.9$\,$^{\prime\prime}$  seeing are rejected from these stacks and final  5$\sigma$ point source limiting magnitudes are 25.7, 24.6, 24.5, 24.0 and 23.5\,mag in Z, Y, J, H and Ks respectively. In D02 we use only YJHKs for our imaging catalogues and opt to use the HSC z-band data to be consistent with D10. However, as noted above, in D03 we do not have HSC coverage and therefore use the VIDEO Z-band data, which due to an updated strategy implemented during the VIDEO survey is incomplete and shallower (see Figure \ref{fig:imagesD03}). For details of the VIDEO survey and data, see \cite{Jarvis13}.

\subsection{Mid Infrared}

\subsubsection{S-COSMOS and Spitzer-SPLASH (D10)}

The COSMOS $Spitzer$ survey \citep[S-COSMOS,][]{Sanders07}, is a 3-160\,$\mu$m survey of the COSMOS region using all seven bands of the $Spitzer$ Infrared Array Camera (IRAC) and the Multiband Imaging Photometer for Spitzer (MIPS) instruments. The IRAC instrument covers four NIR-MIR filters centred at  3.6, 4.5, 5.6, 8.0\,$\mu$m over a 27 arcmin$^{2}$ field of view with native 1.2\,$^{\prime\prime}$  pixel$^{-1}$ resolution. MIPS covers three MIR-FIR filters centred at  24, 70 and 160\,$\mu$m with native resolution of 1.2, 4.0 and 8.0 $^{\prime\prime}$  pixel$^{1}$ respectively. 

Following the S-COSMOS campaign, and during $Spitzer's$ `warm' mission phase once the cryogens had been exhausted, the COSMOS field was observed by the Spitzer Large Area Survey with Hyper-Suprime-Cam (SPLASH) survey at 3.6 and 4.5\,$\mu$m increasing the depth and area coverage in these bands \citep[see][]{Lin16, Laigle16}. Both SPLASH and S-COSMOS IRAC data have been resampled to 0.6\,$^{\prime\prime}$  pixel$^{-1}$ resolution by their respective teams, with typical PSF FWHM of 1.7, 1.7, 1.9 and 2.0\,$^{\prime\prime}$  for the IRAC bands respectively. 

Here we use the superior SPLASH IRAC 3.6 and 4.5\,$\mu$m data reaching a 5$\sigma$ point source limiting magnitude of 24.9 in both bands, and S-COSMOS IRAC 5.6 and 8.0\,$\mu$m, and MIPS 24 and 70\,$\mu$m reaching a 5$\sigma$ point source limiting magnitude of 22.4, 22.3, 19.3, and 14.2\,mag respectively. We do not use the MIPS 160\,$\mu$m data in favour of the deeper $Herschel$ data that exist in this region (see below).

\subsubsection{SERVS and SWIRE (D02/D03)}

Both D02 and D03 are covered by the $Spitzer$ Extragalactic Representative Volume Survey \citep[SERVS,][]{Mauduit12} survey at 3.6 and 4.5\,$\mu$m (IRAC1 and IRAC2). SERVS is an 18\,deg$^{2}$  medium-deep survey with the post-cryogenic Spitzer Space Telescope reaching $\sim$2\,mJy (AB=23.1\,mag) depth of five deep astronomical fields (ELAIS-N1, ELAIS-S1, Lockman, ECDFS and XMMLSS). We use the DR1 release of the SERVS data products which reach a 5$\sigma$ point source limiting magnitudes of 22.48 and 22.07 in S36 and S45, respectively.  

At longer wavelengths, both fields are covered by the $Spitzer$ Wide-area InfraRed Extragalactic \citep[SWIRE,][]{Lonsdale03} survey at 3.6, 4.5, 5.8, 8.0, 24,70 and 160\,$\mu$m (IRAC1234, MIPS123). SWIRE has surveyed $\sim$50\,deg$^{2}$ centred on the same deep fields as SERVS, but to a shallower depth. Here we use the DR5 SWIRE images at 5.8, 8.0, 24 and 70\,$\mu$m, opting to use the deeper SERVS data for IRAC 1 and 2 and $Herschel$ data for 160\,$\mu$m. The DR5 data reaches 5$\sigma$ point source limiting magnitudes of 19.7, 19.96, 18.0 at 13.26\,mag at 5.8, 8.0, 24 and 70\,$\mu$m, respectively. Both SERVS and SWIRE data are obtained from the NASA/IPAC Infrared science archive\footnote{\url{https://irsa.ipac.caltech.edu/frontpage/}}.

\subsection{Far Infrared}

\subsubsection{$Herschel$-PACS Evolutionary Probe (D10)}

The PACS (Photodetector Array Camera and Spectrometer) Evolutionary Probe \cite[PEP,][]{Lutz11} was a $Herschel$ key program covering a number of deep fields at 100 and 160\,$\mu$m, including COSMOS. The PACS instrument has a field of view of 1.75 $\times$ 3.5 arcmin,  and 1.2 and 2.4\,$^{\prime\prime}$  pixel$^{-1}$ resolution at 100 and 160\,$\mu$m, respectively.   

In this paper we use the first PEP data release\footnote{\url{http://www.mpe.mpg.de/ir/Research/PEP/DR1}} which imaged the COSMOS region reaching a 5$\sigma$ point source limiting magnitude of 14.1 and 13.3\,mag at 100 and 160\,$\mu$m, with a PSF FWHM of 7.4 and 11.3\,$^{\prime\prime}$ , respectively.

\subsubsection{$Herschel$-HerMES (D10/D02/D03}

The $Herschel$ Multi-Tiered Extragalactic Survey \citep[HerMES,][]{Oliver12} is a far-IR survey conducted using $Herschel's$ Spectral and Photometric Imaging Receiver
(SPIRE). As part of HerMES, the COSMOS region was observed at 250, 350 and 500\,$\mu$m with a pixel scale 6.0, 8.3 and 12.0\,$^{\prime\prime}$  pixel$^{-1}$ and FWHM of 18.15, 25.15 and 36.3\,$^{\prime\prime}$ , respectively. The XMMLSS and ECDFS regions were also observed by HerMES and cover both PACS 100 and 160\,$\mu$m and SPIRE 250, 350 and 500\,$\mu$m.      

In this work we use the second HerMES data release\footnote{\url{http://hedam.lam.fr/HerMES/}} which reaches 5$\sigma$ point source limiting magnitudes of 12.4, 11.6, 14.1, 14.4\,mag and 13.9 at 100, 160, 250, 350 and 500\,$\mu$m, respectively.  Construction of the HerMES images is described in detail in \cite{Levenson10} and \cite{Viero13}.

Note, for the rest of this paper we will predominantly only show figures and quote numbers for the D10 region. The reason for this is two-fold, firstly the D10 region has superior existing photometric catalogues with which to compare to, and secondly, for ease of description. However, we note that our process is identical in the D02/D03 regions.

\begin{table*}

\begin{center}
\caption{Overview of multi-wavelength data in the DEVILS regions used in this photometric analysis. For further details see sections \ref{sec:D10}. Nominal Depth is the depth quoted by the respective survey, while Measured Depth is the depth to galaxy sources derived from the turn-over point in the galaxy number counts (see Section \ref{sec:counts}) For the common data in D02 and D03 the Measured Depth column displays D02 "/" D03 depths.}
\label{tab:D10photom}
\begin{tabular}{l l l l l l l l l l}
Field &  Common & Facility & Survey & Band & Central & Nominal & Measured & Zero- & Ref   \\
 & Name(s) & & & & Wavelength &  Depth & Depth & point &  \\
  & & & & & ($\mu$m) & (5$\sigma$ AB) & (AB) & (AB) &  \\
\hline
\hline
    
D10 & COSMOS &GALEX& GALEX-DIS & FUV  & 0.154 & 26.0  & 28.42 & 18.82 &  \cite{Zamojski07}  \\
& & & &  NUV & 0.231 & 25.6  & 27.37 & 20.08 & \dittotikz \\
   
& & CFHT & CFHT-COSMOS & u & 0.379 & $>$26.4  & 26.36 &  30.00 & \cite{Capak07}  \\        
   
& & Subaru & HSC-SSP (DUD) & g & 0.474  &27.3  & 26.09 & 27.80 &  \cite{Aihara19} \\ 
& &  &  & r & 0.622  &26.9 & 25.82   & 27.70 & \dittotikz \\ 
& &  & & i & 0.776  &26.7  & 25.09 & 27.60  &  \dittotikz \\  
& &  & & z & 0.893  &26.3  & 25.41 & 27.80  & \dittotikz  \\  
& &  & & y & 1.00    &25.3  & 24.90 & 26.20  &  \dittotikz \\

 &  & VISTA & UltraVISTA & Y & 1.02  & $>$24.7  & 25.46 & 30.00 & \cite{McCracken12} \\
  &  &  & & J &1.26  & $>$24.5  & 25.22 & 30.00  &  \dittotikz \\
&  &  & & H &1.65  & $>$24.1 & 24.98 & 30.00  &\dittotikz\\
&   & & & Ks & 2.16  & $>$24.5  & 24.75 & 30.00 & \dittotikz\\

&  & Spitzer & SPLASH & S36 & 3.53  & 24.9  & 24.66 & 21.58 & \cite{Laigle16}  \\
&  &  & & S45 &4.47  & 24.9   & 24.77 & 21.58 & \dittotikz \\  
&  &   & S-COSMOS  & S58 & 5.68  & 22.4 & 23.86   & 21.58 &  \cite{Sanders07} \\  
&  &   &  & S80 & 7.75  & 22.3 & 23.90   & 21.58 & \dittotikz\\  
&    & &   &  MIPS24 & 23.5  &  19.3 & 18.53   & 20.15 & \dittotikz \\   
&   &  &  &  MIPS70 & 70.4  & 14.2   & - & 17.53 &  \dittotikz \\

&  & $Herschel$ & PEP  & P100 & 98.9  &14.1 & 14.93  & 8.90 & \cite{Lutz11}  \\     
&  &   &  PEP & P160 &156    &13.3  & 14.07 &  8.90 & \dittotikz \\    
&  &   & HerMES  & S250 &250 &14.1 & 14.24 & 11.44$^\dagger$ &  \cite{Oliver12} \\    
&  &   & HerMES & S350 & 350  &14.4 & 14.30 & 11.43$^\dagger$ & \dittotikz \\    
&  &   & HerMES & S500 & 504  & 13.9  & 14.64 & 11.44$^\dagger$ & \dittotikz \\ 

\hline

D02 &  XMMLSS &GALEX& multiple & FUV  & 0.154 & $\sim$25.0  & 27.32/27.04 & 18.82 &  multiple \\
\& D03 & \& ECDFS & & &  NUV & 0.231 & $\sim$25.0   & 26.70/27.02 & 20.08 & \dittotikz \\

&  & VISTA & VIDEO & Y & 1.02  & 24.6  & 24.65/24.82  & 30.00 &  \cite{Jarvis13}   \\
&  &  & & J &1.26  & 24.5  & 24.34/24.45 & 30.00 &  \dittotikz \\
&  &  & & H &1.65  & 24.0  & 24.05/24.14 & 30.00 & \dittotikz\\
&   & & & Ks & 2.16  & 23.5  & 23.84/23.9 & 30.00 & \dittotikz\\

&  & Spitzer & SERVS & S36 & 3.53  & 22.48 & 23.45/23.97 & 21.58 &  \cite{Mauduit12} \\
&  &  &  & S45 &4.47  & 22.07 & 23.08/23.98  & 21.58  & \dittotikz \\  
&  &  & SWIRE  & S58 & 5.68  & 19.70 & 22.31/22.64  & 21.58 & \cite{Lonsdale03} \\  
&  &   &  & S80 & 7.75  & 19.96 & 22.33/22.70 & 21.58 & \dittotikz\\  
&    & &   &  MIPS24 & 23.5  & 18.00 & 18.26/18.21  & 20.15 & \dittotikz \\   
&   &  &  &  MIPS70 & 70.4  & 13.26 & -/-  & 17.53 &  \dittotikz \\

&  & $Herschel$ & HerMES  & P100 & 98.9  &12.4 &13.77/13.67 & 8.90 & \cite{Oliver12} \\     
&  &   &  HerMES & P160 &156    &11.6 &13.07/12.98 & 8.90 & \dittotikz \\    
&  &   & HerMES  & S250 &250 &14.1& 13.99/13.99 & 11.44$^\dagger$ &  \dittotikz \\    
&  &   & HerMES & S350 & 350  &14.4& 14.03/14.07 & 11.43$^\dagger$ & \dittotikz \\    
&  &   & HerMES & S500 & 504  & 13.9& 14.27/14.29  & 11.44$^\dagger$ & \dittotikz \\ 

\hline
   
D02 & XMMLSS & CFHT & CFHT-LS & u & 0.379 & $>$25.3  & 25.68 & 30.00 & CFHTLS  \\        
   
 &  & Subaru & HSC-SSP (DUD) & g & 0.474  &27.3  & 25.50 & 27.80 &  \cite{Aihara19} \\ 
& &  &  & r & 0.622  &26.9  & 25.19 & 27.70 & \dittotikz \\ 
& &  & & i & 0.776  &26.7  & 24.94 & 27.60 &  \dittotikz \\  
& &  & & z & 0.893  &26.3  & 24.75 & 27.80 & \dittotikz  \\  
& &  & & y & 1.00    &25.3  &  23.97 &  26.20 &  \dittotikz \\     

\hline

D03 & ECDFS  & VST & VOICE & u & 0.379  & 25.3  & 25.76  &30.00 &  \cite{Vaccari16} \\ 
& &  &  & g & 0.474  &26.4  & 25.58 & 30.00 &  \dittotikz \\ 
& &  &  & r & 0.622  &26.1  & 25.33 & 30.00 & \dittotikz \\ 
& &  & & i & 0.776  &25.2  &  24.89 & 30.00 &  \dittotikz \\  
  
 &   & VISTA & VIDEO & Z & 0.893 & 25.7 & 25.65 & 30.00 &  \cite{Jarvis13}   \\

\vspace{2mm}

\end{tabular}

$^\dagger$ SPIRE maps are in units of Jansky per Beam and to generate these zero-points we have added a factor $2.5 \log_{10}(B/N^2)$ where B is the beam size given as 373, 717, and 1494 sq $^{\prime\prime}$  and N is the pixel size given as 6.0, 8.3 and 12 $^{\prime\prime}$  in 250, 350 and 500$\mu$m, respectively.
\end{center}
\end{table*}

\section{The need for a new COSMOS (D10) photometric catalogue}

There have been a number of multi-wavelength photometric catalogues produced in the COSMOS region over the past two decades, such as those outlined in \cite{Capak07},  \cite{Laigle16} and for the central 1\,deg$^2$ discussed in \cite{Andrews17} and \cite{Driver18}. However, as noted in Section~\ref{sec:into}, these catalogues have relied on software that causes some known issues when deriving total photometry catalogues across a broad range of source types and sizes. They have relied on table-matching of photometry derived by different teams using varied techniques, and/or do not contain the most up-to-date imaging in the region (such as UltraVISTA DR4 and HSC). As such, we now derive a new photometry catalogue which uses a state-of-the-art photometry code, designed to overcome issues in current catalogues. This enables us to measure photometry consistently from the UV-FIR with a single package on homogenised data, and include the most up-to-date imaging campaigns in this region. 

Through the process of producing the G10/COSMOS catalogue outlined in \cite{Andrews17} and, using a similar method, the GAMA photometric catalogues \citep[see][]{Driver16, Wright16}, we identified a number of issues when determining total source photometry. Briefly, within both of these surveys initial source detection was undertaken using \textsc{Source Extractor} \citep{Bertin96} to define elliptical apertures. However, it was quickly identified that these apertures are prone to some issues, such as bright galaxy fragmentation into a number of sub-apertures, and highly erroneous apertures, largely due to the aperture following an isophotal bridge to loop around a nearby bright star (these issues mostly arise because of the difficulty in simultaneously measuring fluxes for both very bright and very faint sources). For details of these issues see \cite{Robotham18} and \cite{Davies18}. These errant apertures led to a significant fraction ($\sim$10\%) of sources having poorly measured photometry. To overcome these issues in the G10/COSMOS sample and in GAMA \citep[see][respectively]{Andrews17, Wright16}, we undertook a systematic, and time-consuming, visual inspection and manual fixing of the errant apertures both within the teams and via a citizen science project (AstroQuest \footnote{\url{https://astroquest.net.au/}}).

Motivated by the level of manual fixing required in the GAMA and G10/COSMOS data, we developed a new source finding and photometry code, \textsc{ProFound} \citep[][]{Robotham18}. Briefly, the \textsc{ProFound} code has three fundamental differences to its approach over \textsc{Source Extractor}. Firstly, \textsc{ProFound} does not measure source photometry based on elliptical apertures, but identifies and retains the source isophote, allowing for both simple and irregular shapes (called segments). Secondly, these initial segments are dilated until the flux converges providing a pseudo-total flux. Thirdly, \textsc{ProFound} applies a watershed deblending approach where during the dilation, segments cannot overlap ($i.e.$, all the flux in any one pixel is allocated to one galaxy only). This is different to \textsc{Source Extractor}, where hierarchical or nested-deblending allows for overlapping elliptical apertures and the possibility of flux being double counted. While the merits and failings of these approaches can be argued - through numerous tests we have found that the \textsc{ProFound} approach is far less likely to go catastrophically wrong and produces more robust photometric measurements.  For full details on \textsc{ProFound} and numerous tests, we refer the reader to the code description paper \cite{Robotham18} and four recent applications to the DEVILS input catalogue \citep{Davies18}, GAMA photometry \citep{Bellstedt20}, the VISTA VIKING datasets \citep{Koushan21} and evolution of the high redshift stellar mass function \citep{Adams21}. 

In this work we apply the \textsc{ProFound} software to the state-of-the-art UV-FIR photometric data in the DEVILS regions. This is done in two modes, firstly using the vanilla \textsc{ProFound} multi-band approach to the UV-MIR data (described in Section \ref{sec:vanilla}), and secondly applying a new Bayesian technique to extract photometry in the unresolved MIR-FIR regime (described in Section \ref{sec:FIRphotom}). A similar approach has been applied in our shallower updated GAMA photometry, as outlined in \cite{Bellstedt20}.  We note that we do not present new photometric measurements to the medium- and narrow-band Subaru Suprime-cam data in the D10 (COSMOS) region. This is largely due to the fact that we are aiming for consistent photometric catalogues across all of the DEVILS fields. As comparable data does not exist in the D02 and D03 regions, we do not include it here. We also note that in our SED fitting analysis of the D10 sample, presented in \cite{Thorne21}, we find little improvement/changes to measurements of stellar mass and star-formation rate when including the medium- and narrow-band data. Where this data is of significant benefit is in the derivation of photometric redshifts where finer wavelength sampling allows for greater redshift precision. However, we do not derive new photometric redshifts here as the COSMOS2015 sample already obtains robust photometric redshift measurements for these systems using the medium-band and narrow-band, and it is unlikely that we could significantly improve upon this.

\begin{figure}
\begin{center}
\includegraphics[scale=0.58]{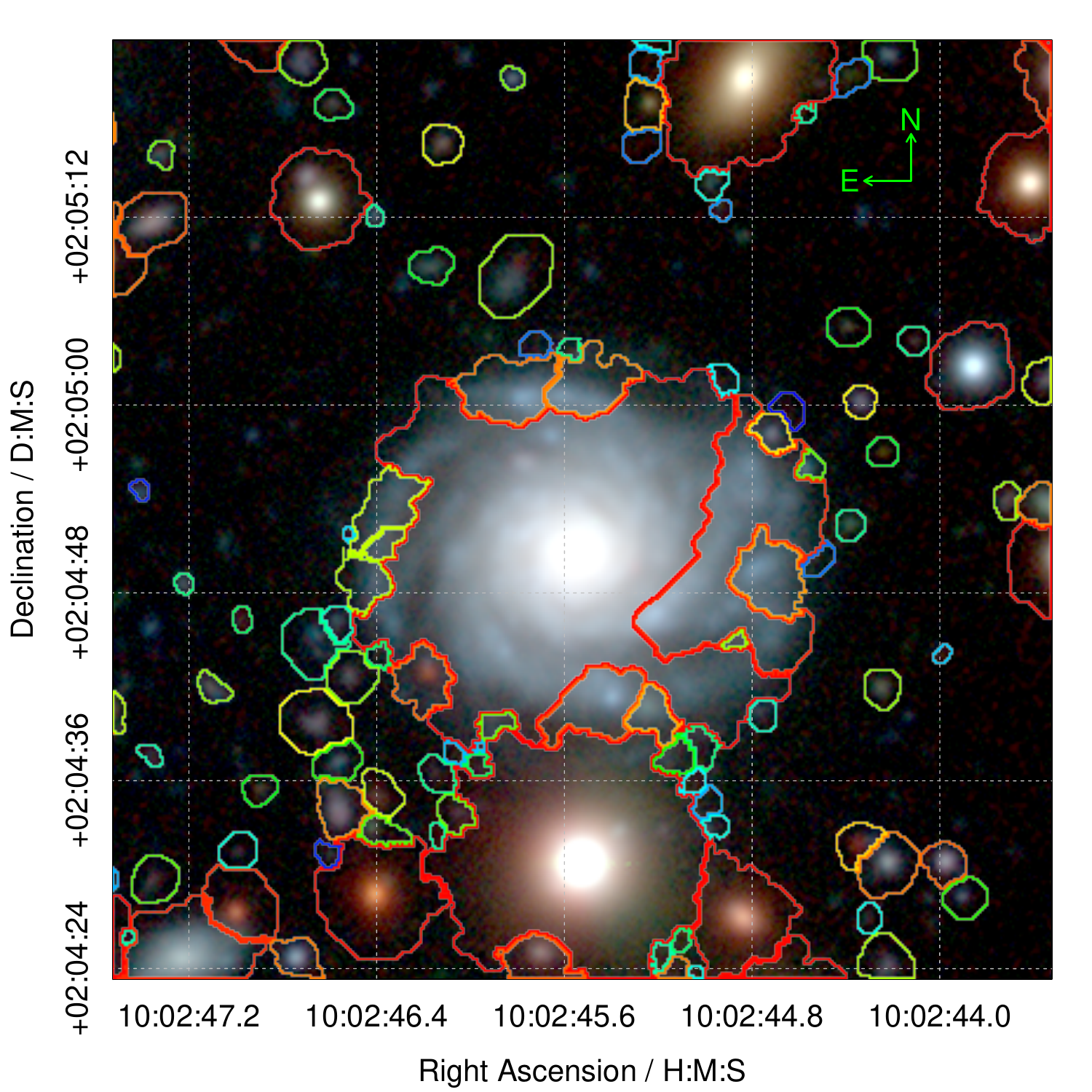}
\includegraphics[scale = 0.58]{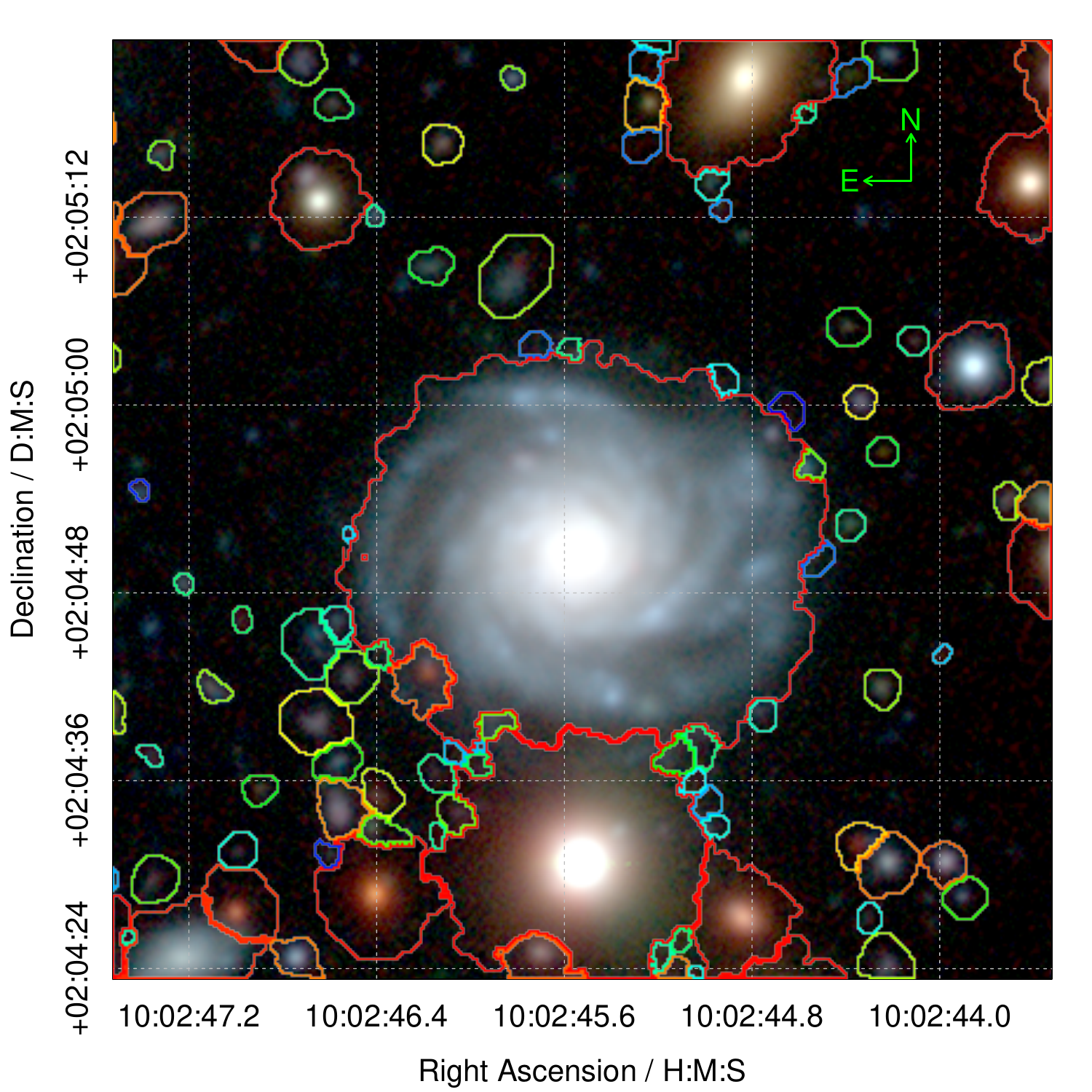}
\caption{Example of a fragmented galaxy that has been rebuilt. Image shows the \textsc{ProFound} segmentation map for a galaxy that has been split into multiple segments. Segments are bounded by colours assigned by the segments flux from blue (faint) to red (bright). The top panel shows the galaxy before regrouping and the lower panel shows the segments after regrouping.}
\label{fig:SegRegroup}
\end{center}
\end{figure}

\begin{figure}
\begin{center}
\includegraphics[scale=0.48]{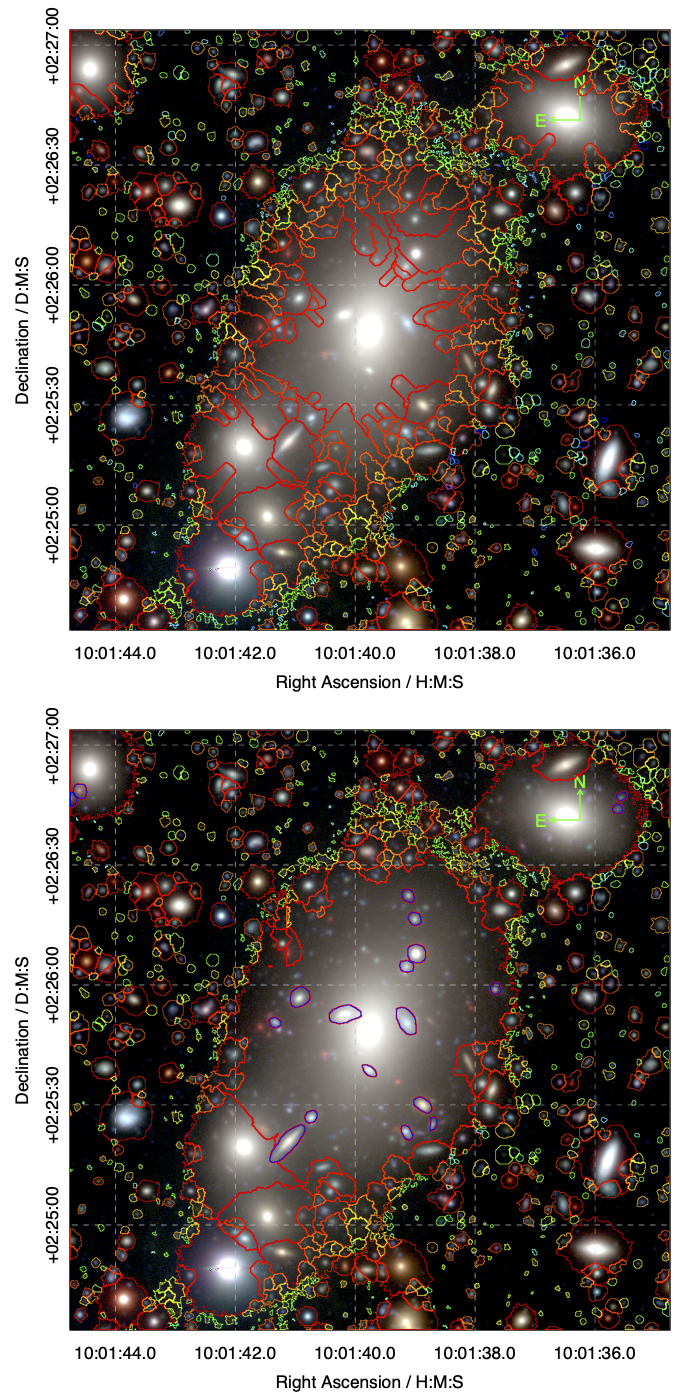}
\caption{Example of a complex region that required manual intervention to fix segments. The top panel shows the region before manual fixing and the lower panel shows the final segments after manual fixing. Segments which have been manually added, after some regrouping, are bordered by dark blue and red lines and largely fall in the centre of the complex.}
\label{fig:ungroup}
\end{center}
\end{figure}

\section{UV-MIR Photometric Measurements}
\label{sec:vanilla}
\subsection{\textsc{ProFound} Source Detection, Rebuilding Fragmented Galaxies and Splitting Merged Systems}
\label{sec:group}

To use \textsc{ProFound} to first identify target sources, we produce an inverse variance weighted stacked image from the UltraVISTA (D10) and VIDEO (D02/D03) Y, J, and H band images. Note that we do not use Ks because the UltraVISTA Ks image does not cover the same spatial extent as the other bands in D10 and has a small region of the DEVILS field missing. In D02/D03 we use the VIDEO Y, J, and H images only for consistency. This stacking allows us to maximise the depth in the initial detection image and remove any artefacts associated with a single band. This is a similar process to that outlined in \cite{Davies18} for the DEVILS input catalogues. This stacking process is handled internally by \textsc{ProFound}  using the \textsc{profound-MultiBand} command, allowing the user to specify one or more bands for source detection. If more than one band is supplied the images are inverse variance weighted stacked.  We note here that one could opt to add additional bands in our stacking procedure ($i.e.$ at bluer wavelengths) to both increase the source identification depth and potentially include bluer sources which are undetected in the NIR. However, we do not do this for a number of reasons: i) We wish to have a DEVILS survey selection band which is close to a single band measurement of stellar mass ($i.e. >$4000\AA\ in the rest-frame) for all galaxies to $z\sim1$. This necessitates the NIR bands. We use our stack for source detection and then Y-band magnitude to select our spectroscopic sample. As such, we would not target faint blue galaxies that are undetected in the Y-band. ii) The NIR data used here is already far deeper than that which we would use for science. So the faint blue sources that are undetected in the NIR are unlikely to be included in our science samples. This is especially true when performing a stellar mass-based selection \citep[$i.e.$ as in][]{Thorne21}, and which will be done for all DEVILS core science, and iii) we wish to have consistency in the detection filters across all DEVILS fields. As noted previously, Z-band data does not exist over all of the D03 region and the I-band data comes from different facilities (HSC and VST) in different fields. 

Once the stack is produced, \textsc{ProFound} then identifies all sources of flux in this stacked image producing a segmentation map designed to separate flux from individual sources \cite[see][ for further details]{Robotham18}.  \textsc{ProFound} produces two different segmentation maps, first a colour-optimised segmentation that covers the central high-signal-to-noise region of sources prior to the \textsc{ProFound} segment dilation process, and a total-optimised segmentation that dilates the colour-optimised segments in an attempt to extract total flux measurements associated with a source. The colour-optimised segments are essentially tight detection segments (akin to a small fixed-size aperture - but tuned to the shape of the source's central distribution of light), which do not encompass all of galaxy flux. This will best-represent the overall SED shape, but not normalisation. This means that the colour-optimised segments are best used for scientific analyses that do not require total flux measurements ($i.e.$ photometric redshifts), while total-optimised segments are best used for the converse ($i.e.$ stellar mass and SFR measurements, etc).  Both of these segmentation maps are used in our analysis. For running our \textsc{ProFound} source detection we use the  Y, J, H band images trimmed to the DEVILS spectroscopic region, but with a 0.1\,deg$^2$ buffer to minimise any issues with sources that overlap the edge of the field. Finally, due to memory issues, \textsc{ProFound} can not be run over a full DEVILS field in a single process. To overcome this, we split the field in to a number of overlapping subregions and undertake our photometric detection/measurements in each region independently (8 regions in D02 and also 4 regions in D10 and D03). These are later combined in a process described in Section \ref{sec:multi}. For example, in total \textsc{ProFound} identifies $\sim951,000$ unique segments across our four sub-regions in D10.       

While the \textsc{ProFound} segmentation map successfully identifies individual sources and robustly maps their extent, a known problem with most automated source finding algorithms (including \textsc{ProFound} --- but to a lesser degree than $e.g.$ SExtractor) is the fragmentation of bright galaxies. This problem has been explored in the derivation of new photometry in GAMA \citep{Bellstedt20} and a number of enhancements to \textsc{ProFound} were made to assist with fixing this issue. The first two enhancements were the introduction of the \textsc{ProFound-} \textsc{reltol} and \textsc{cliptol} parameters in the source finding mode. \textsc{Reltol} allows \textsc{ProFound} to keep very extended and flocculent spiral galaxies intact, while having little negative impact on the fainter source deblending that parameters will tend to be optimised for (since this is where almost all of our survey sources exist). \textsc{Cliptol} specifies the saddle point flux above which segments are always merged, regardless of competing criteria. This is useful for very bright objects with complex image artefacts ($e.g.$ in the regions around bright stars), and allows for proper reconstruction, which might otherwise be hugely fragmented into many small segments. We do however utilise a bright star mask (see Section \ref{sec:mask}), which means that the resulting segments around bright stars are somewhat cosmetic. These new parameters are described in more detail in \cite{Bellstedt20}. 

Despite the inclusion of these new \textsc{ProFound} parameters, some low-level fragmentation of bright galaxies still occurs. To fix this, and regroup segments belonging to the same bright object, an in-house interactive tool was developed that allows users to view a thumbnail of an object and click on segments to be regrouped. This tool is available through the \texttt{profoundSegimFix} function within \textsc{ProFound}, and is applied to our total-photometry segments.

As part of its output, \textsc{ProFound} provides information on $group$ statistics, where a group is defined as  touching segments in a $cluster$ of segments around a source, or multiple sources \cite[see][]{Robotham18}. These $group$ statistics allow us to easily identify potentially fragmented galaxies in our segmentation maps. Rather than visually inspecting every group, we determine the groups that need to be inspected based on the following criteria: \\

\noindent i) The group of touching segments must contain at least three segments (this  mean that there might be cases where a single galaxy has been split in two and it would not be selected as an object to fix, but because the field is so deep and there are lots of segments there are very few instances where this is the case). \\

\noindent ii) We only inspect groups with a total group magnitude (across the three stacked detection bands) between 12 and 20.5\,mag to reduce the number of groups that needed to be inspected. \\

\noindent iii) We also only inspect groups where the brightest pixel is at least 500 pixels away from the edge of the frame (this removes groups in our sub-region overlap areas).\\

\noindent  iv) Finally, we also exclude masked objects such as stars and artefacts from the visual re-grouping process, because the majority of touching segments reside in the ghosting surrounding stars. \\

For example, in the D10 region this resulted in 12,354 groups to visually inspect. Of these, 1,272 required some manual intervention (10.3 per cent).  An example of an object whose segments have been regrouped in this way can be seen in Figure \ref{fig:SegRegroup}, where the bottom panel shows the resulting segmentation map after merging. For each of the four input sub-regions, one output file was produced that recorded which segments (as determined by the detection phase of \textsc{ProFound}) needed to be regrouped.  

In addition to the segment regrouping, upon visual inspection of the \textsc{ProFound} outputs, we found that a small number of large, highly clustered sources were merged into a single \textsc{ProFound} segment - and thus required $ungrouping$.  The \texttt{profoundSegimFix} function also allows for segments to be manually created using a GUI tool where the new segments are drawn onto an image. In order to fix all erroneously grouped regions, we first ignore stars, artefacts and masked regions (see below) and then select the superset of segments which are either: i) in the top 100 Y-band brightest segments in their subregion, ii) in the top 100 Y-band largest segments in their subregion, and iii) objects where the brightest Y-band pixel within a segment in $>4$\,$^{\prime\prime}$  away from the flux-weighted central pixel of the segment (both provided by \textsc{ProFound}). This process visually selected all erroneously grouped regions in our subregions and lead to $\sim$30-60 regions to visually inspect per subregion ($i.e.$ there is significant overlap in the selections). Figure \ref{fig:ungroup} shows one of the extremely complex regions in the D10 field that requires both segment regrouping and ungrouping. The top panel show the \textsc{ProFound} segmentation map prior to manual fixing, while the bottom panel shows the region following manual fixing. Segments that have been manually included in the ungrouping process are shown bounded by blue/red lines.  We note here, that while this process requires manual intervention, it is far less time-consuming that the manual fixing process discussed in \cite{Andrews17} and \cite{Wright16}.

Once complex regions have been both manually regrouped and ungrouped, where appropriate, we generate a new segmentation map for each subregion and use this for the remainder of the photometry pipeline. 

\begin{figure*}
\begin{center}
\includegraphics[scale=0.52]{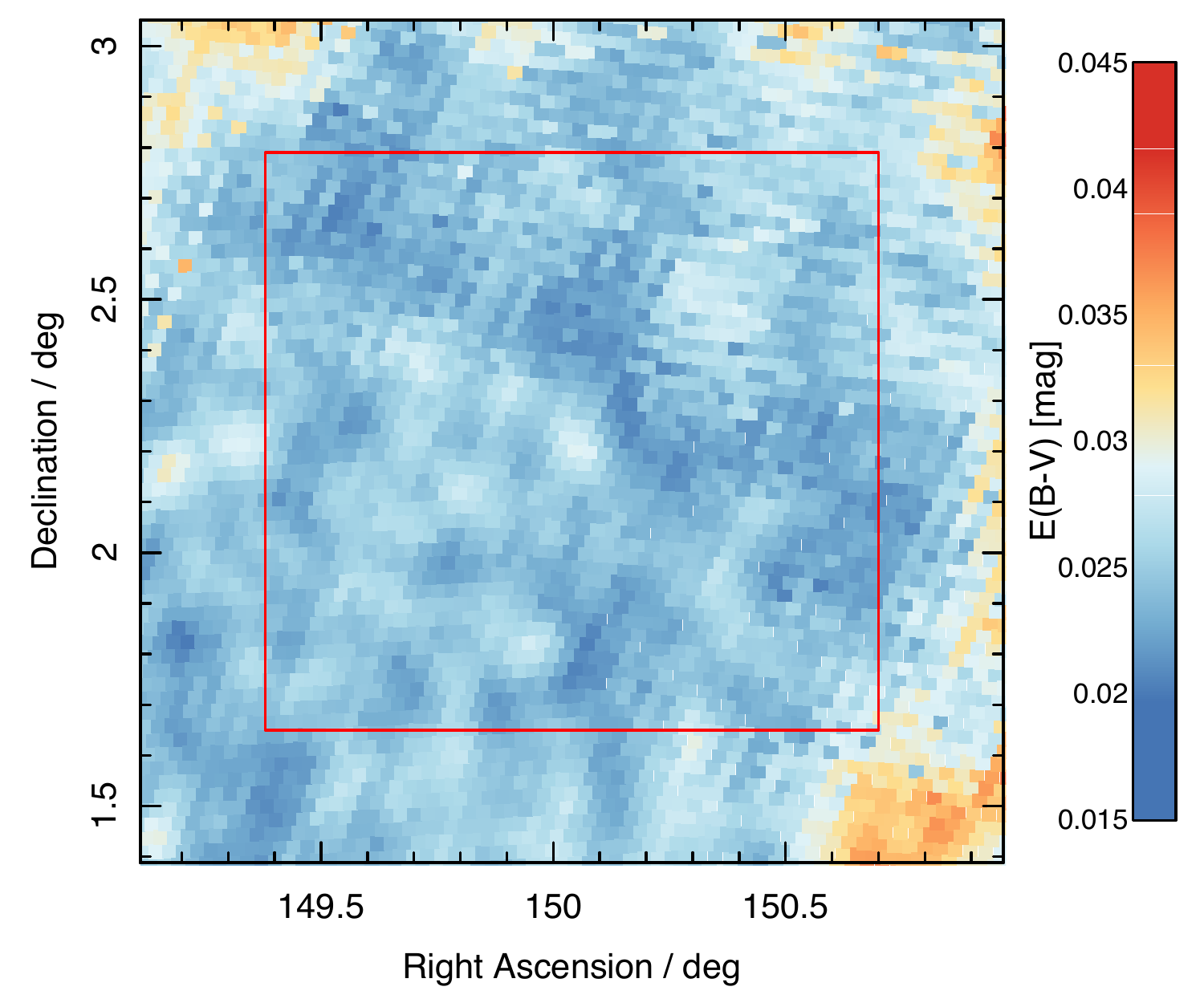}
\includegraphics[scale=0.52]{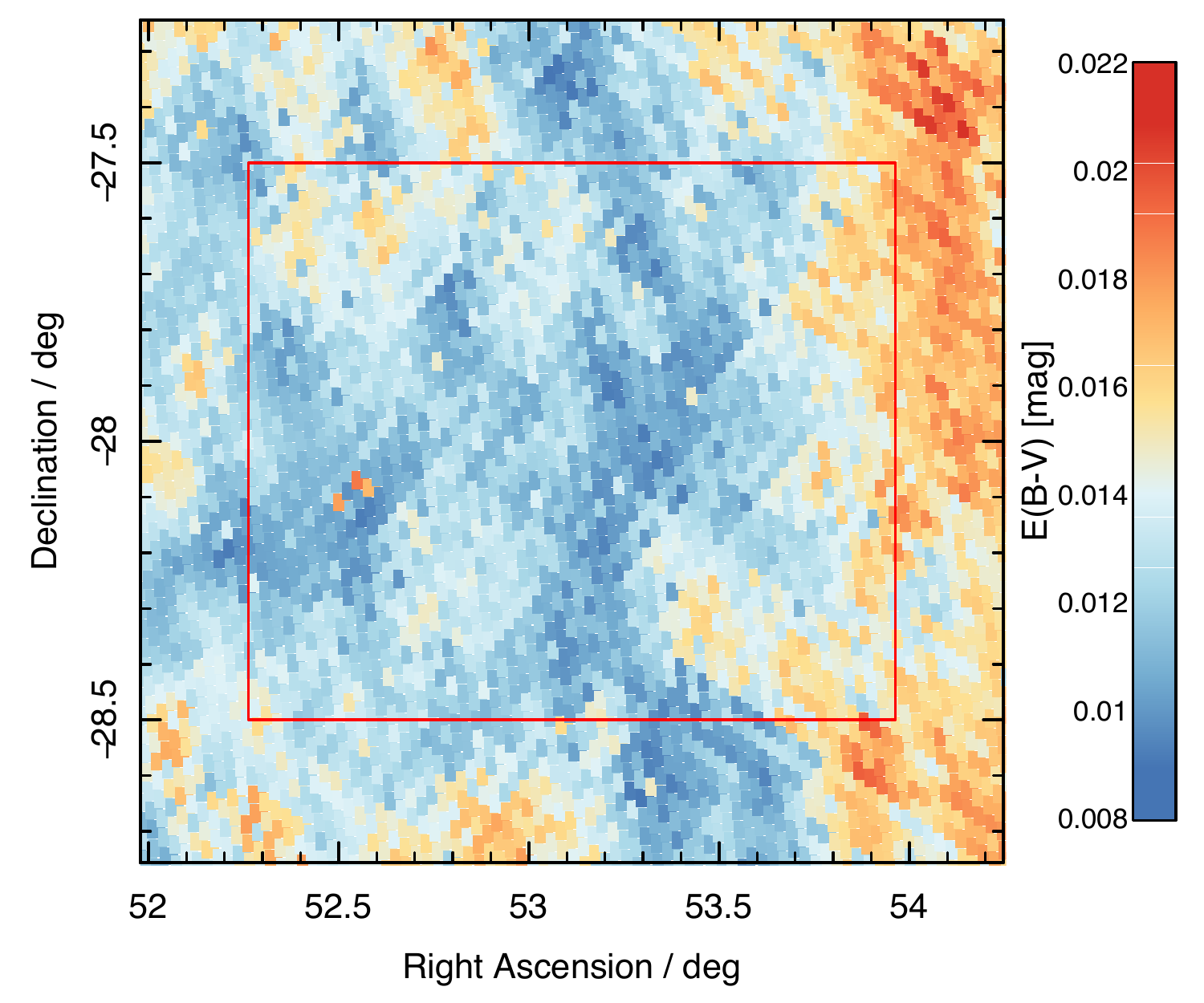}
\includegraphics[scale=0.7]{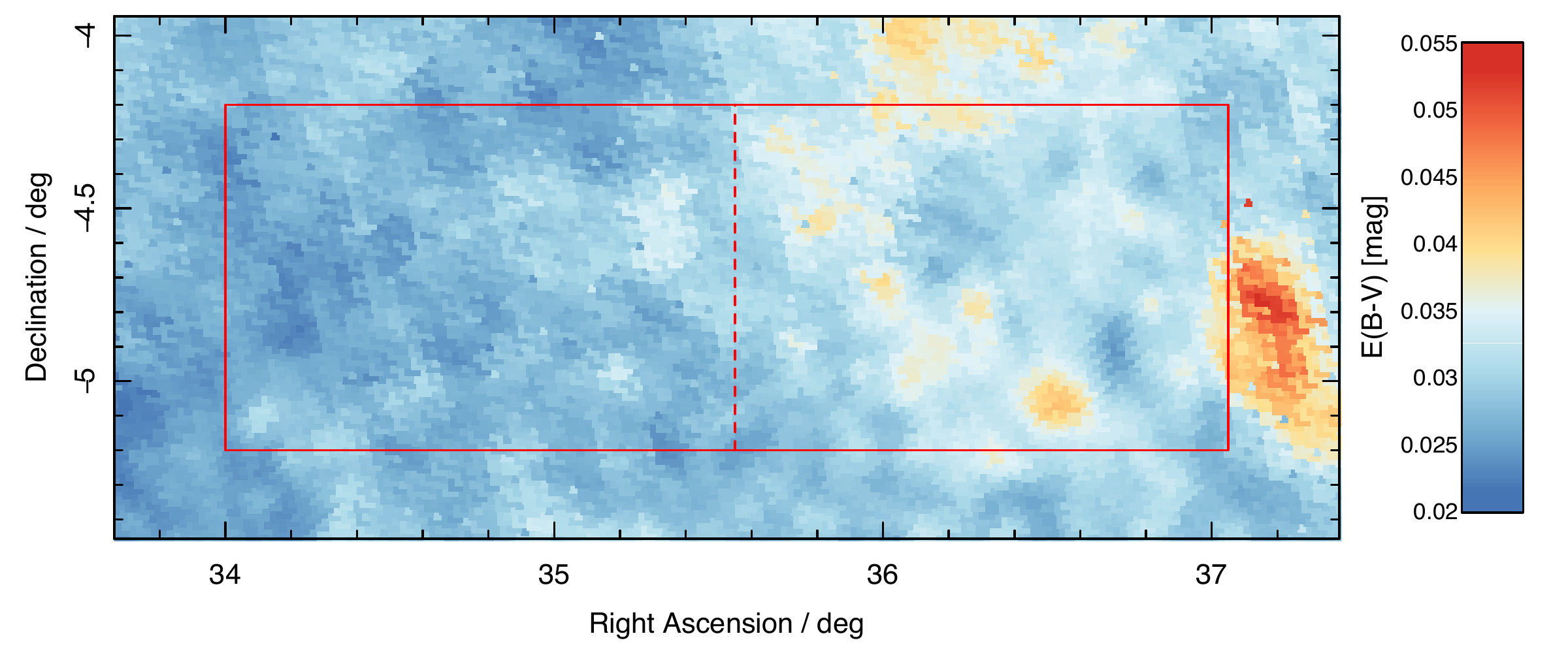}
\caption{Galactic Extinction in the DEVILS regions based on the Planck E(B-V) maps. Top left = D10, Top right = D03, Bottom = D02. Note that we use a different colour bar scaling for each field as the extinction is low, but significantly different in each region. Thus if the same scale is used, all detail would be removed.  The dashed vertical line in D02 displays the separation between the D02A (left) and D02B (right) regions. }
\label{fig:extinction}
\end{center}
\end{figure*}

\subsection{Multiband Photometry with \textsc{ProFound} and Sub-Region Combination}
\label{sec:multi}

After the initial segmentation map has been defined and manual fixing applied in each of our subregions, these segments are then used to measure both total- and colour-optimised photometry in the UV-MIR bands (GALEX-FUV to $Spitzer$-IRAC 8.0\,$\mu$m). Our initial process is only applied to the UV-MIR, where the pixel-scale/seeing are comparable and source blending/confusion between bands is low. The process for measuring our photometry at longer wavelengths using PSF fitting is described in Section \ref{sec:FIRphotom}. Note that in the IRAC bands it is somewhat ambiguous as to which method would be preferred, ProFound segmentation versus PSF extraction. As such, we initially perform both approaches for our IRAC photometry and compare to existing MIR measurements for these galaxies. We find that for all of the IRAC bands our ProFound segmentation approach provides closer consistency between our new photometry and existing catalogues. Hence, we opt to switch between methods at the IRAC24/MIPS24 boundary. As is shown later, this choice produces tighter MIR colours that the existing photometry catalogues. 

Prior to photometric measurements, all imaging bands must first be transformed to the same pixel scale and field size. To do this we use the \textsc{SWARP} package, matching all bands to the same pixel scale and area as the stacked Y, J, and H band image (0.15\,$^{\prime\prime}$  pixel$^{-1}$ for UltraVISTA in D10 and 0.2\,$^{\prime\prime}$  pixel$^{-1}$ for VIDEO in D02/D03). \textsc{ProFound} then initially applies the colour-segmentation maps (central high signal-to-noise region) to each band. As the PSF of our imaging varies significantly from band to band, we can not use fixed segments for each filter (as we would miss flux in the lower resolution data). As such, \textsc{ProFound} takes the colour-optimised segmentation map as a starting, source detection, point and then allows the segments to further dilate in an attempt to encompass more flux until convergence is reached \cite[see][for details]{Robotham18}. This process accounts for the varying PSF, without having to apply detailed modelling of the PSF shape in each band.  \textsc{ProFound} then outputs both colour and total flux measurements, errors and segment statistics (such a radii, number of pixels, sky background, etc) in each band.            

Following measurement in each band, catalogues from our sub-regions are combined to produce a photometry catalogue for the full DEVILS field. To do this, we combine all sub-regions to a single catalogue, and remove duplicates in the overlapping regions, retaining the source which is closest to its sub-region field centre. We then trim the catalogue to the extent of the DEVILS spectroscopic field. As an example, this results in $\sim777,000$ unique segments in the D10 region.

\subsection{Extinction Corrections}

To correct our measured magnitudes for Galactic extinction we use the Planck $E(B-V)$ map\footnote{HFI CompMap ThermalDustModel 2048 R1.20.fits, \url{https://irsa.ipac.caltech.edu/data/Planck/release_1/all- sky- maps/previews/HFI_CompMap_ThermalDustModel_2048_ R1.20/index.html}}, converting HEALpix values to RA and Dec positions across the field. Figure \ref{fig:extinction} displays the Planck $E(B-V)$ values across the DEVILS regions. For each \textsc{ProFound} source we then identify the closest Planck $E(B-V)$ pixel assigning the corresponding extinction value. We then correct all magnitudes, surface brightnesses and fluxes based on this extinction value. We determine the attenuation correction for each band in the traditional manner ($i.e.$ $A_{x} =[A_{x}/E(B-V)] \times E(B-V))$ using the extinction coefficients listed in Table \ref{tab:ext} \citep[which implicitly use the Galactic extinction law from][]{Schlafly11}. Note that we do not apply any corrections long-ward of the IRAC bands, where emission then arises from dust itself, and would be subtracted as part of the background.

  \begin{table}

\begin{center}
\caption{Attenuation values used in conjunction with Planck $E(B - V )$ map, consistent with the local GAMA analysis.   }
\label{tab:ext}
\begin{tabular}{c c}
Filter &  $[A_{x}/E(B-V)]$   \\
\hline
\hline
FUV & 8.24152 \\
NUV & 8.20733\\
u & 4.81139\\
g&3.66469\\
r&2.65460\\
u&2.07472\\
Z&1.55222\\
Y&1.21291\\
J&0.87624\\
H&0.56580\\
Ks&0.36888\\
S36&0.20124\\
S45&0.13977 \\   
S58& 0.10094 \\
S80& 0.09818 \\

\end{tabular}

\end{center}
\end{table}

\begin{figure*}
\begin{center}
\includegraphics[scale=0.55]{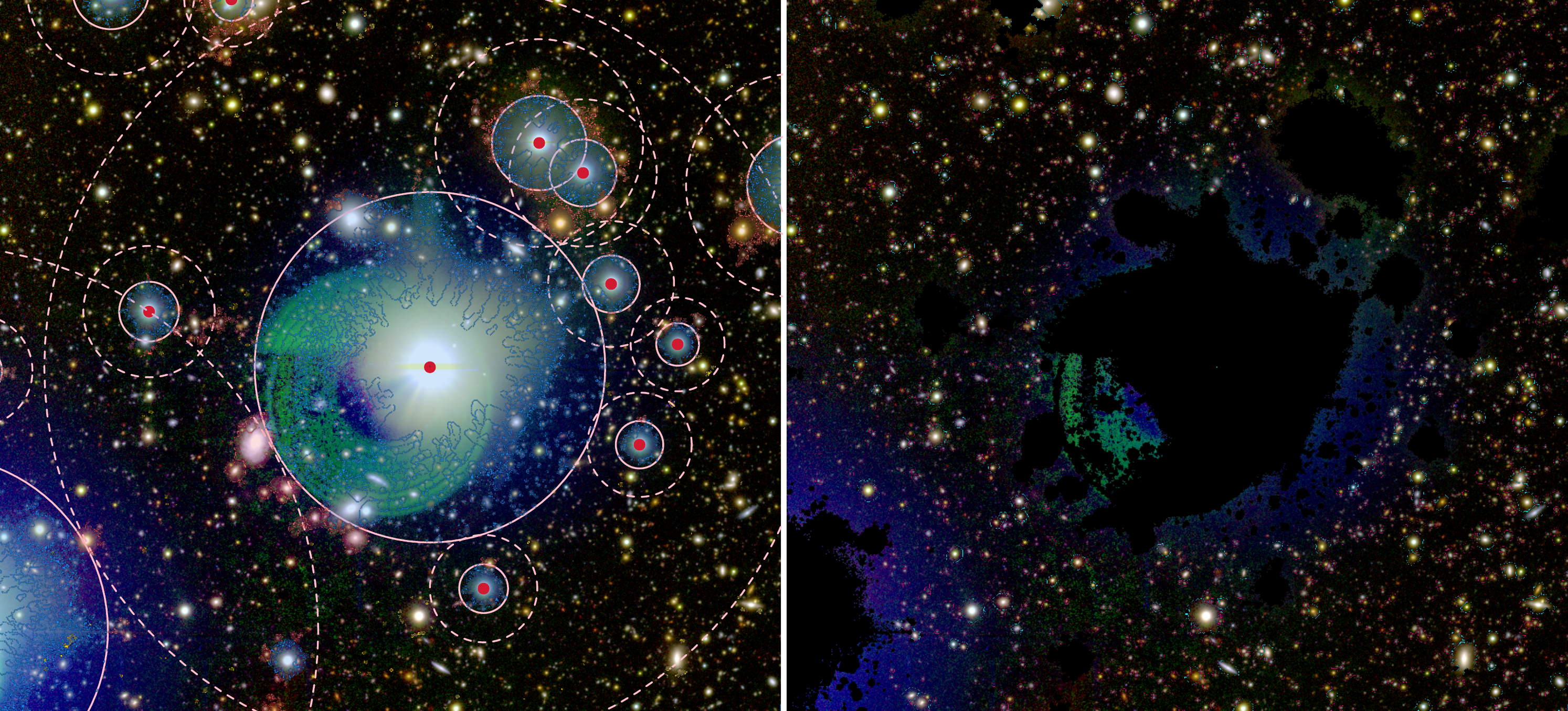}
\vspace{0.5mm}

\includegraphics[scale=0.55]{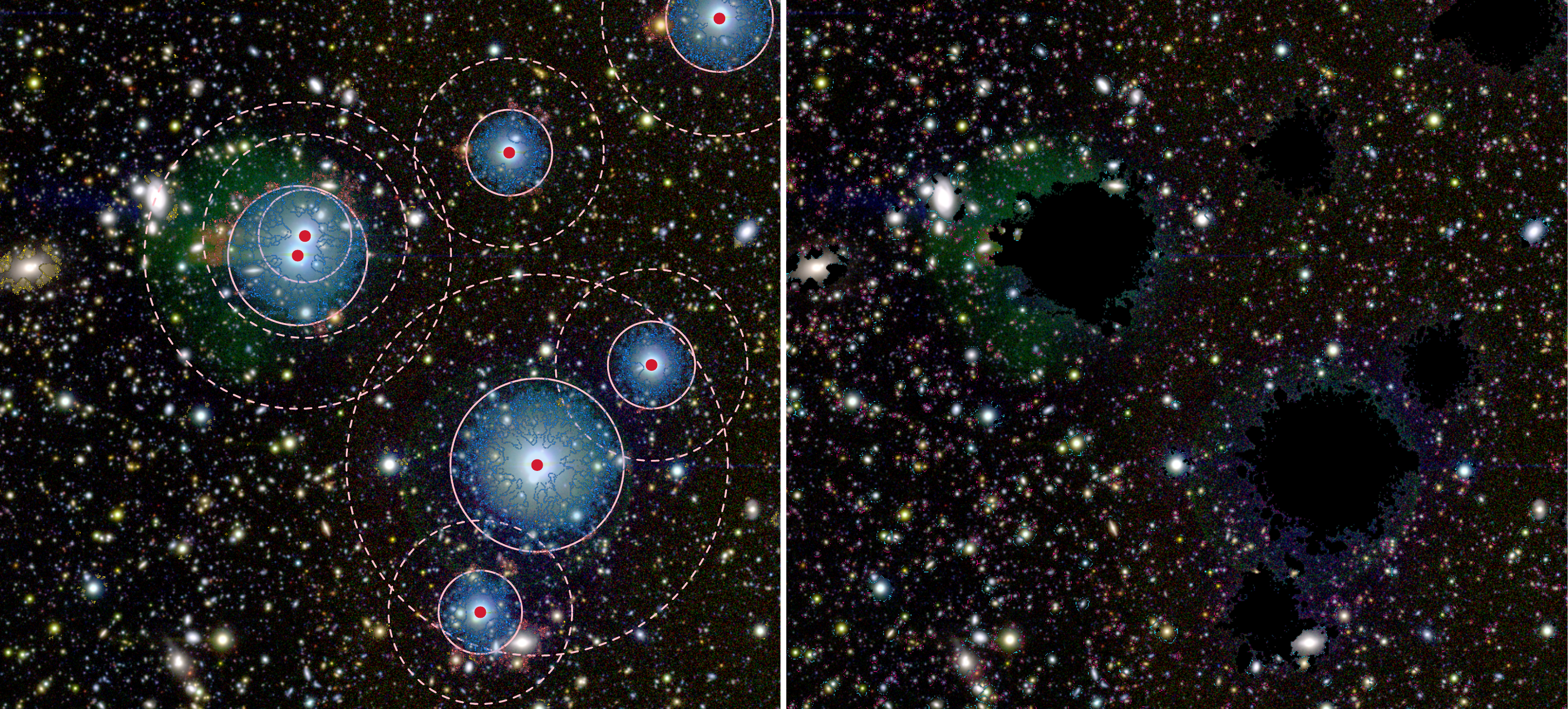}
\caption{Example of the masking procedure overlaid on a H (VISTA), Y (VISTA), r (HSC) RGB image (note that only the H and Y bands are used in our source detection stacked image, so blue extended emission does not contribute to our source detection). Left: \textsc{ProFound} segments are either masked or identified as artefacts using three different methods. Firstly a GAIA star mask is used to define an exclusion radius around bright stars (pink solid circle). This radius is scaled to the GAIA g-band magnitude (see Section \ref{sec:mask}). All segments within this radius are flagged as being masked (blue-segments). Secondly, we define a region of 2.2 $\times$ the GAIA radius (pink dashed circle). Segments within this region that are contained within the same \textsc{ProFound} group as the central star and are fainter than the DEVILS spectroscopic limit (Y$>$21.2) are also flagged as masked (red-segments). Finally, we also identify additional artefacts using the procedure outlined in Section \ref{sec:class} - which are coloured orange. Right: The results of masking process. Here all masked segments are coloured black. We highlight all Y$<$24.0\,mag segments in magenta, and Y$<$21.2\,mag (the DEVILS spectroscopic limit) segments in blue, showing that the masking process only retains segments where robust photometry can be derived.}
\label{fig:masking}
\end{center}
\end{figure*}

\subsection{Masking}
\label{sec:mask}

As can be seen in the top left panel of Figures \ref{fig:masking} (and Figures \ref{fig:images}, \ref{fig:imagesD02} and \ref{fig:imagesD03}),
 our input images have significant issues with ghosting around bright stars. This ghosting is dependent on both position within the focal plane and the brightness of individual stars. In \cite{Davies18} we developed a method for masking bright stars for the DEVILS input catalogue. However, in this work we have improved our masking procedure to better exclude erroneous sources near bright stars and minimise the area lost to our mask. This process is carried out in a number of stages, which are visually described in Figure \ref{fig:masking}. 

Initially, following the method outlined in \cite{Bellstedt20} we select all GAIA DR2 stars with GAIA g-band magnitude brighter than 15.75\,mag in the DEVILS regions, and define a star exclusion radius which is scaled to the GAIA g-band magnitude, with: 

\begin{equation}
r_{star}[\mathrm{\prime}]=10^{(1.795-0.158g_{\mathrm{GAIA}})} \, \, \mathrm{and} \,[r<1.85^{\prime}, g_{\mathrm{GAIA}}<15.75\,mag]
  \end{equation}  

However, unlike \cite{Bellstedt20} who use a large conservative radius, our radius is tuned to $just$ encompass the ghosting for all stars in our field, excluding the minimum region possible (solid pink circles in Figure \ref{fig:masking}). All \textsc{ProFound} segments within this minimum radius are classified as masked (blue segments in Figure \ref{fig:masking}). While this masks out the majority of segments in the ghosted region, it does miss some outlying faint segments caused by the ghost. Following this, we then match the GAIA DR2 source positions to our \textsc{ProFound} segmentation maps, finding the closest segment centre to the GAIA position (red dots in Figure \ref{fig:masking}) and defining this as the central stellar segment. As noted previously, \textsc{ProFound} fragments very bright sources, and this also occurs for the bright stars. As such, this central segment does not encompass all of the star's flux (or the ghost). We therefore select all segments which are within a $group$ (see Section \ref{sec:group}) with the central GAIA-matched segment and also within $2.2 \times r_{\mathrm{star}}$ (dashed pink circles in Figure \ref{fig:masking}). Segments that match this criteria but are fainter than the DEVILS spectroscopic limit are also masked (red segments in Figure \ref{fig:masking}). Finally, we class all segments defined as artefacts in the process outlined in Section \ref{sec:class} as being masked. The right panels of Figure \ref{fig:masking} displays two example regions with all masked segments coloured black, showing that our process effectively removes all erroneous sources of flux in segments across the region. While some remnant of the stellar halo can still be seen, this is not associated with any sources. We then define our masked region as the area contained in masked segment areas. This process results in $\sim$1-3\% of each field being masked, with our final unmasked areas given in Table \ref{tab:mask}. For comparison \cite{Bellstedt20} report 5.4\% of the GAMA area is masked using their process.  Note that the masked percentage within D10 is larger as the deep UltraVISTA data has larger haloes around bright stars (see Y-band panels in Figures \ref{fig:images}, \ref{fig:imagesD02} and \ref{fig:imagesD03}). We retain all \textsc{ProFound}-detected segments in our final catalogues, but assign a \textsc{mask=1} flag to sources that have been masked.

\begin{table}
    \centering
    \caption{Final unmasked areas in each of our DEVILS regions in the Y-band} 
        \label{tab:mask}
    \begin{tabular}{cccc}
        Field  & Total  & Unmasked & Percentage  \\
          & (deg$^2$)  & (deg$^2$)  & masked \\
        \hline
        \hline
        D02 & 3.04 &  3.00 & 1.3\%   \\
	D03 & 1.50 &  1.49 &  0.7\% \\
	D10 & 1.52 &  1.47  &  3.4\% \\
        \hline
    \end{tabular}
\end{table}

\subsection{Object Classification}
\label{sec:class}

\subsubsection{Star, galaxy and ambiguous classification}
\label{sec:starGal}

\begin{figure*}
\begin{center}
\includegraphics[scale=0.7]{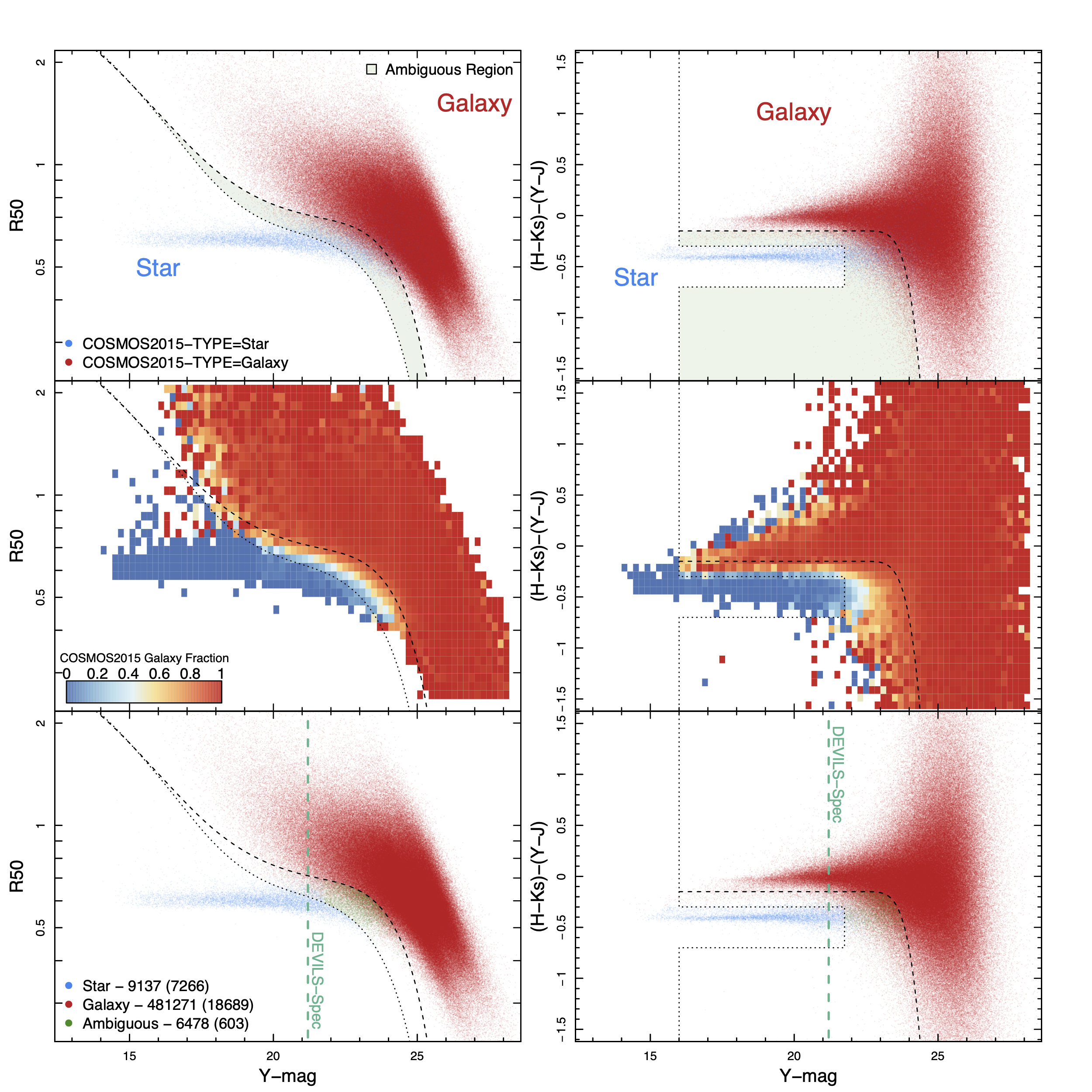}
\caption{Star-galaxy separation using measured half-light radius versus Y-band magnitude (left) and (H-Ks)-(Y-J) colour versus Y-band magnitude (right) in the D10 region. The top panels display the distribution of sources in our photometric catalogue colour coded by the source type assigned via the photometric redshift fitting code \textit{Le Phare} in the COSMOS2015 catalogue. The middle panels show the same data, but binned and colour-coded by the faction of galaxies in each bin. We use these figures to define selection boundaries between galaxy/ambiguous (dashed lines) and stars/ambiguous (dotted line). We then determine the stars, galaxies and ambiguous sources, based on the dual selection in both planes (bottom row). These lines and the relevant selections are described in Section \ref{sec:starGal}. The DEVILS spectroscopic limit is shown by the dashed vertical line in the bottom row. Values in the legend are the number of sources in each category in the D10 region for the full samples, and just to the DEVILS spectroscopic limits in brackets. $i.e.$ to the DEVILS spectroscopic limit $\sim27\%$ of sources are stars and just $\sim2\%$ are ambiguous. These numbers are similar in D02 and D03. }

\label{fig:starGal}
\end{center}
\end{figure*}

In order to separate stars and galaxies, we perform a multi-stage process using the \textsc{ProFound}-derived outputs. This process is detailed in Figure \ref{fig:starGal} - showing just sources for the D10 region. First we consider two separate parameter spaces, Y-band R50 (the radius containing 50\% of the source's Y-band flux in arcsec) vs. Ymag, and (H-Ks)-(Y-J) vs. Ymag, displayed in the left and right columns of Figure \ref{fig:starGal}, respectively. In the top panel of Figure \ref{fig:starGal} we colour-code each of our sources by the SED-derived source type, as estimated from the COSMOS2015 catalogue via the photometric redshift fitting code \textit{Le Phare} (where available), with the middle panels displaying the gridded galaxy fraction. Using these we define an ambiguous region where galaxies and stars are largely indistinguishable in each parameter space (bounded by the dashed and dotted lines) - $i.e.$ there are roughly equal numbers of galaxies and stars. Sources that are above the dashed line in either parameter space are classed as galaxies, and sources below the dotted line in either parameter space are classed as stars.  Sources that fall in between these two classification in both planes are flagged as `ambiguous', and sources that are classified as a star in one plane and a galaxy in the other are also classified as `ambiguous'. The equations that define the lines are:               

\vspace{2mm}

\noindent Galaxies (dashed lines): 
\begin{align*}
R_{50}> -28.49 + 9.52 Y -1.03 Y^2+ 0.05 Y^3 \\
-0.0011 Y^4+ 8.4e^{-6} Y^5,\\
\\ 
(H-Ks)-(Y-J)>-0.15 \,\,\,\,\,\, \mathrm{if \,\,\, Y<22.2},\\
(H-Ks)-(Y-J) > -0.15-0.03(Y-22.2)^5 \,\,\,\,\,\, \mathrm{if \,\,\, Y>22.2}
\end{align*}

\vspace{2mm}

\noindent  Stars (dotted lines): 
\begin{align*}
R_{50}< -161.33 + 44.89 Y -4.75 Y^2+ 0.24 Y^3 \\
-0.0061 Y^4+ 5.9e^{-5} Y^5, \\
\\
-0.7<(H-Ks)-(Y-J)<-0.15  \,\,\&  \,\, Y<21.75 \\ 
  \mathrm{OR} \,\,Y<16
\end{align*}

The bottom panels of each column in Figure \ref{fig:starGal} show our final classifications as either star, galaxy or ambiguous in the D10 region, with the number of sources in each category displayed in the legend.  Within our final catalogues these classification are delineated by the \textsc{StarClass} value as either 0 (galaxy), 1 (star), or 2 (ambiguous). The same selection is applied to sources in both D02 and D03. 

\subsubsection{Artefact Flags}
\label{sec:Artefact}

In addition to flagging of galaxies and stars, we also identify potential artefacts in our photometry catalogues, where flux is not associated with an astronomical source, in a similar method to \cite{Bellstedt20}. This is done in a number of different ways with artefacts flagged as segments that meet one of the following criteria:  i) the source is only detected in one of the g, r or i bands and not the others; ii) the source is only detected in one of the z, Y, J, H, Ks bands and not the others; iii) the source's Y-band R50 is $<0.15^{\prime\prime}$ (smaller than half a pixel); or iv) the source's r-z colour is $< -0.75$\,mag (unphysical colour). These sources are kept in our final catalogues but delineated with \textsc{artefactFlag=1} (all other sources have \textsc{artefactFlag=0}). As an example of the artefacts in our data, \textsc{artefactFlag=1} segments are displayed in orange in the left column of Figure \ref{fig:masking}.

\begin{figure}
\begin{center}
\includegraphics[width = 0.85 \linewidth]{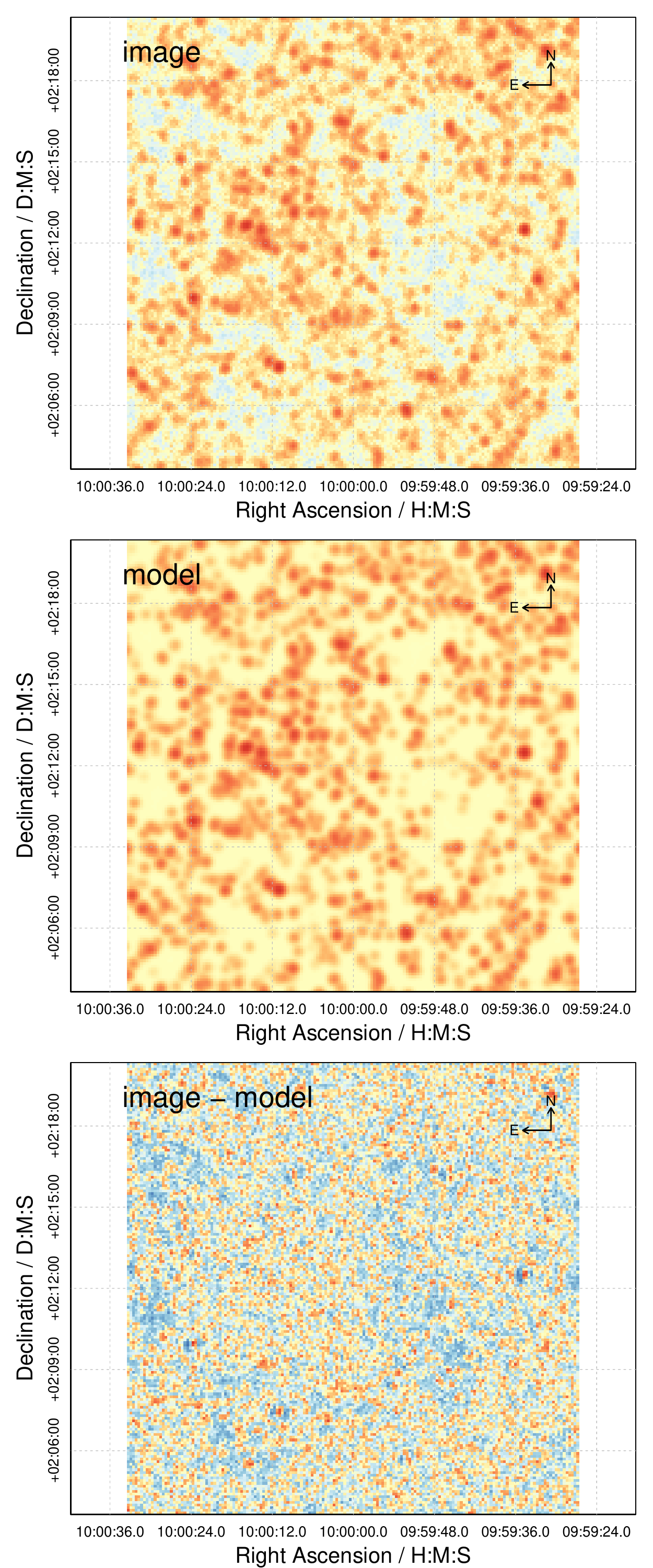}
\caption{Example of the photometry extraction in the SPIRE 250 band. The top panel shows the original image, and the middle panel shows the \textsc{ProFound}-produced model of the objects. The residual when subtracting the model off the image is shown in the bottom panel.}
\label{fig:FIRmodel}
\end{center}
\end{figure}

\subsection{Differences to the original DEVILS target input catalogue}

In \cite{Davies18} we have already discussed the use of ProFound to generate the initial spectroscopic targeting catalogue for DEVILS in the NIR. While our approach is very similar here (but expanded to a broader wavelength range), there are notable changes designed to make our photometric measurements more robust. The process outlined in \cite{Davies18} was somewhat conservative and designed to produce robust photometry for the bright (Y$<21.2$\,mag) spectroscopic sample for DEVILS at the start of spectroscopic observations. Here we have improved upon this process in order to both provide more robust measurements for our spectroscopic input catalogue and to produce photometric measurements to faint magnitudes for additional science.   

Firstly, we now use updated versions of both the UltraVISTA and VIDEO data that were made available following \cite{Davies18}. Secondly, we have performed the segmentation fixing outlined in Section \ref{sec:group} to fix larger, extended sources. This was not applied in \cite{Davies18} and has improved the photometric measurements for sources at the bright end. Thirdly, we have improved our masking process based on the experience of \cite{Bellstedt20} in producing the GAMA photometry catalogues. Within \cite{Davies18} we used a masking procedure based on the \textsc{ProFound} skymean measurements. This was found to miss some halo regions around bright stars and over-mask other regions. Our new method improves upon this and reduces the fraction of the fields that required masking but still removes erroneous sources. Finally, we also improved our star-galaxy separation method. Previously we only used a simple NIR colour separation between stars and galaxies. While this works well for bright systems (which we are targeting spectroscopically) it fails at fainter magnitudes, hence we apply a more complex method here. However, it is worth noting that these methods are essentially identical at the bright (Y$<21.2$\,mag) end.  

While there are significant changes between the method outlined in \cite{Davies18} and the method used here, this has in fact very little impact on the photometry for our spectroscopic sample (modulo the addition of some new un-masked objects). The spread ($\sigma$) of Y-band photometric difference between the target catalogue outlined in \cite{Davies18} and that described here is $\sim0.1$\,mag. However, in order to use the best available photometry for our spectroscopic sample, we now update to our new photometry and have transitioned to the new input catalogue for our observations based on the catalogues discussed here.

\begin{table}
    \centering
    \caption{\textsc{ProFound} Options used for FIR photometry for each instrument. For details of each of these parameters and how they affect source detection, see \citet{Bellstedt20}.} 
        \label{tab:FIRSettings}
    \begin{tabular}{ccccc}
        \textsc{ProFound} Option  & MIPS 24/70 & PACS  & SPIRE  \\
        \hline
        \hline
        fit$\_$iters  &  5 &  5 &  5 \\
         pixcut  &  3 &  3 &  3\\
         skycut  & 2 & 2 & 2\\
         ext  & 1 & 1 & 1 \\
         redosky  & FALSE  & FALSE  & FALSE  \\
         psf$\_$redosky  & TRUE & TRUE & TRUE \\
         iters  & 4 & 4 & 4\\
         tolerance  & 1/0 & 0 & 0  \\
         sigma  & 2 & 2 & 0 \\
         magdiff & 5 & 5 & 5 \\
        \hline
    \end{tabular}
\end{table}

\section{MIR-FIR Photometric Measurements}
\label{sec:FIRphotom}

Once our UV-MIR catalogues and flags are in place, we then extract photometry in the MIR-FIR using, once again, a similar method to that of \cite{Bellstedt20}. Due to the shallower depth, poorer resolution and the existence of FIR bright sources that are undetected in the optical-NIR, it is not appropriate to directly apply our NIR-defined segmentation map to the MIR-FIR data in a similar manner to the shorter wavelength bands. To overcome this, \textsc{ProFound} has an inbuilt \texttt{profoundFitMagPSF} function specifically defined for the measurement of photometry in low-resolution, shallow data ($i.e.$ the MIR-FIR). This method uses the known source positions, here provided from our optical-NIR catalogues, to fit the PSF of each MIR-FIR band to the images, iterating by applying an expectation maximisation function. This process is described extensively in \cite{Bellstedt20} but briefly, in each band independently the function extracts flux at the positions of an object catalogue, and then iterates over the flux of each object to ensure that all non-sky flux is accounted for in the image and sources are not over extracted. For each band, the locations of our NIR-selected objects provide the coordinates at which to place a PSF and extract flux in the longer wavelength data. However, it is inappropriate to include all of our NIR-selected sources to faint magnitudes becasue i) Y-flux is not well correlated with the expected emission at longer wavelengths and ii) the lower resolution of the MIR-FIR data means that sources will be heavily confused with multiple input galaxies occupying the same resolution element. As such, we must first apply appropriate cuts to out NIR-selected catalogue to provide sensible input positions for our MIR-FIR analysis.  

For this, we first run \texttt{profoundFitMagPSF} on the MIPS24 images, using a Y$<$21.2\,mag sample from our UV-MIR catalogue, excluding masked objects and objects classified as stars or artefacts (for example, this results in 17,209 objects for D10). \texttt{profoundFitMagPSF} not only measures source photometry for the input sample, but also identifies any additional sources for flux from objects not aligned with the input catalogue ($i.e.$ NIR-faint systems). In the MIPS24, a large fraction of sources (28\% in D10) are not associated with a Y$<$21.2\,mag source, these sources are then matched back to the full UV-MIR catalogue using both on-sky positional offset (with a maximum matching radius of 5$^{\prime\prime}$) and expected Y-24$\mu$m colour. To estimate the expected Y-24$\mu$m colour we take all Y$<$21.2\,mag matches and find the typical Y-24$\mu$m colour to be 3\,mag. We then perform a 3D spherical match between MIPS24 sources and our UV-MIR catalogue using on-sky separation and (Y-24$\mu$m)-3 colour offset. $i.e.$ for each UV-MIR source within 5$^{\prime\prime}$ of a MIPS24-detected position, we assign a probability of correct match based on the sky separation and the expected  Y-24$\mu$m colour, and take the most likely match. As such, we implicitly assume that all MIPS24 sources are contained within our UV-MIR catalogue, and that all sources should have the typical Y-24$\mu$m colour. This process will potentially fail for AGN-like sources which may have atypical Y-24$\mu$m colours, but these are likely to only form a small fraction of the source sample. We note that not all MIPS24-detected sources are matched to a UV-MIR catalogue source using this method, these are retained as true NIR-faint and MIPS24-bright sources.            

To build our input source positions for longer wavelengths, we combine our Y$<$21.2\, mag sample with the MIPS24-detected matched sample as described above and the additional sources which are unmatched in the UV-MIR catalogue. Using these input objects, we run \texttt{profoundFitMagPSF} on all of the 70-500\,$\mu$m bands.  To account for potential previous over-subtraction of the sky in the data reduction phase of the FIR imaging by the original teams (possibly due to the confused nature of sky pixels), we also include a second phase of extraction with an explicit sky subtraction using the sky value as measured by \textsc{ProFound} in the first phase. 

For completeness the \texttt{profoundFitMagPSF} settings used in our analysis are shown in Table \ref{tab:FIRSettings}. \texttt{profoundFitMagPSF} accepts an empirical PSF provided by the user in various forms. Here we use the observed PSF kernels from PEP for the PACS bands\footnote{\label{note1}\url{http://www.mpe.mpg.de/resources/PEP/DR1_tarballs/readme_PEP_global.pdf}}, and a Gaussian PSF for the SPIRE bands using the FWHM as given in \cite{Oliver12} (18.15, 25.15, 36.3 arcsec for the 250, 350 and 500\,$\mu$m bands, respectively). For the MIPS bands, we use the 100\,K PSFs from \cite{Gordon08} and re-scale the PSF to the image pixel scale. While measuring flux, we apply multiplicative aperture corrections of 1.15 for MIPS 70 \citep{Frayer09}, 1.50 for PACS 100 and 1.477 for PACS 160, and an additional multiplicative high-pass correction of 1.12 and 1.11 for PACS 100 and 160, respectively, as suggested by the PACS data release\textsuperscript{\ref{note1}}. 

The fraction of Y$<$21.2\,mag objects with detections (arbitrarily defining a detection as having a measured magnitudes of $<30$\,mag) varied in each band depending on the depth of the imaging. For example, in D10 the $M24/M70/P100/P160/S250/S350/S500$\,$\mu$m bands, $81.1/44.4/55.8/57.1/62.5/44.4/28.3$ per cent of objects had detections, respectively. 1494 objects had detections in all seven bands, corresponding to 8.6 per cent of the Y$<$21.2\,mag sample. No objects were undetected in all of the seven bands.

The resulting uncertainties, due to the expectation maximisation mixture modelling process, accurately reflect the inherent uncertainties in this process. If two optically-detected sources are close in projection, then the uncertainty will reflect the potential confusion between these two sources. In Figure \ref{fig:FIRmodel} we show an example of an original image, final modelled image, and resulting residual in the SPIRE 250\,$\mu$m band in D10. 

In addition to the objects detected in the FIR with optical/MIR counterparts, a small fraction of FIR objects were detected with no clear counterpart - i.e. objects which are not detected in MIPS24 but are detected at longer wavelengths - likely high redshift FIR-bright systems. The outputs for these additional sources are saved, but not included in our DEVILS photometry catalogue (as they are not associated with a DEVILS target). However, we do add these sources into the catalogues used to generate the deep number counts presented in this work.

\begin{figure*}
\begin{center}
\includegraphics[scale=0.48]{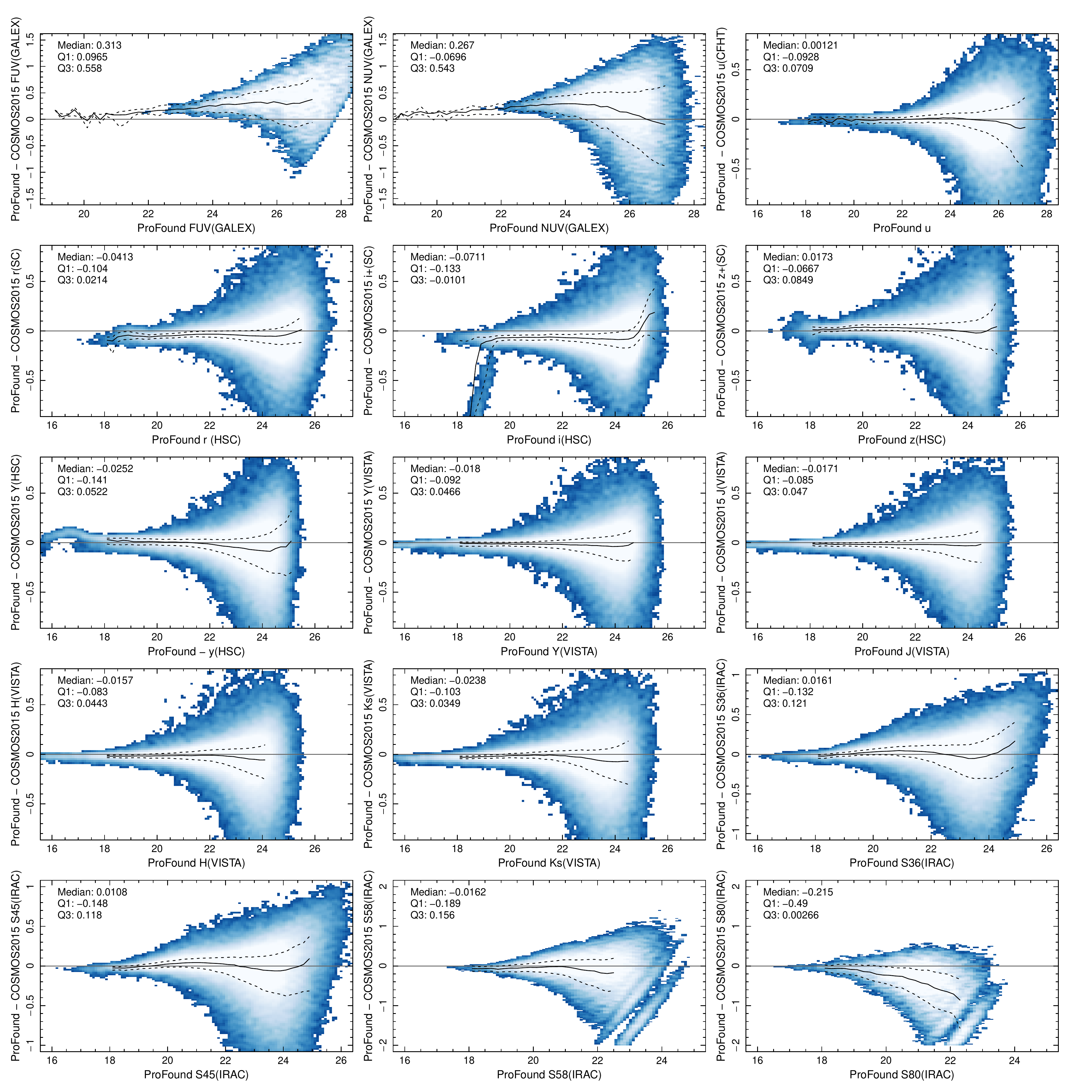}
\caption{Comparison between our new \textsc{ProFound} photometry and COSMOS2015 catalogue. Each panel displays \textsc{ProFound}-COSMOS2015 for 2$^{\prime\prime}$ matched sources. Note that in r, i and z bands we compare HSC with Subaru SuprimeCam, and in VISTA Y, J, H and Ks bands we compare UltraVISTA DR4 with DR2. Solid and dashed lines show the running median and interquartile range of the data, respectively. The full sample median and ranges are noted in each panel. Note that these do not have MW extinction corrections applied to be comparable to the public COSMOS2015 catalogue. Note that COSMOS2015 does not provide g-band measurements, and as such no comparison is shown here. Note, the diagonal feature in the i-band for both this figure and Figure \ref{fig:compG10} are cause by saturated sources in the old SuprimeCam data used in both COSMOS2105 and G10/COSMOS.}
\label{fig:compCOSMOS2015}
\end{center}
\end{figure*}

\begin{figure*}
\begin{center}
\includegraphics[scale=0.48]{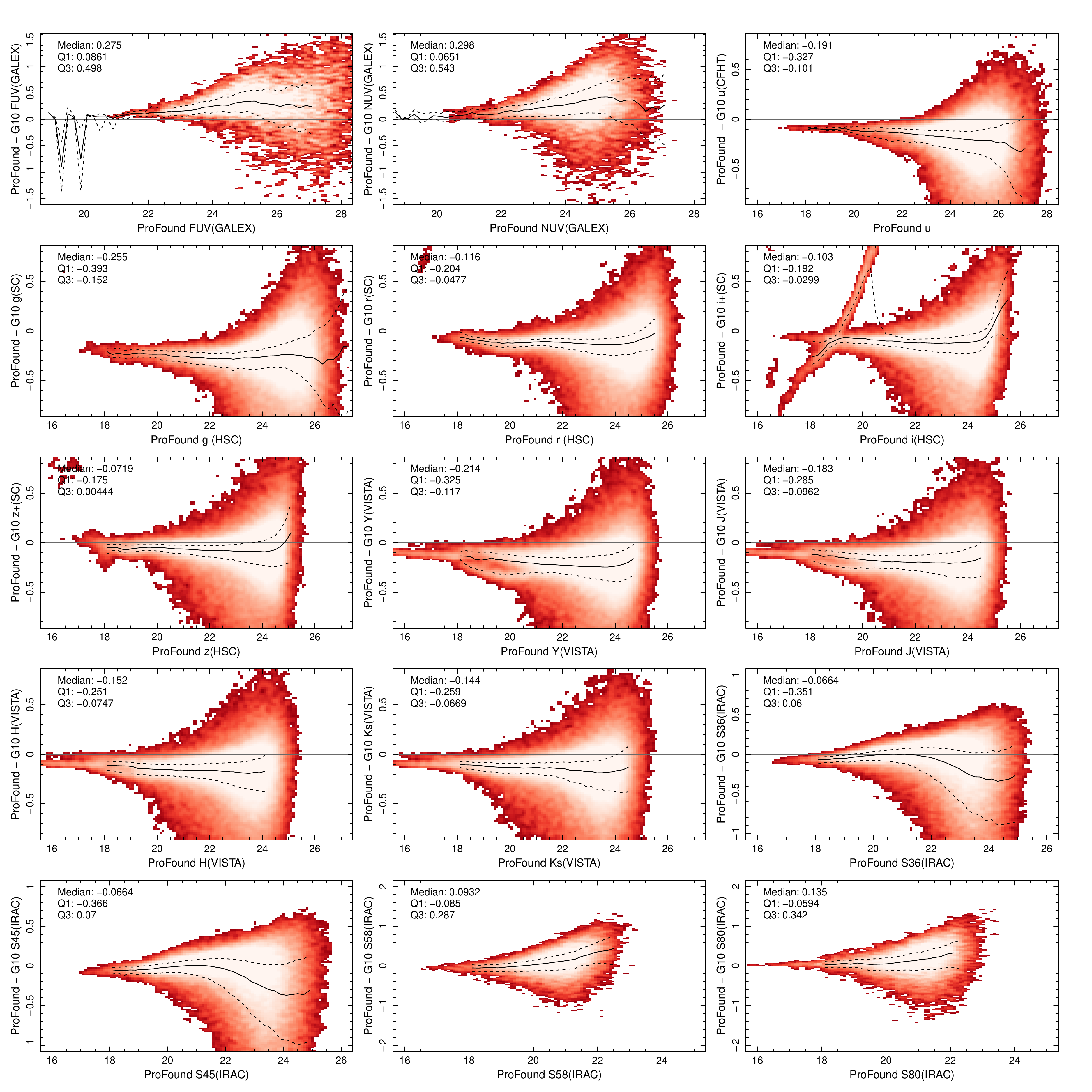}
\caption{Same as Figure \ref{fig:compCOSMOS2015} but for the G10/COSMOS catalogue. However, these have MW extinction corrections applied to be comparable to the public G10/COSMOS photometry. Note that in r, i and z bands we compare HSC with Subaru SuprimeCam, and in VISTA Y, J, H and Ks bands we compare UltraVISTA DR4 with DR2}
\label{fig:compG10}
\end{center}
\end{figure*}

\begin{figure*}
\begin{center}
\includegraphics[scale=0.48]{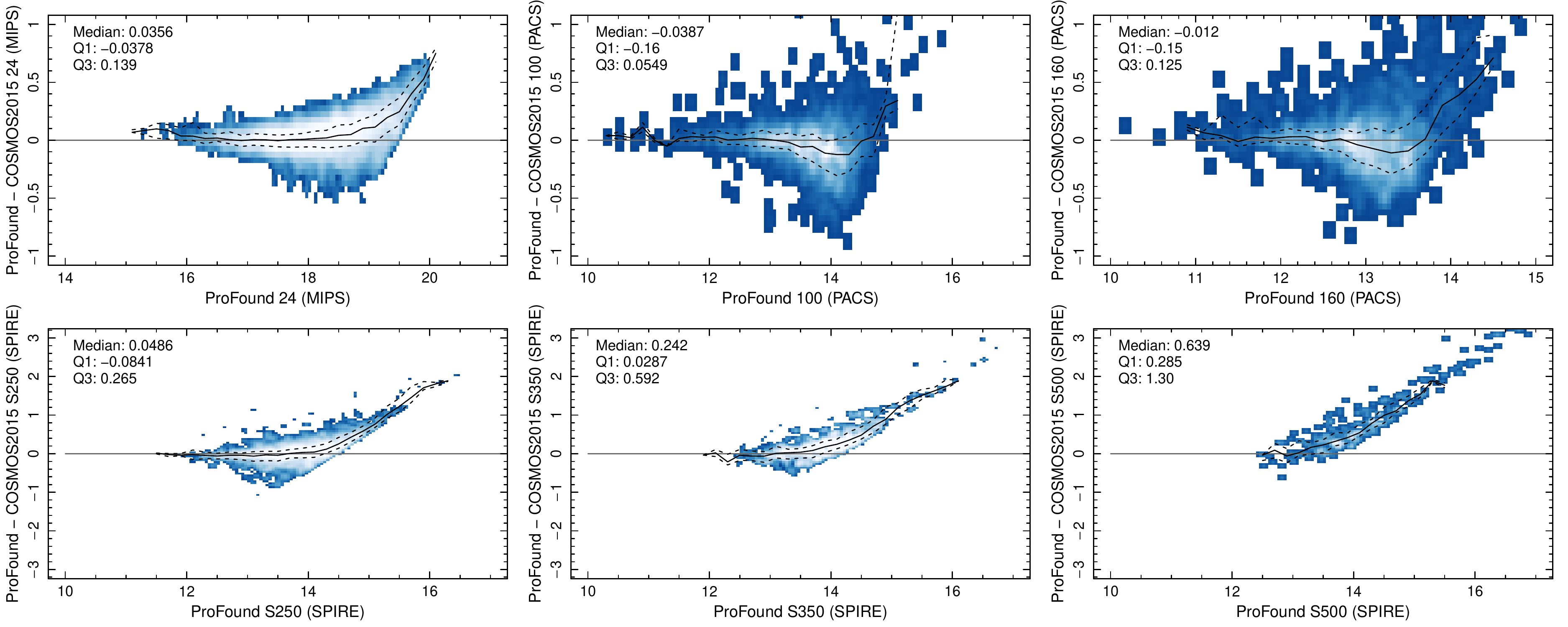}\\
\includegraphics[scale=0.48]{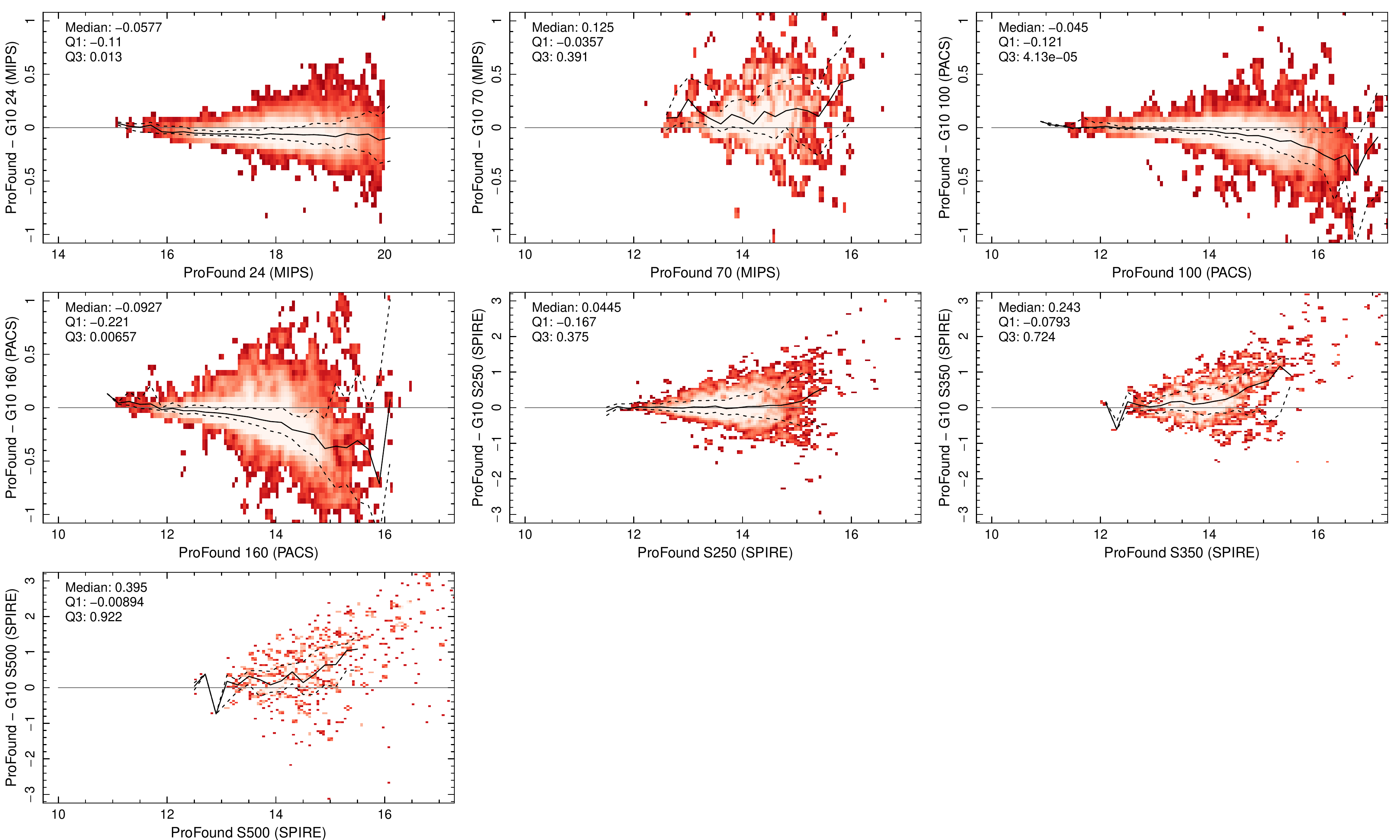}\\
\includegraphics[scale=0.48]{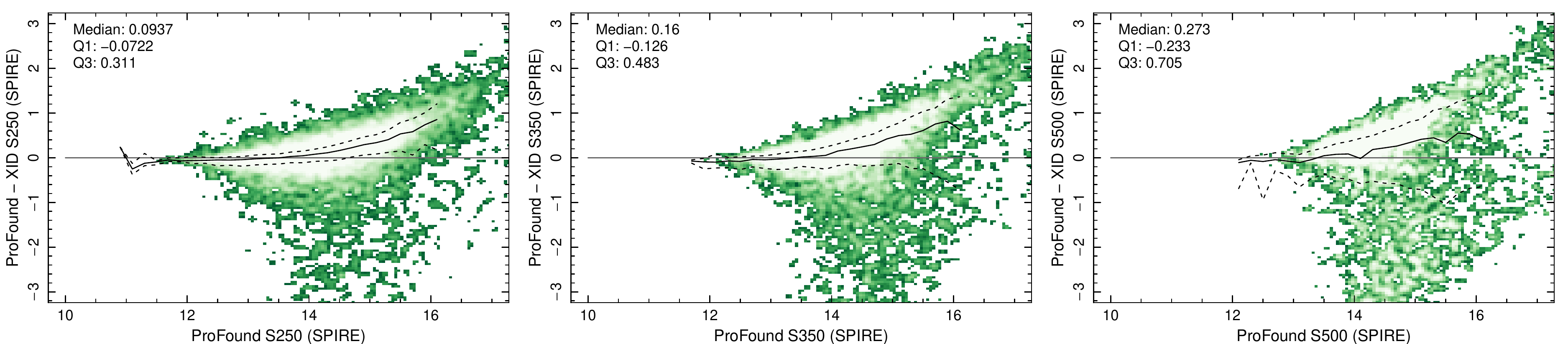}\\

\caption{Same as Figure \ref{fig:compCOSMOS2015} but showing comparisons between our new MIR-FIR \textsc{ProFound} photometry and COSMOS2015 (blue), G10/COSMOS (red) and XID+ (green) catalogues, where available, in the MIR-FIR regime. }
\label{fig:compFIR}
\end{center}
\end{figure*}

\section{Verification of Photometry}

In the following section we carry out a number of verification checks to assess the validity of our current photometry catalogue. Here we present these comparisons just in the D10 region, where more extensive existing data is available, to highlight the validity of our process. However, similar checks are performed on all regions with similar results. These tests are intended to provide the reader with the confidence that our new photometry is robust, and can be used for future science projects.    

\subsection{Comparisons to Previous Photometry in the COSMOS Field}

First we compare our total photometry measurements to existing photometry catalogues in the COSMOS region. We do not assume any particular catalogue is the ground truth, but simply highlight the consistency (or lack thereof) between photometric measurements using different techniques. 

In Figure \ref{fig:compCOSMOS2015} we compare our UV-MIR photometric measurements with those from the COSMOS2015 catalogue of \cite{Laigle16}. Briefly, \cite{Laigle16} uses SExtractor \citep{Bertin96} to derive total and fixed aperture photometry from the Optical-NIR multi-wavelength data in COSMOS. These are then catalogue matched to the UV and longer wavelength data available in the region - please refer to \cite{Laigle16} for further details. Here we first compare to the total (\textsc{MAG\_AUTO}) measurements where available. We note that the imaging used in this work and \cite{Laigle16} catalogue is identical except for the following: we have used newer HSC data in g, r, i, z bands in comparison to the Subaru SuprimeCam data used in COSMOS2015, and we use UltraVISTA DR4 while COSMOS2015 uses UltraVISTA DR2. In Figure \ref{fig:compCOSMOS2015} we also compare photometry prior to galactic extinction correction being applied, as this is how the raw COSMOS2015 photometry is provided. We find only very small median offsets in the majority of bands ($<$0.1\,mag), showing consistency between our measurements, with the exception of the GALEX bands that show a normalisation offset of $\sim0.25-0.3$\,mag. This is discussed further in Section \ref{sec:offsets}. However, we find that our photometry and that derived in COSMOS2015 is largely consistent despite using different methods, and in some cases, different data sets. However, our new photometry will not suffer from some of the issues discussed previously \citep[and extensively in][]{Robotham18} and is more robust for extracting total photometry measurements.

Next we compare our UV-MIR photometry to the G10/COSMOS catalogue of \cite{Andrews17} in Figure \ref{fig:compG10}. The G10/COSMOS catalogue uses the Lambda Adaptive Multi-wavelength Analysis for R \citep[LAMBDAR,][]{Wright16} code with identical imaging datasets to  COSMOS2015. The LAMBDAR software starts with base apertures defined by SExtractor, but within \cite{Andrews17} they go through an extensive processes of manually fixing any erroneous apertures. Firstly, in Figure \ref{fig:compG10} we see comparable median offsets to the COSMOS2015 catalogue in the GALEX UV bands, once again this is discussed further in Section \ref{sec:offsets}. Through the u-Ks bands, we see normalisation offsets of up to $\sim$0.3\,mag, particularly in the VISTA NIR bands. These are consistent with the offsets between G10/COSMOS and COSMOS2015 presented in \cite{Andrews17}. In \cite{Andrews17} it was argued that these offsets are likely due to the initial aperture definitions and/or choice of selection band. However, given that our new \textsc{ProFound} photometry does not rely on apertures for photometry, and we agree more closely with the COSMOS2015 catalogue in the NIR, this is unlikely to be the case and the G10/COSMOS measurements are simply missing flux and/or have a zero-point error. We do agree very well with IRAC MIR measurements from G10/COSMOS with $<0.1$\,mag median offset in all bands.

For our MIR-FIR photometry, we again compare to COSMOS2015 and G10/COSMOS, but also to the XID+ SPIRE catalogues of the $Herschel$ Extragalactic Legacy Project (HELP) team \citep[see][]{Hurley17}. All comparisons are presented in Figure \ref{fig:compFIR}. We find good agreement with all existing photometry in the \textit{Spitzer} and, $Herschel$ PACS and SPIRE data - suggesting our photometry is consistent with existing catalogues. We highlight that while we are consistent with these existing catalogue across their individual regimes, the benefit here is that we are using a single consistent package to extract photometry across the full wavelength range, removing some of the issues associated with catalogue matching of data. We have also optimised our catalogues to robustly measure total flux measurements across all sources, which are required for robust measurements of galaxy properties such as SFRs and stellar masses \citep[see][]{Thorne21}.

\subsection{Photometric Offsets}
\label{sec:offsets}

As discussed in the previous section, while our photometric measurements are largely consistent with existing catalogues we do see a number of small median zero point offsets. These are most significant in the GALEX-UV bands. However, the methods used to calculate the photometry in these bands is vastly different between our new work and the existing catalogues, and a detailed assessment of the previous photometry finds a number of seemingly arbitrary zero-point corrections are applied to the photometry in these bands. As such, we do not know which catalogue now presents the ground truth. 

\begin{figure}
\begin{center}
\includegraphics[scale=0.7]{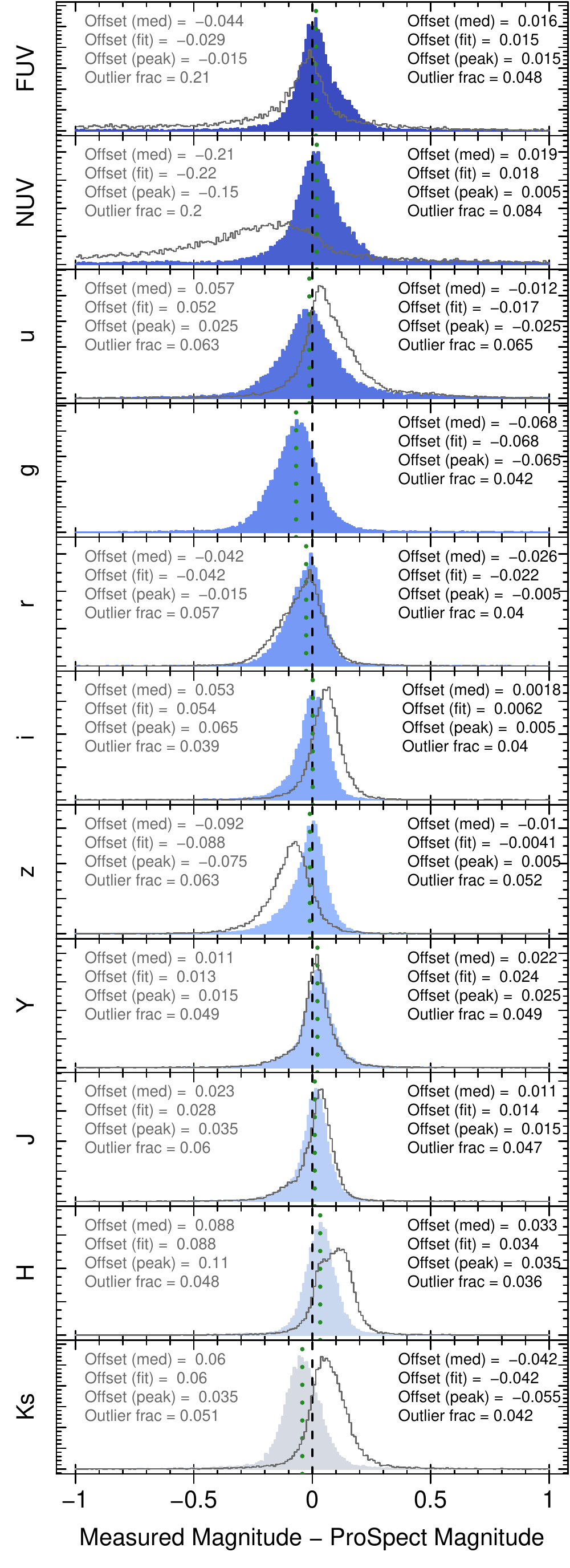}
\caption{Individual band offsets between measured photometry for a random sample of 24,000 D10 galaxies and the \textsc{ProSpect} best fit SED in FUV - Ks. Coloured filled histograms and right-hand legend show our new photometry. Grey lines and left hand legend show COSMOS2015 photometry. In the legends we display the offset values measured in three different ways, where large absolute values indicate a larger discrepancy with the ProSpect SED models (see text for details). In most bands our new photometry has smaller offsets using all methods. We also show the outlier fraction (see text for details), where our new photometry has comparable or smaller values in every band. The dashed vertical line display the zero offset line, while the dotted green line shows the peak of the \textsc{ProFound}-derived distribution. }
\label{fig:FilterOffsets1}
\end{center}
\end{figure}

\begin{figure}
\begin{center}
\includegraphics[scale=0.7]{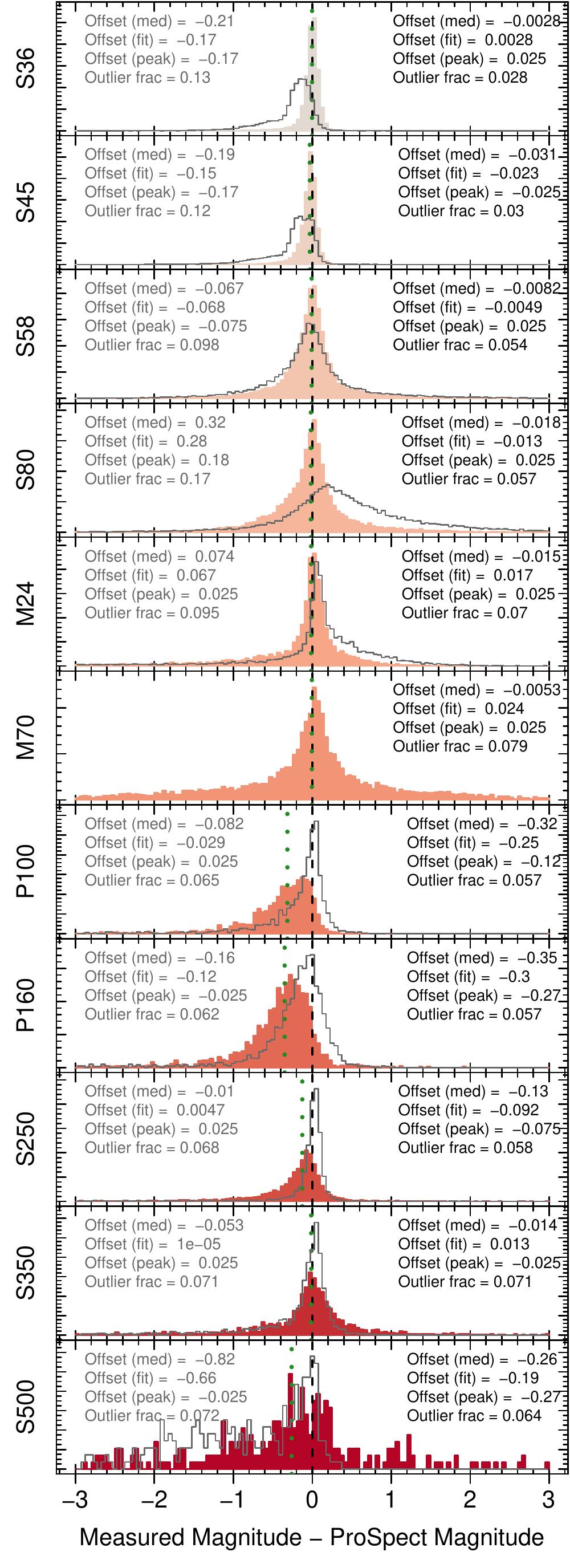}
\caption{The same as Figure \ref{fig:FilterOffsets1} but for the MIR-FIR (S36 - S500) bands. }
\label{fig:FilterOffsets2}
\end{center}
\end{figure}

To test this, we use the \textsc{ProSpect} \citep{Robotham20} spectral energy distribution (SED) fitting results outlined in \cite{Thorne21}. Thorne et al fit all DEVILS D10 galaxies using our new photometry to produce a robust SED fit and derive star-formation histories, stellar masses, SFRs, etc. For our tests we then also fit the existing COSMOS2015 photometry (which shows the similar offsets as G10/COSMOS - particularly in the UV bands) for a random sample of $\sim24,000$ DEVILS galaxies with spectroscopic redshifts at $z<1$, and using the same \textsc{ProSpect} parameters as Thorne et al. Here we use the best-fit \textsc{ProSpect} SED and compare to the input photometric measurements, for both \textsc{ProFound} total photometry and the COSMOS2015 photometry. For COSMOS2015 we use the 2$^{\prime\prime}$ aperture photometry measurements to maximise robust SED shape fits, $i.e.$ aiming to obtain the best possible fits. We also note here that the COSMOS2015 sample already apply per-band zero-point offsets using the \textsc{LePhare} SED fitting code in order to improve the accuracy of their photometric redshifts. As such, their photometry should be tuned to closely match the overall SED shape of existing galaxy templates. However, systematic offsets between the \textsc{ProSpect} fit and input photometry can highlight potential zero-point offsets in a given photometric band, assuming the suite of SED models is representative of the true galaxy population. Figures \ref{fig:FilterOffsets1} and \ref{fig:FilterOffsets2} display the input photometric measurement minus the \textsc{ProSpect} best-fit SED measurements for all bands. In each panel we note the median offset of each distribution, the offset from the peak of a Gaussian fit to the distribution and the peak position offset for both photometry catalogues as a measure of any systematic differences between the photometric catalogues and \textsc{ProSpect} SED models. We find no significant residual offset between our \textsc{ProFound} photometry and the \textsc{ProSpect} best-fit SED in the GALEX bands, or in fact any others. There is a slight offset in the HSC g-band ($\sim0.07$\,mag), which we believe is due to the 4000\AA\ break falling within this band at the redshift of a typical DEVILS galaxy and the \textsc{ProSpect} models not completely encompassing the true range of galaxy types at this point. When the full \cite{Thorne21} sample (including photometric redshifts) is fit using a broader redshift range, this offset disappears.  However, we do see numerous offsets for the COSMOS2015 photometry when fitted using \textsc{ProSpect}, in multiple bands in Figure \ref{fig:FilterOffsets1} in the NUV, z, H, Ks and IRAC channels. This suggests that our new photometry is more consistent with realistic galaxy SEDs (at least those probed by \textsc{ProSpect}). We also display the outlier fraction, defined as the fraction of sources that are outside the offset peak value from \textsc{ProFound}/COSMOS2015 $\pm$ 2 $\times$ the standard deviation of the \textsc{ProFound} distribution. We find that the COSMOS2015 photometry has larger outlier fractions in almost all bands (modulo u, i, Y and S500 where the offsets are comparable). This further suggests that measured scatter in our new photometry is smaller when compared to the \textsc{ProSpect} SED models. In combination, these results suggest that any differences in zero point between our new photometry and the COSMOS2015 catalogues is unlikely to be due to errors in our current approach.

\begin{figure}
\begin{center}
\includegraphics[scale=0.6]{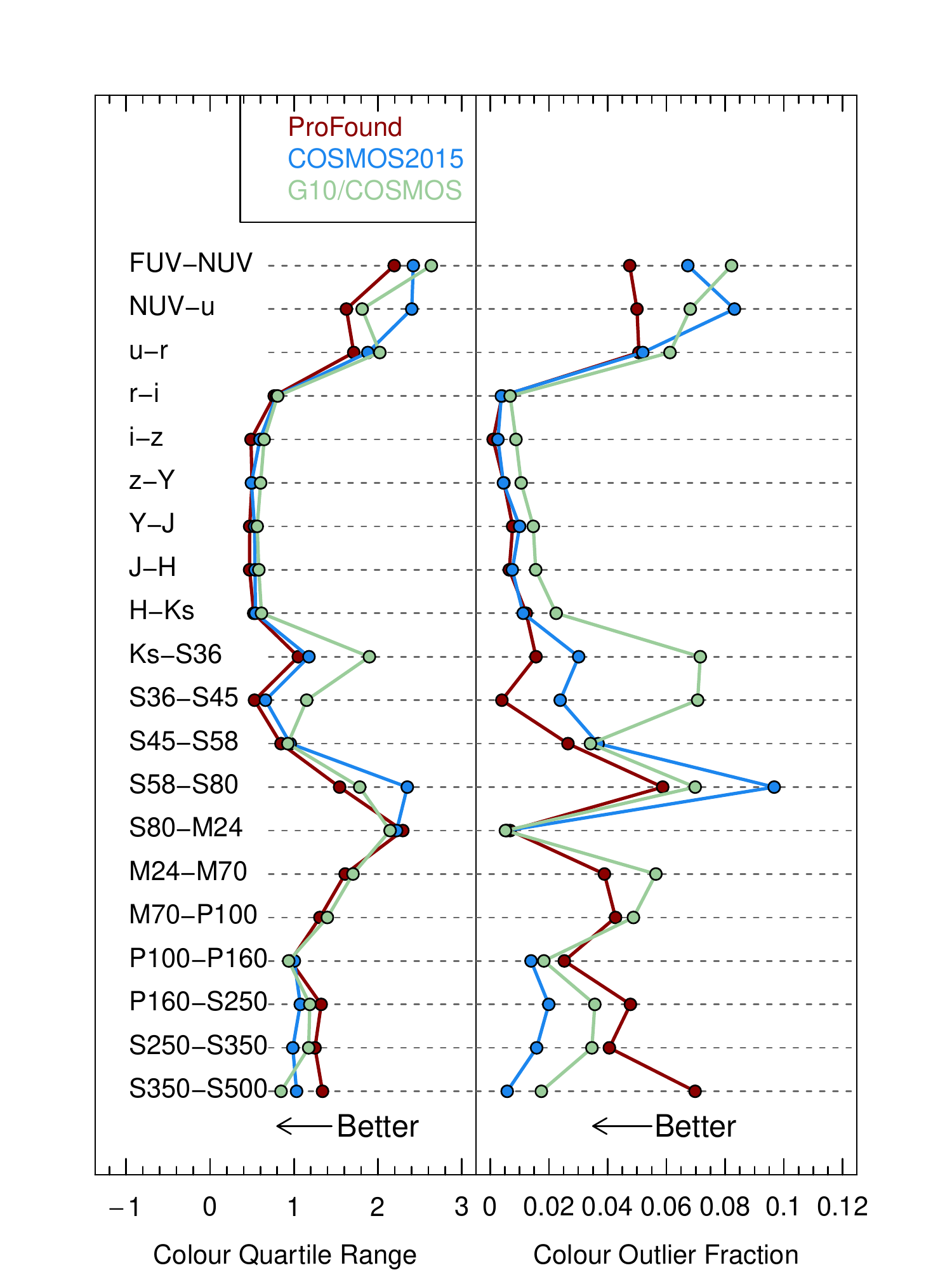}
\caption{Comparison of colour histograms in adjacent filters for a matched sample of D10 \textsc{ProFound}, COSMOS2015 and G10/COSMOS galaxies. The left panel displays the interquartile range and right panel the outlier fraction. A smaller value in each of these figures is desirable. }
\label{fig:colourStat}
\end{center}
\end{figure}

\subsection{Colour Comparisons}

While the figures displayed in previous sections are useful for diagnosing zero-point and linearity problems ($i.e.$ magnitude-dependent offsets), they do not give an objective assessment as to which data set is more robust in terms of the photometry measurement process. One potential avenue to explore this is through galaxy colours in adjacent filters. A sample of galaxies will have some intrinsic colour distribution, which is then convolved with error distributions introduced by the instrumentation, observing conditions, photometric data reduction methods and photometric measurement error. A narrower colour distribution and a lower outlier fraction thus indicating a narrower photometric error distribution are, hence, likely more reliable. 

To explore the colour distributions of our sample we first measure the colours in adjacent bands for a 2$^{\prime\prime}$ matched sample of Y$<$24.5\,mags galaxies in the D10 region, for FUV-S80, and sources brighter than the 5$\sigma$ detection limit, for MIPS24-S500, for \textsc{ProFound}, COSMOS2015 and G10/COSMOS. Here we use the colour-optimised photometry from \textsc{ProFound} and fixed 2$^{\prime\prime}$ (colour-optimised) photometry from COSMOS2015 (for G10/COSMOS only total photometry is provided). For each distribution we measure the interquartile range and outlier fraction (defined as the number of sources which are $>$0.5\,mag outside of the interquartile range) for each catalogue and in each band. 

Rather than showing the raw distributions, to better illustrate these comparisons, Figure \ref{fig:colourStat} displays the interquartile range and outlier fraction for all colours, where a smaller value in each of these figures is desirable. Our new \textsc{ProFound} photometry does better (or comparable) to both existing datasets in all colours at FUV-160\,$\mu$m wavelengths. We do see a significantly larger outlier fractions in our FIR SPIRE measurements, which are consistent with the \textsc{ProSpect} residual distributions shown in Figure \ref{fig:FilterOffsets2}, where the COSMOS2015 photometry displays a more peaked distribution in the FIR. We have explored this discrepancy in a number of ways, and have been unable to improve the tightness of the colour distributions in our SPIRE photometry. However, for the sake of consistency between both our multi-wavelength photometry and the approach used in GAMA, we opt to retain our method for deriving the FIR photometry. We also note that only a small fraction of the DEVILS spectroscopic sample ($\sim15\%$) have a detection in all of the SPIRE bands, and as such the DEVILS sample is unlikely to be significantly affected by these small differences in photometry.

\begin{figure*}
\begin{center}
\includegraphics[scale=0.58]{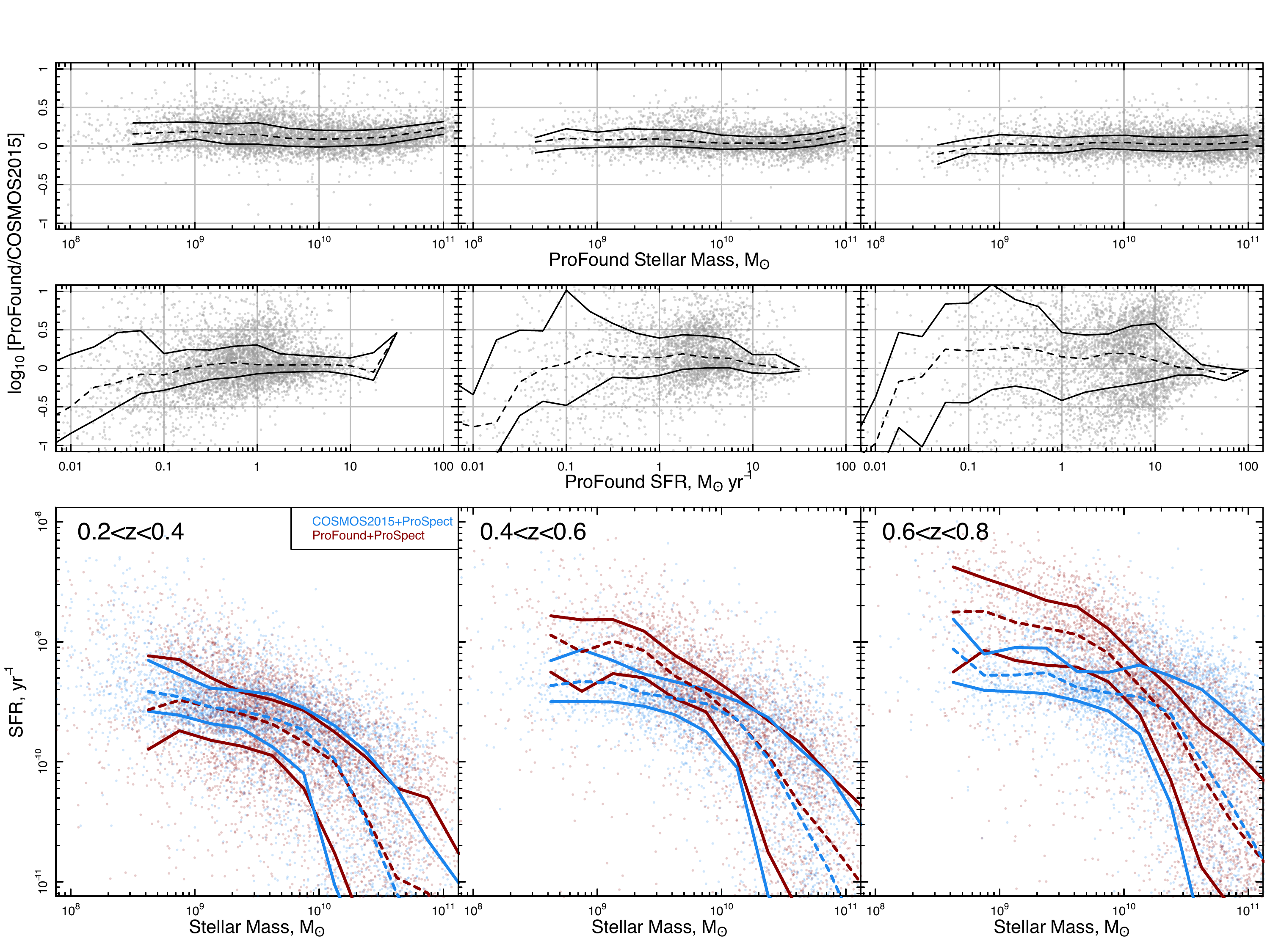}
\caption{Comparison of the physical properties derived using \textsc{ProSpect} for a common sample of galaxies but using both the COSMOS2015 photometry and our new \textsc{ProFound} photometry. We split the sample  into three redshift bins between $0.2<z<0.8$. The top row displays the difference between stellar masses derived using the COSMOS2015 and \textsc{ProFound} photometry. Middle row show the same but for SFRs, while the bottom row shows the full sSFR-M$^{*}$ relation for this sample. Stellar masses are consistent but with a 0.1-0.2\,dex offset and the \textsc{ProFound} photometry produces systematically larger SFR which goes with sSFR. Combined this would lead to a distinctly different measurement of the evolution of the sSFR-M$^{*}$ relation at M$^{*}<10^{10}$M$_{\odot}$.} 
\label{fig:MScomp}
\end{center}
\end{figure*}

\subsection{Impact on derived physical properties}
Following on from the above residuals and colour comparisons, it is also beneficial to assess how our new photometry catalogue impacts the derived properties of galaxies, when removing any other differences in the method for measuring these properties ($i.e.$ how does the photometry alone change our galaxies physical properties and hence scientific conclusions). To do this we compare the SED-derived properties of a subsample of galaxies using the same method, but differing photometric measurements.  

While we leave a full analysis of the DEVILS \textsc{ProSpect} stellar mass and SFR measurements to \cite{Thorne21}, Figure \ref{fig:MScomp} shows a comparison of the \textsc{ProSpect}-derived stellar masses and sSFR produced using the analysis/sample outlined in Section \ref{sec:offsets} ($i.e$ for an identical fitting procedure with \textsc{ProSpect} but using both the COSMOS2015 photometry and our new \textsc{ProFound} photometry for an identical set of galaxies). We split the sample into three redshift bins between $0.2<z<0.8$ \cite[see][for a full description of this sample and redshifts]{Thorne21} and display the difference between the resultant \textsc{ProFound} and COSMOS2015 stellar masses (top row), SFRs (middle row) and along the sSFR-M$^{*}$ relation (bottom row). We find that derived stellar masses are consistent between the two samples, but with a systematic 0.1-0.2dex offset, but SFRs show a systematic bias to higher SFRs in \textsc{ProFound} in particular galaxy populations. This effect becomes more pronounced as we move to higher redshifts. This is most clearly apparent in the comparison of the sSFR-M$^{*}$ relation where our \textsc{ProFound} and COSMOS2015 photometry produce a comparable relation at low redshift (median and scatter in sSFR are consistent at all stellar masses), but the samples begin to diverge significantly for M$^{*}<10^{10}$M$_{\odot}$ galaxies at higher redshifts. Importantly this shows that \textit{only} changing the photometric measurements for relatively bright galaxies, can have a significant impact on the measured evolution of the sSFR-M$^{*}$ relation and associated conclusions. This just represents a single example where different photometric measurements can strongly impact derived scientific conclusions.  

While it is still difficult to assess which sample would be deemed `correct' through this analysis, we note that through the photometry measurements alone, a larger stellar mass and SFR (as is the case systematically from \textsc{ProFound}), suggests that one is measuring more true flux from the system, and therefore more robustly capturing the total emission from the galaxy. It is somewhat difficult to get the converse (systematically lower measurements) erroneously when only considering photometric measurement differences for a broad population of galaxies across a range of epochs. This, coupled with the analysis in Section \ref{sec:offsets}, suggests that our new \textsc{ProFound} measurements are robustly capturing the total galaxy flux across a range of wavelengths, and thus can be used to measure galaxy properties \citep[$i.e.$ see][]{Thorne21}

\subsection{Astrometric Accuracy}

\begin{figure}
\begin{center}
\includegraphics[scale=0.4]{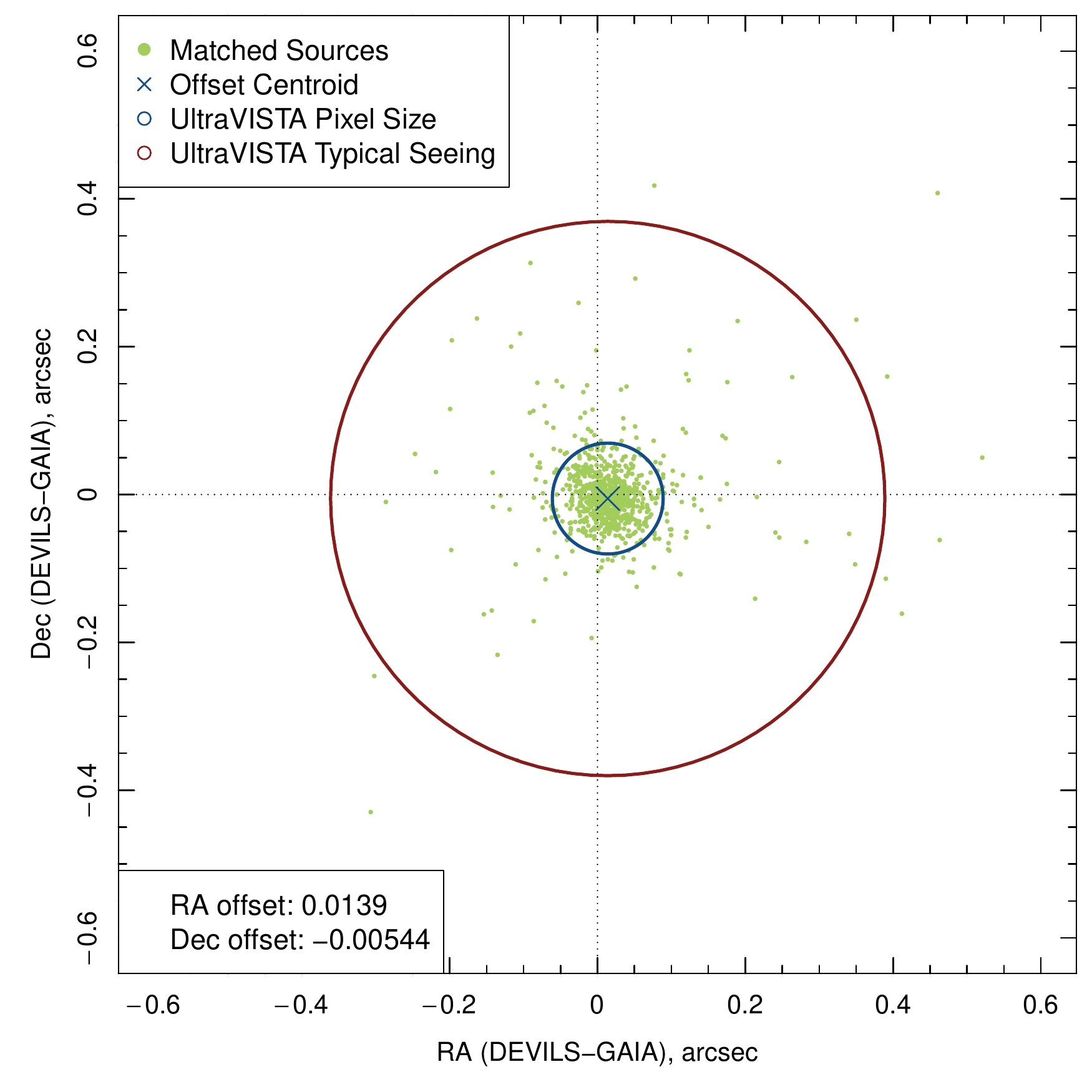}
\includegraphics[scale=0.4]{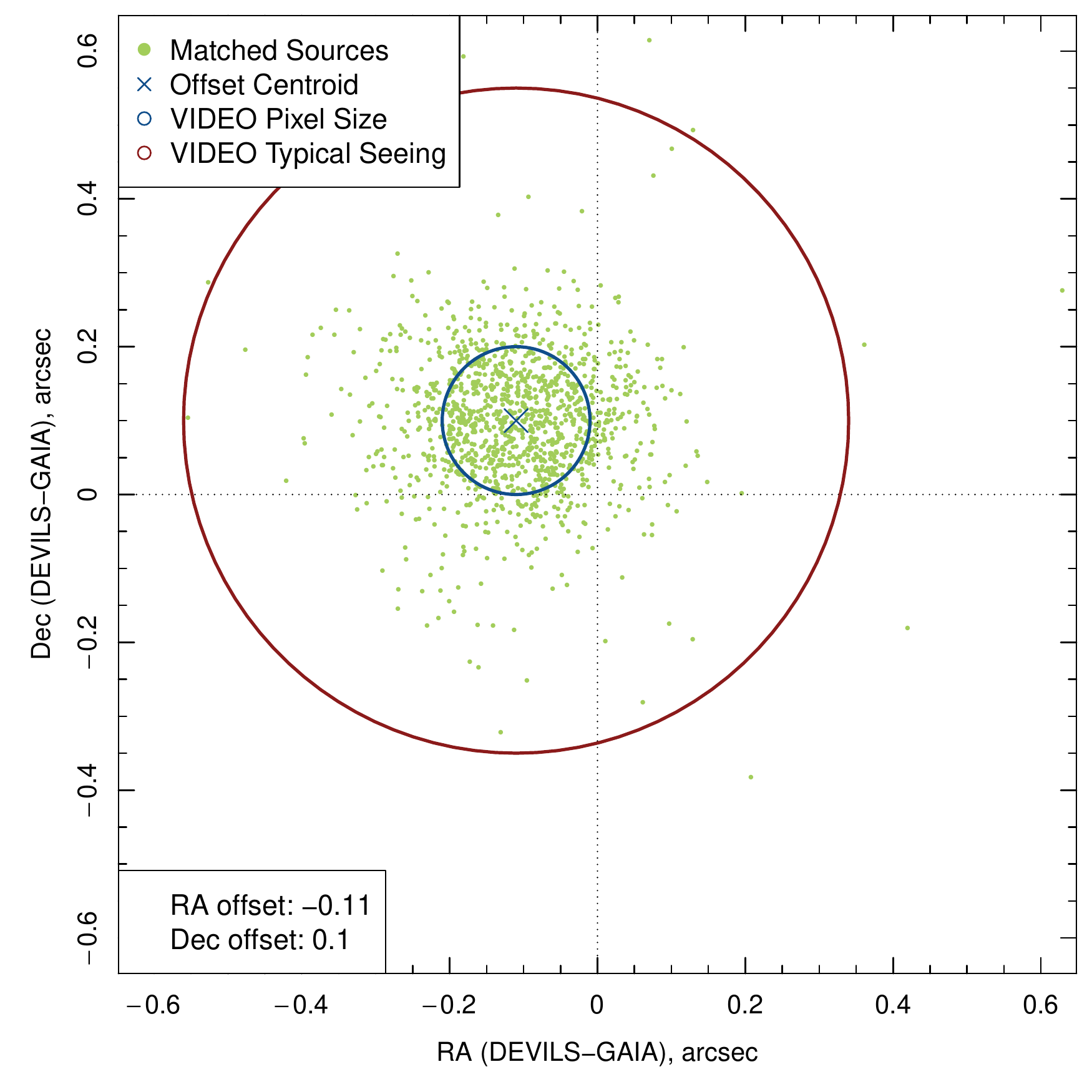}
\includegraphics[scale=0.4]{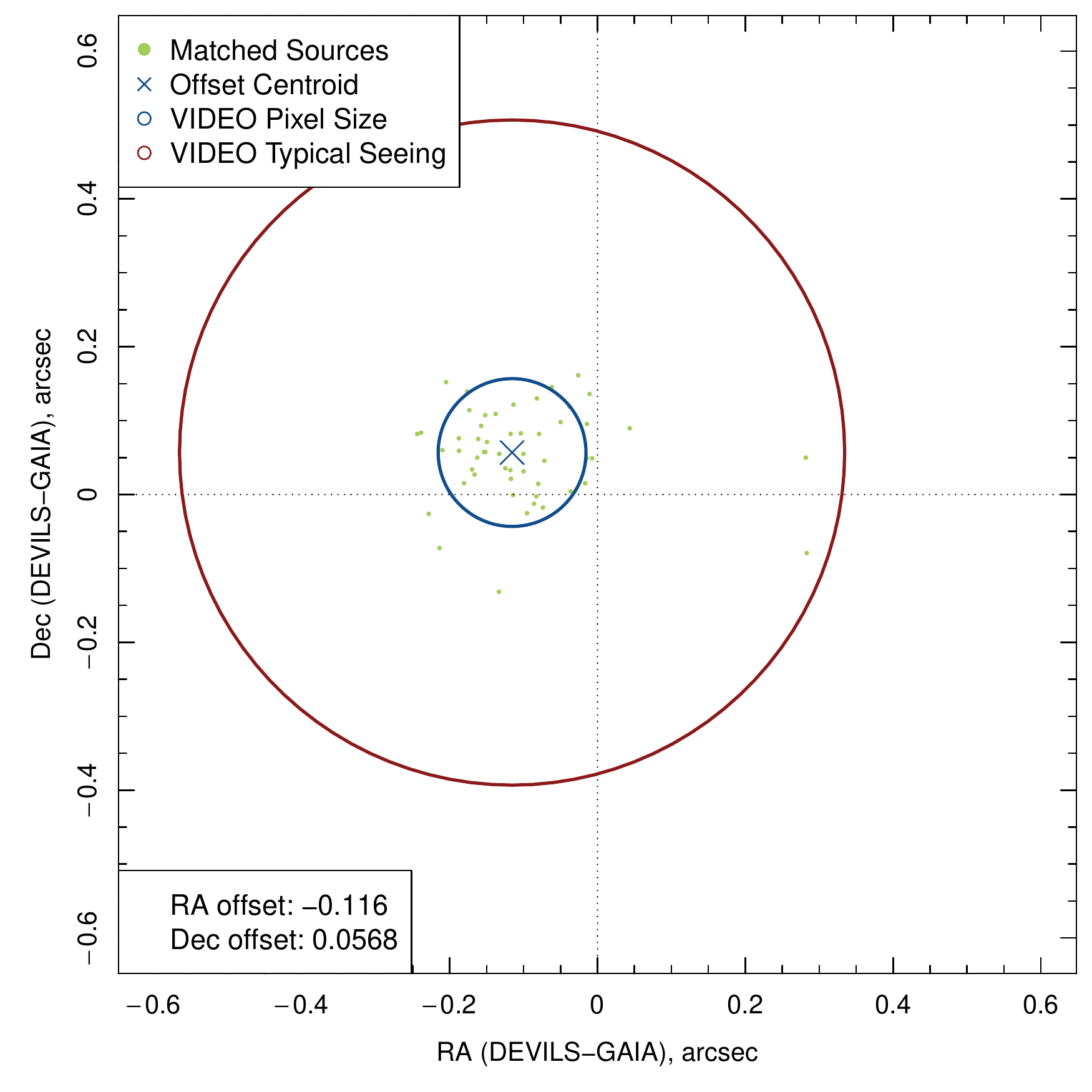}
\caption{Astrometric accuracy of new catalogues against GAIA DR2 in each of the DEVILS regions (Top= D10, Middle=D02, Bottom=D03). The offsets in D02/D03 are driven by offsets in the VIDEO data astrometry.}
\label{fig:Astrometry}
\end{center}
\end{figure}

Finally, to verify the astrometric accuracy of our catalogue, we perform a $2^{\prime\prime}$ match to $12<g_{\mathrm{GAIA}}<18$\,mag sources from the GAIA DR2 catalogues \citep{Gaia18, Lindegren18}. Figure \ref{fig:Astrometry} displays the R.A. and Dec offsets with the median offset marked as the blue cross. Circles indicate the UltraVISTA/VIDEO pixel size and typical seeing, showing that our final catalogues have sub-pixel astrometric accuracy in comparison to GAIA in D10. While we do see small ($\sim$1\,pixel) offsets in D02/D03 we do not opt to correct for this astrometry mis-match. Instead we retain our catalogues with the astrometry derived from the DEVILS input catalogue photometry ($i.e.$ derived from the UltraVISTA and VIDEO data directly). We note that the VIDEO data we use here has not been aligned with GAIA, but newer iterations and public releases will be.

\begin{figure*}
\begin{center}
\includegraphics[scale=0.41]{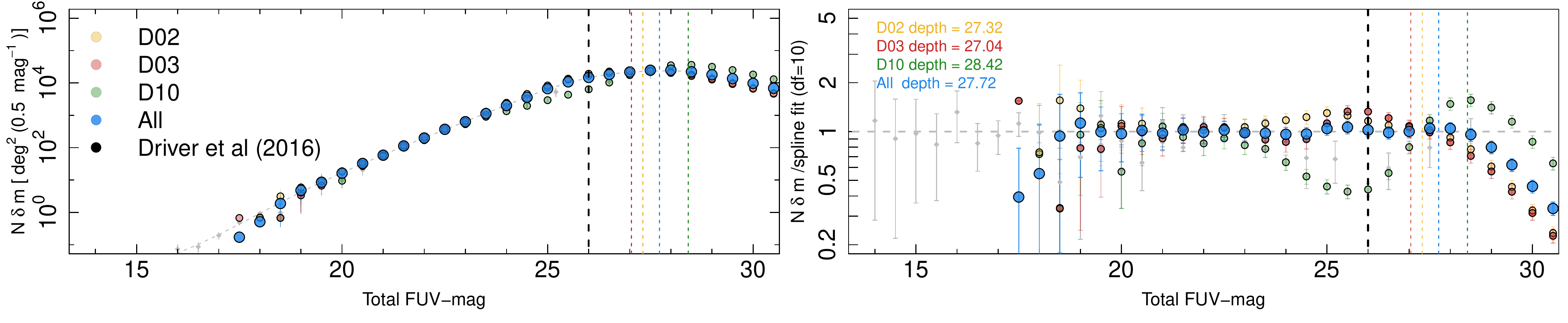}
\includegraphics[scale=0.41]{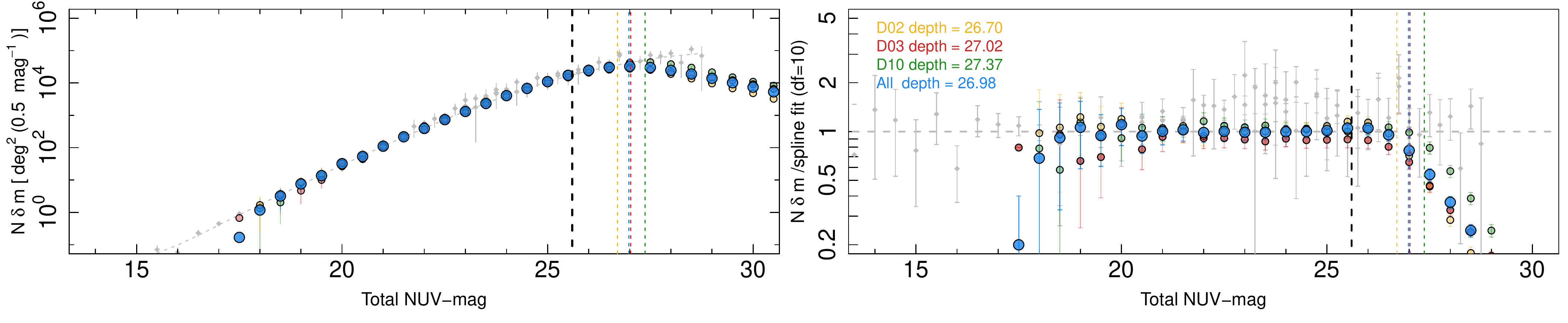}
\includegraphics[scale=0.41]{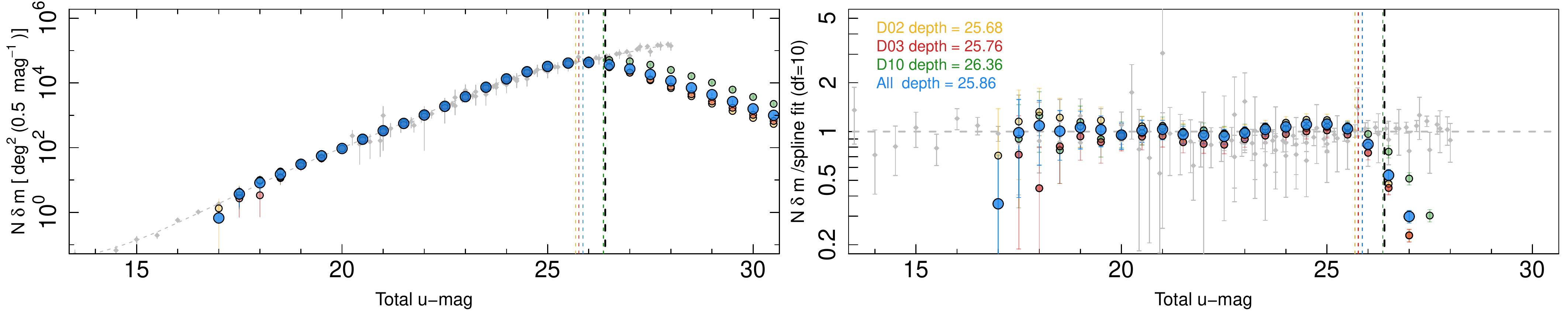}
\includegraphics[scale=0.41]{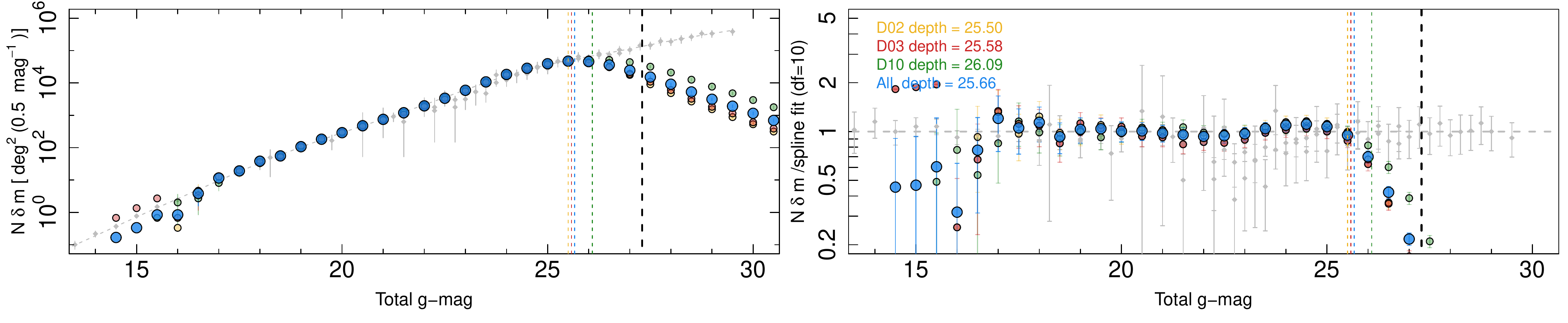}
\includegraphics[scale=0.41]{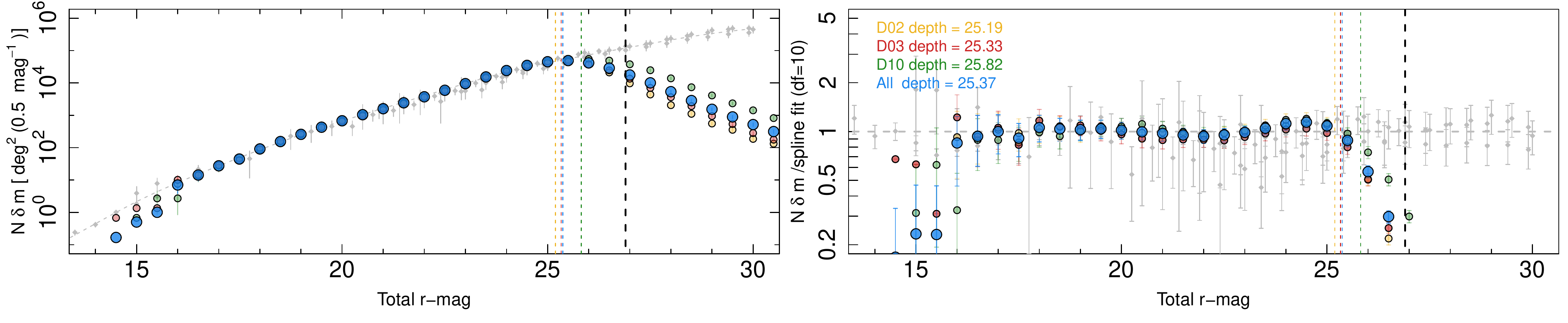}
\includegraphics[scale=0.41]{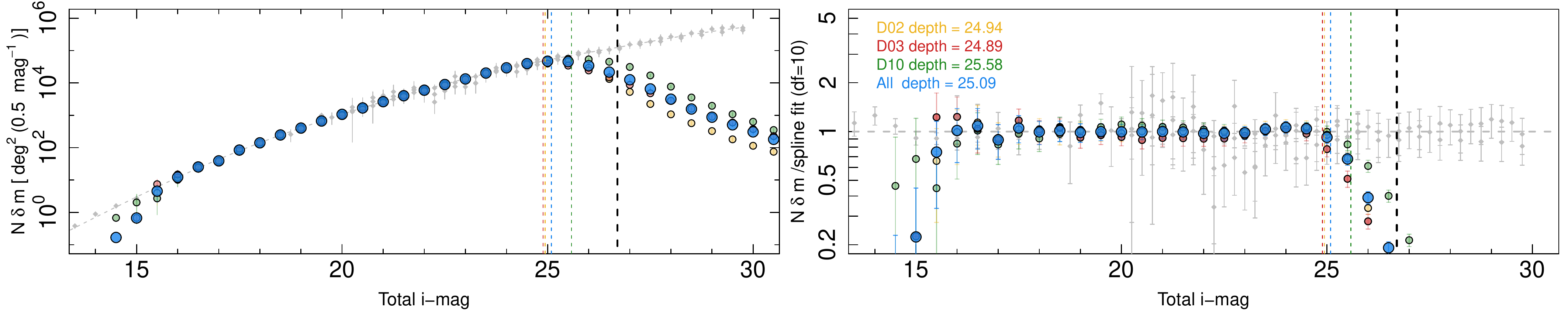}

\caption{Deep galaxy number counts. We show each field individually for D02 (gold), D03 (red) and D10 (green), and for the full DEVILS sample (blue). Error bars (for Poisson+cosmic variance errors) are plotted, but in most cases are smaller than the data point. We also show the compendium of data from \citet{Driver16} as faint grey points. The left panel displays the number counts directly, while the right panel shows the number counts with a spline fit to the data removed (grey dashed line, see Section \ref{sec:counts} for details). Also shown are the nominal survey quoted depth (black vertical dashed) and measured galaxy completeness depth for each individual field and the full sample (dashed vertical coloured lines).  These depths are also noted in the left panel legend and Table \ref{tab:D10photom}.}
\label{fig:NumberCounts}
\end{center}
\end{figure*}

\begin{figure*}
\begin{center}

\includegraphics[scale=0.41]{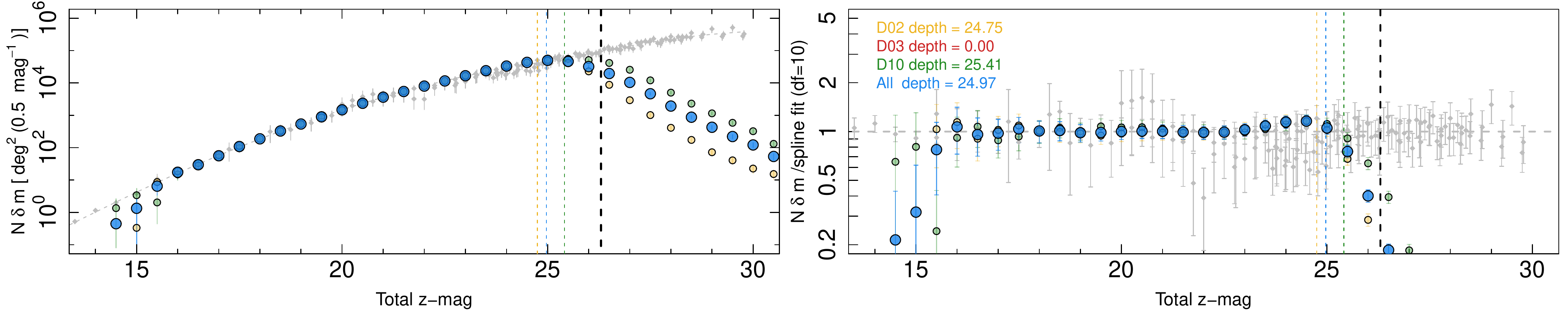}
\includegraphics[scale=0.41]{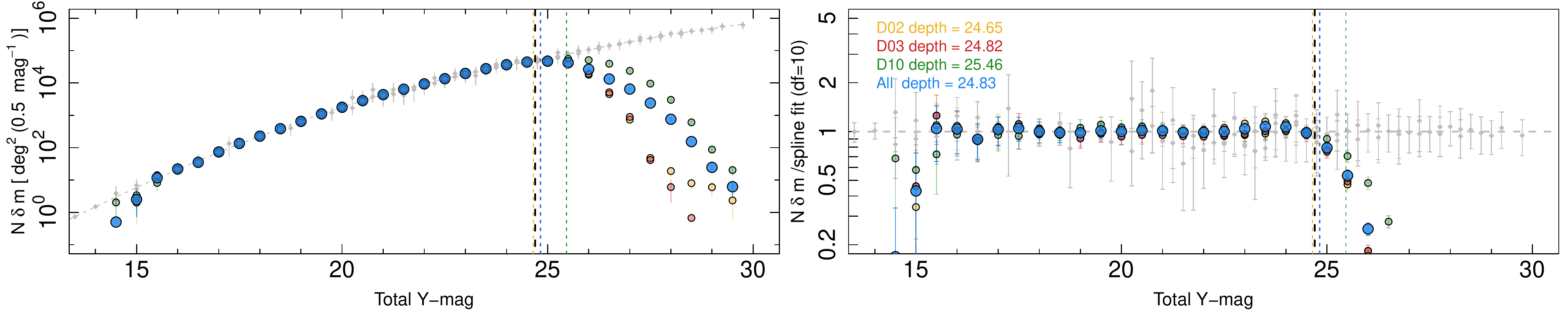}
\includegraphics[scale=0.41]{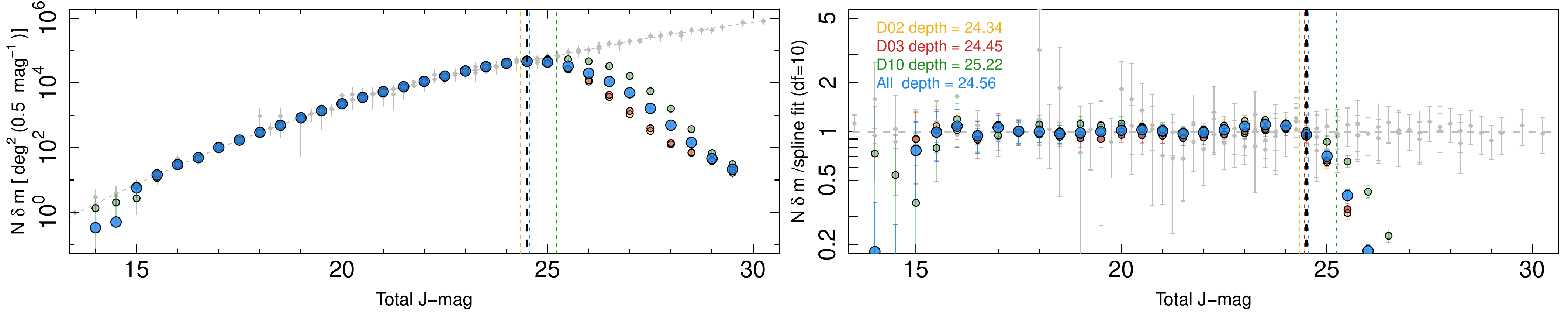}
\includegraphics[scale=0.41]{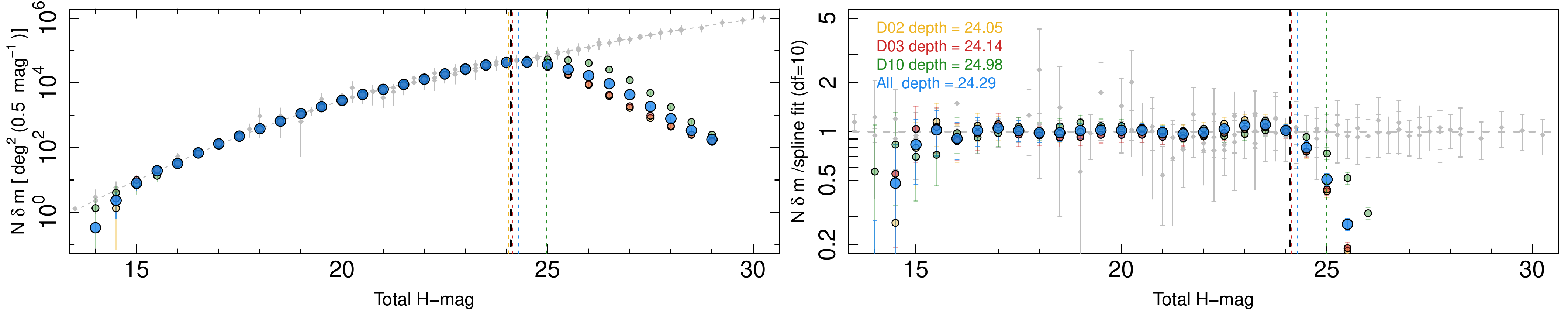}
\includegraphics[scale=0.41]{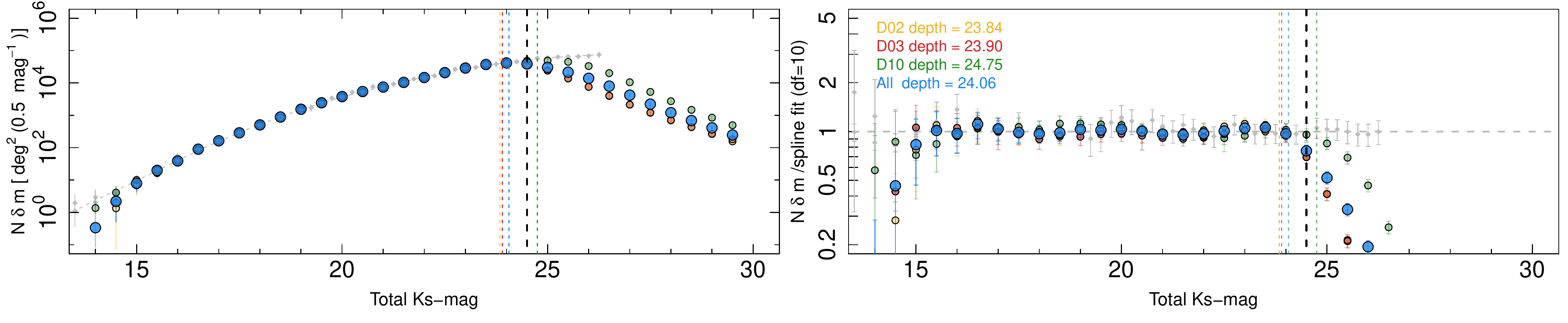}
\includegraphics[scale=0.41]{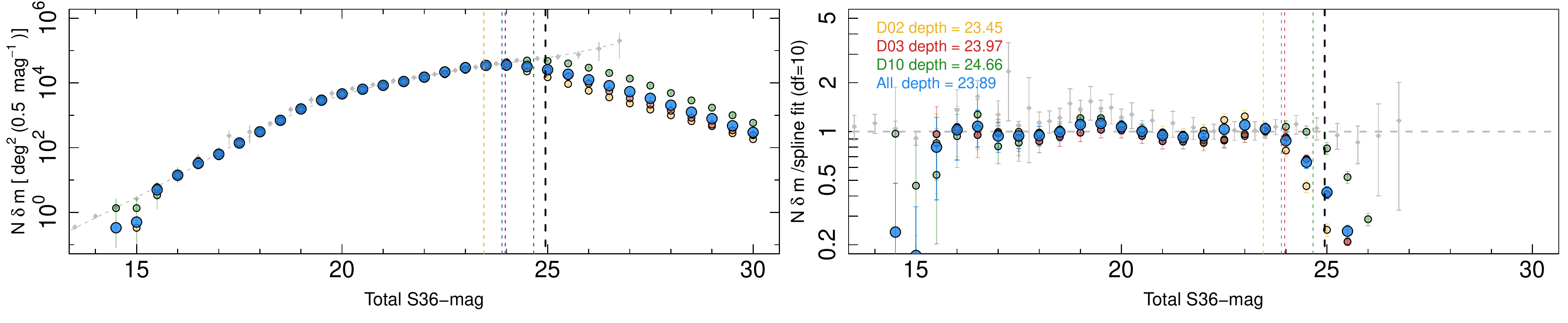}

\caption{Figure \ref{fig:NumberCounts} continued.}
\label{fig:NumberCounts2}
\end{center}
\end{figure*}

\begin{figure*}
\begin{center}

\includegraphics[scale=0.41]{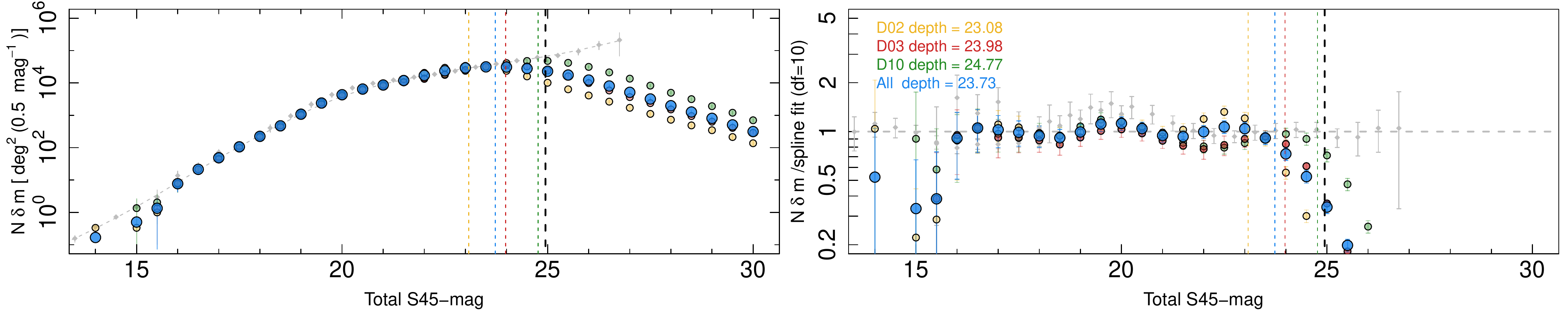}
\includegraphics[scale=0.41]{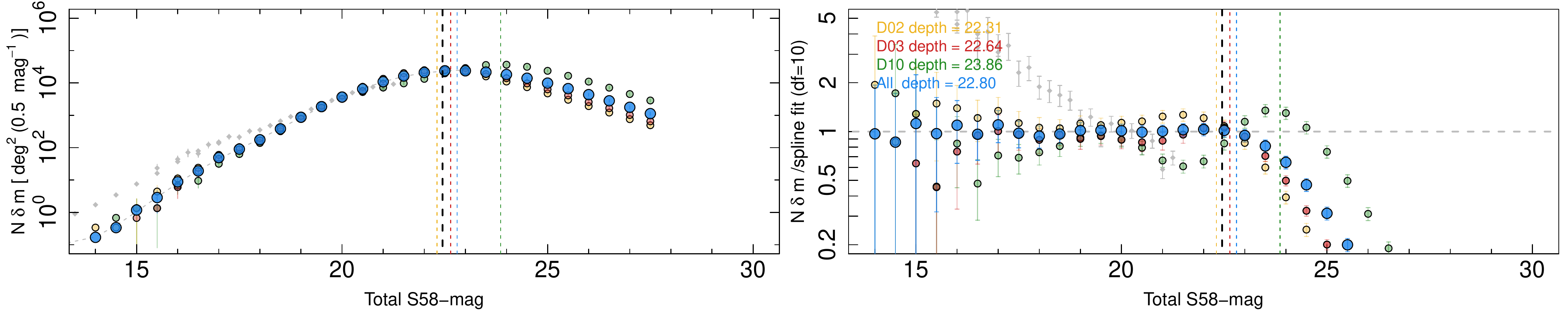}
\includegraphics[scale=0.41]{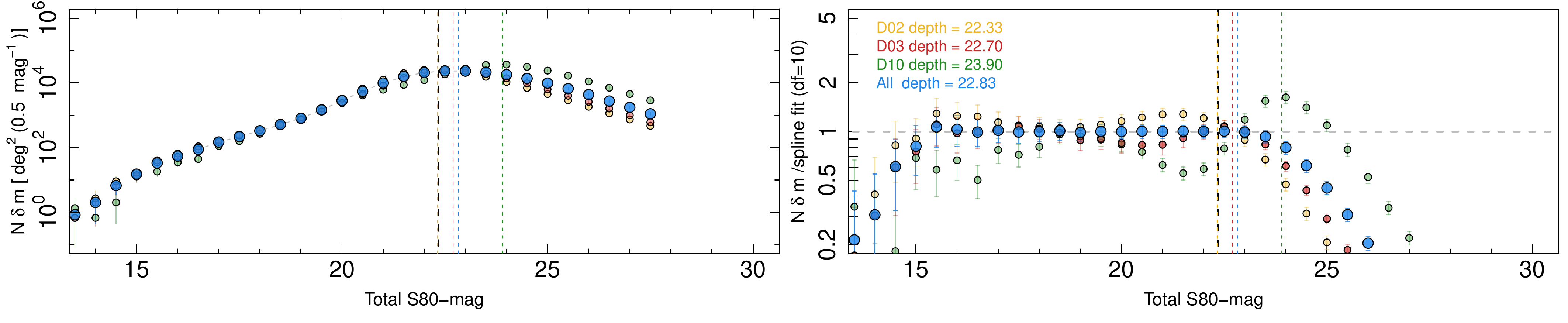}
\includegraphics[scale=0.41]{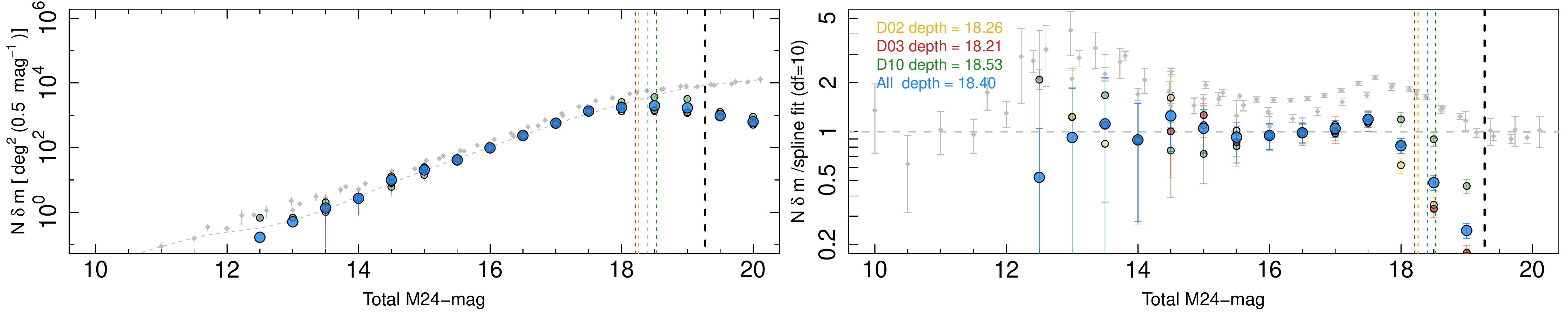}
\includegraphics[scale=0.41]{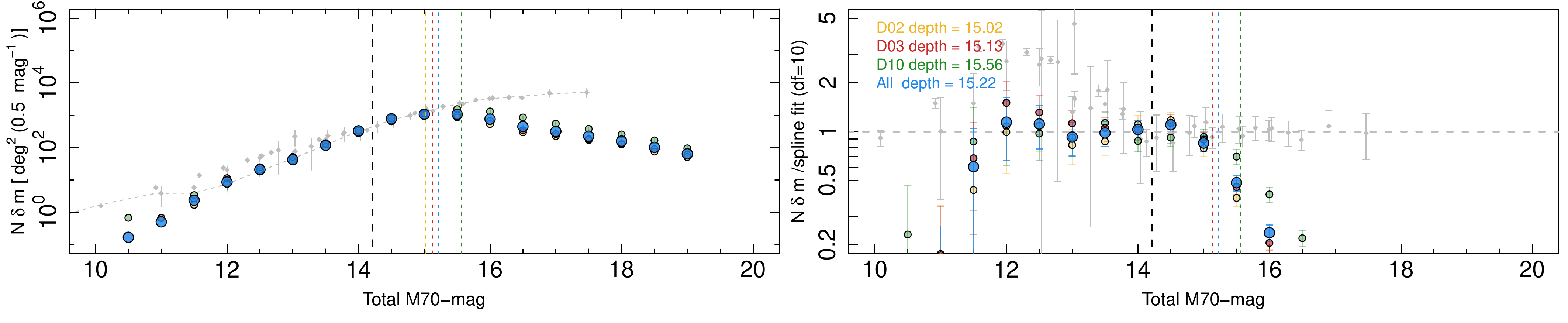}
\includegraphics[scale=0.41]{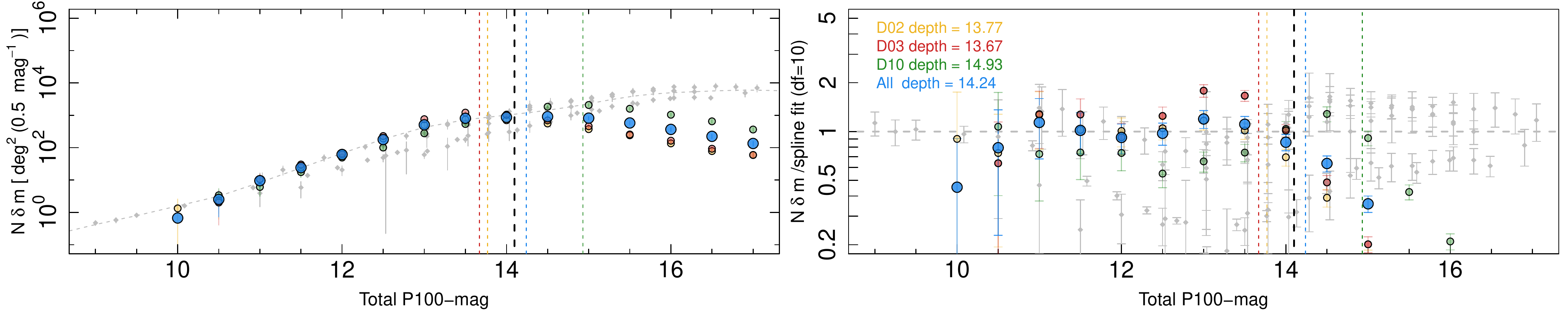}

\caption{Figure \ref{fig:NumberCounts} continued.}
\label{fig:NumberCounts3}
\end{center}
\end{figure*}

\begin{figure*}
\begin{center}

\includegraphics[scale=0.41]{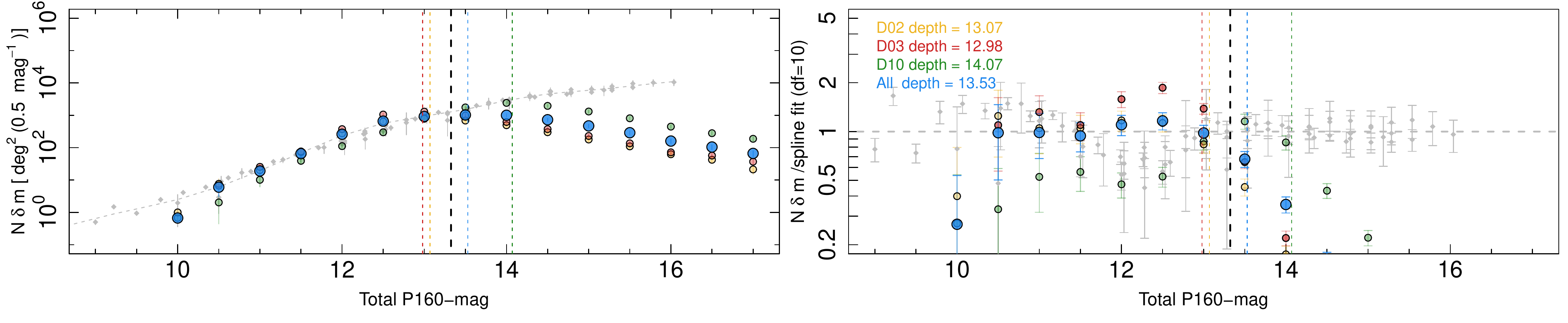}
\includegraphics[scale=0.41]{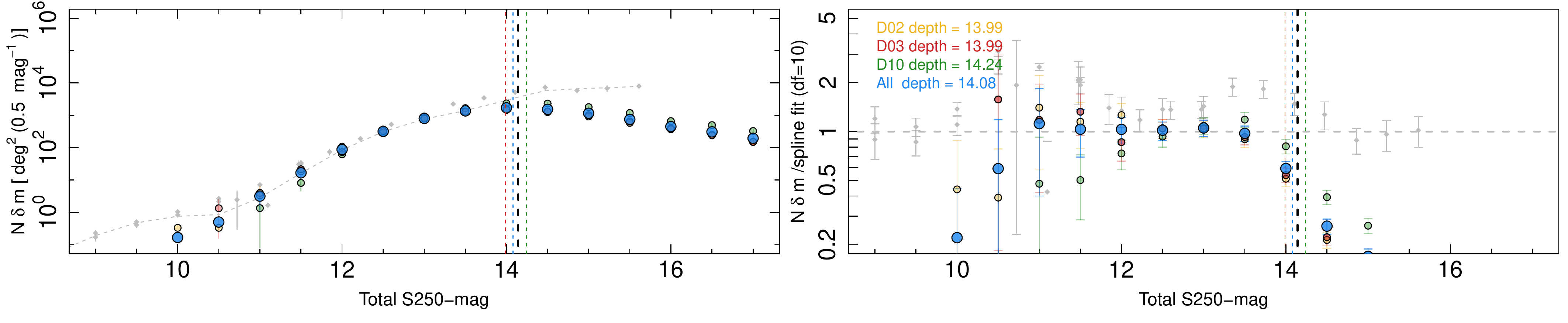}
\includegraphics[scale=0.41]{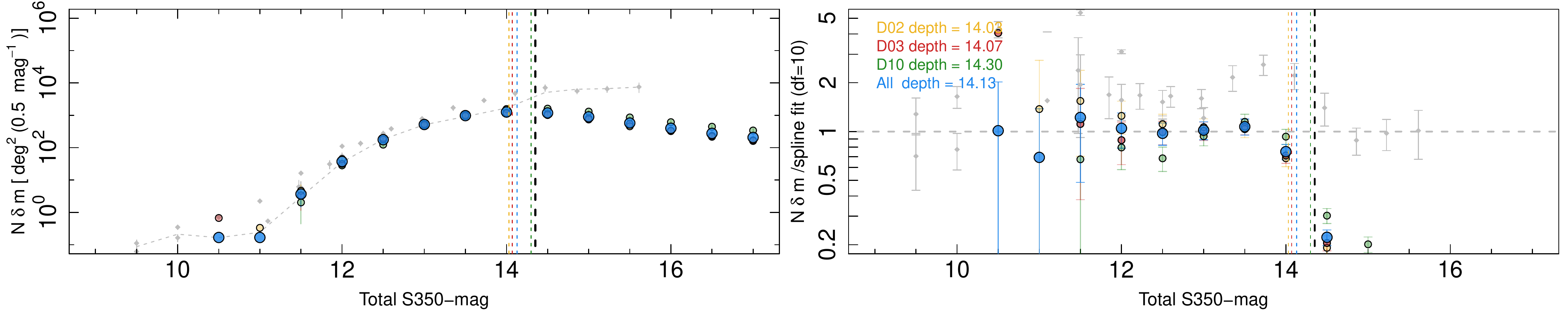}
\includegraphics[scale=0.41]{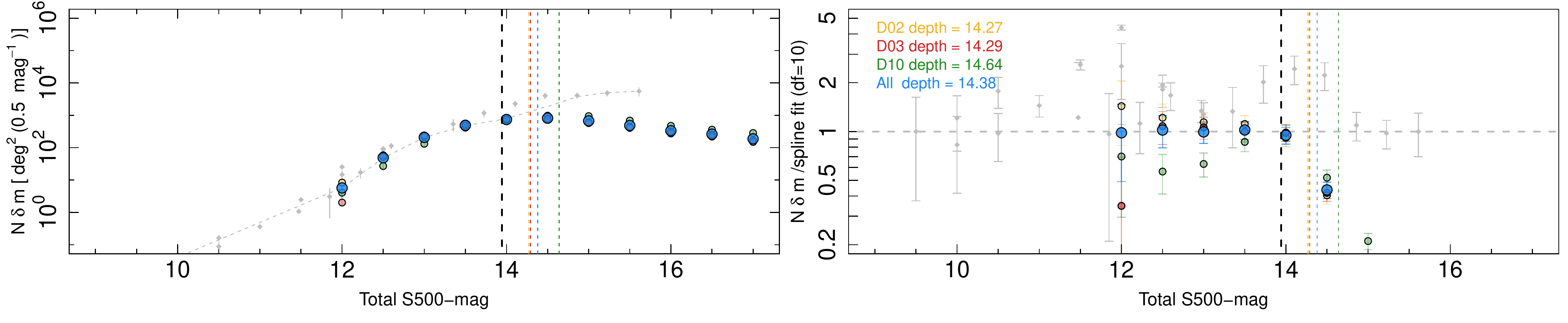}

\caption{Figure \ref{fig:NumberCounts} continued.}
\label{fig:NumberCounts4}
\end{center}
\end{figure*}

\section{Deep Number Counts}
\label{sec:counts}

While our photometric catalogues are produced primarily to allow scientific analysis in a number of areas, they also provide some of the deepest photometric catalogues consistently measured over a broad wavelength range and over $>$1\,deg$^2$ area scales. Here we use these catalogues to produce deep galaxy number counts as an additional photometric validation, for the purposes of exploring the overall galaxy population in broader studies, and for tuning numerical simulations of the galaxy distribution. These number counts have recently been used in \citet{Koushan21} to measure the Extragalactic Background Light (EBL) from the UV-MIR. Hence, we do not go into detail regarding the scientific implications of these distributions, but simply produce them and make available to the wider community. 

Figures \ref{fig:NumberCounts}, \ref{fig:NumberCounts2}, \ref{fig:NumberCounts3} and \ref{fig:NumberCounts4} display the number of galaxies in our unmasked regions per unit sky area (left column). We take all sources identified as galaxies using our selections described in Section \ref{sec:starGal}, and bin in 0.5\,mag ranges in each band and DEVILS field independently. Here we include both our NIR-selected catalogues and, in the FIR, include the `extra sources' identified using \textsc{ProFound} (described in Section \ref{sec:FIRphotom}). We display the number counts for each DEVILS region independently as small gold, red and green circles and the combined counts for the full DEVILS sample as the larger blue circles.  Error bars are derived from the Poisson error combined linearly with cosmic variance errors, estimated using the method outlined in \cite{Driver10}. In most cases, these error bars are smaller than the data points. For comparison, we also display the compendium of photometric data from \cite{Driver16}, and find remarkable agreement in the UV-MIR but now with smaller errors (the Driver et al. points essentially use just the COSMOS region at these depths and thus have larger Poisson and cosmic variance errors). 

On these figures we also display the nominal quoted depth of the input imaging as the dashed vertical black lines (also given in Table \ref{tab:D10photom}), and the measured depth to galaxies in each field individually and for the full sample as the vertical coloured lines. These values are calculated as the point where each of the number count distributions turn over. These limits are displayed in the right column of these Figures and also in Table \ref{tab:D10photom}.

In order to highlight the small differences between each DEVILS region (and the literature values), in the right column of Figures \ref{fig:NumberCounts}, \ref{fig:NumberCounts2}, \ref{fig:NumberCounts3} and \ref{fig:NumberCounts4} we also show these number counts with a 10$^{th}$ order spline fit removed. The spline is fit to combined counts for the full DEVILS sample (blue circles) at magnitudes below the combined counts depth (vertical blue dashed lines). At fainter magnitudes we use the literature values from \cite{Driver16}. We note that the choice of 10 degrees of freedom is largely arbitrary, but that changing this value does not significantly change our results.    

We note that our deep number counts are measured consistently across all bands, with only small differences between fields at faint magnitudes. These counts extend to faint magnitudes over a large area $\sim$6\,deg$^{2}$, making them the state-of-the-art in using galaxy number counts to explore the Universal distribution of flux at a given wavelength \citep{Koushan21}. We therefore provide these for use in the larger community in a tabulated form in Table \ref{Tab:numDense}.       

\newpage

\begin{table*} 
 \begin{center} 
  \caption{Tabulated number densities of sources in each photometric band. Magnitude is the centre of a $\Delta$0.5mag bin. All other values are N(m), 0.5\,mag$^{-1}$\,deg$^{-2}$. These are also made available in a machine-readable form as online data products associated with this paper. } 
  \label{Tab:numDense} 
  \scriptsize 
 \setlength\tabcolsep{3pt} 
 \begin{tabular}{l l l l l l l l l l l}  
Mag.  & FUV  & NUV & u & g & r & i & z & Y & J & H  \\ 
 \hline 
 \hline 
10.0 & - & - & - & - & - & - & - & - & - & - \\ 
10.5 & - & - & - & - & - & - & - & - & - & - \\ 
11.0 & - & - & - & - & - & - & - & - & - & - \\ 
11.5 & - & - & - & - & - & - & - & - & - & - \\ 
12.0 & - & - & - & - & - & - & - & - & - & - \\ 
12.5 & - & - & - & - & - & - & - & - & - & - \\ 
13.0 & - & - & - & - & - & - & - & - & - & - \\ 
13.5 & - & - & - & - & - & - & - & - & - & - \\ 
14.0 & - & - & - & - & - & - & - & - & 0.3$\pm$0.6 & 0.3$\pm$0.6 \\ 
14.5 & - & - & - & 0.2$\pm$0.4 & 0.2$\pm$0.4 & 0.2$\pm$0.4 & 0.4$\pm$0.7 & 0.5$\pm$0.7 & 0.5$\pm$0.8 & 2.3$\pm$1.7 \\ 
15.0 & - & - & - & 0.3$\pm$0.6 & 0.5$\pm$0.8 & 0.7$\pm$0.9 & 1.3$\pm$1.3 & 2.5$\pm$1.8 & 5.7$\pm$2.9 & 8.1$\pm$3.5 \\ 
15.5 & - & - & - & 0.8$\pm$1.0 & 1.0$\pm$1.1 & 4.5$\pm$2.5 & 6.5$\pm$3.1 & 11.7$\pm$4.4 & 14.4$\pm$5.0 & 19.3$\pm$6.0 \\ 
16.0 & - & - & - & 0.8$\pm$1.0 & 7.0$\pm$3.2 & 12.2$\pm$4.5 & 17.4$\pm$5.6 & 21.9$\pm$6.5 & 30.0$\pm$7.9 & 32.6$\pm$8.4 \\ 
16.5 & - & - & - & 3.9$\pm$2.3 & 14.3$\pm$4.9 & 25.3$\pm$7.1 & 29.7$\pm$7.9 & 35.2$\pm$8.8 & 49.0$\pm$11.0 & 68.8$\pm$13.9 \\ 
17.0 & - & - & 0.7$\pm$0.9 & 11.6$\pm$4.3 & 27.5$\pm$7.5 & 39.1$\pm$9.4 & 56.2$\pm$12.1 & 73.9$\pm$14.6 & 101$\pm$18 & 131$\pm$22 \\ 
17.5 & 0.2$\pm$0.4 & 0.2$\pm$0.4 & 3.7$\pm$2.2 & 19.1$\pm$5.9 & 44.6$\pm$10.3 & 84.4$\pm$16.1 & 109$\pm$19 & 135$\pm$23 & 171$\pm$27 & 227$\pm$34 \\ 
18.0 & 0.5$\pm$0.8 & 1.2$\pm$1.2 & 8.2$\pm$3.5 & 38.2$\pm$9.3 & 91.4$\pm$17.0 & 141$\pm$23 & 188$\pm$29 & 227$\pm$34 & 297$\pm$41 & 386$\pm$51 \\ 
18.5 & 1.9$\pm$1.5 & 3.2$\pm$2.1 & 15.1$\pm$5.1 & 55.7$\pm$12.0 & 155$\pm$25 & 247$\pm$36 & 328$\pm$45 & 386$\pm$51 & 494$\pm$62 & 662$\pm$80 \\ 
19.0 & 4.8$\pm$2.6 & 7.6$\pm$3.4 & 30.7$\pm$8.0 & 107$\pm$19 & 258$\pm$37 & 403$\pm$53 & 539$\pm$67 & 652$\pm$79 & 837$\pm$97 & 1137$\pm$126 \\ 
19.5 & 8.6$\pm$3.6 & 13.6$\pm$4.8 & 55.7$\pm$12.0 & 183$\pm$28 & 430$\pm$56 & 668$\pm$80 & 898$\pm$103 & 1109$\pm$124 & 1397$\pm$151 & 1851$\pm$194 \\ 
20.0 & 16.3$\pm$5.4 & 32.0$\pm$8.3 & 94.3$\pm$17.4 & 291$\pm$41 & 678$\pm$81 & 1071$\pm$120 & 1479$\pm$159 & 1776$\pm$187 & 2283$\pm$234 & 2903$\pm$291 \\ 
20.5 & 32.5$\pm$8.4 & 53.6$\pm$11.7 & 182$\pm$28 & 476$\pm$61 & 1047$\pm$118 & 1690$\pm$179 & 2358$\pm$241 & 2848$\pm$286 & 3588$\pm$353 & 4437$\pm$428 \\ 
21.0 & 58.5$\pm$12.4 & 111$\pm$20 & 334$\pm$46 & 751$\pm$89 & 1603$\pm$171 & 2643$\pm$267 & 3643$\pm$358 & 4364$\pm$422 & 5373$\pm$512 & 6371$\pm$599 \\ 
21.5 & 113$\pm$20 & 217$\pm$32 & 563$\pm$70 & 1202$\pm$133 & 2439$\pm$248 & 4034$\pm$393 & 5455$\pm$519 & 6414$\pm$603 & 7660$\pm$712 & 9044$\pm$833 \\ 
22.0 & 199$\pm$30 & 393$\pm$52 & 1023$\pm$115 & 1965$\pm$205 & 3739$\pm$366 & 5989$\pm$566 & 8003$\pm$742 & 9370$\pm$861 & 11227$\pm$1022 & 13116$\pm$1184 \\ 
22.5 & 370$\pm$49 & 737$\pm$87 & 1899$\pm$198 & 3372$\pm$333 & 5944$\pm$562 & 8975$\pm$827 & 11559$\pm$1050 & 13572$\pm$1224 & 16399$\pm$1466 & 19021$\pm$1689 \\ 
23.0 & 635$\pm$77 & 1317$\pm$144 & 3774$\pm$369 & 5939$\pm$561 & 9542$\pm$876 & 13336$\pm$1203 & 16732$\pm$1494 & 19713$\pm$1748 & 23484$\pm$2069 & 26909$\pm$2359 \\ 
23.5 & 1130$\pm$126 & 2335$\pm$239 & 7383$\pm$688 & 10798$\pm$985 & 15371$\pm$1378 & 20295$\pm$1798 & 23943$\pm$2108 & 27671$\pm$2423 & 31971$\pm$2787 & 35937$\pm$3121 \\ 
24.0 & 2002$\pm$208 & 4064$\pm$395 & 13391$\pm$1208 & 18361$\pm$1633 & 24211$\pm$2130 & 29416$\pm$2571 & 33312$\pm$2900 & 36667$\pm$3182 & 40744$\pm$3525 & 43078$\pm$3721 \\ 
24.5 & 3598$\pm$353 & 6808$\pm$638 & 22224$\pm$1962 & 28471$\pm$2491 & 34965$\pm$3039 & 39707$\pm$3438 & 43335$\pm$3743 & 44337$\pm$3827 & 46237$\pm$3987 & 43808$\pm$3783 \\ 
25.0 & 6690$\pm$627 & 11025$\pm$1004 & 32556$\pm$2836 & 39336$\pm$3407 & 45305$\pm$3908 & 47169$\pm$4065 & 50110$\pm$4311 & 47111$\pm$4060 & 44187$\pm$3814 & 36890$\pm$3201 \\ 
25.5 & 10846$\pm$989 & 17263$\pm$1539 & 41211$\pm$3565 & 47325$\pm$4078 & 49256$\pm$4240 & 45640$\pm$3936 & 46263$\pm$3989 & 42220$\pm$3649 & 33304$\pm$2899 & 26126$\pm$2293 \\ 
26.0 & 14840$\pm$1332 & 24679$\pm$2170 & 43026$\pm$3717 & 45773$\pm$3948 & 41809$\pm$3615 & 34209$\pm$2975 & 32160$\pm$2803 & 26500$\pm$2324 & 20150$\pm$1786 & 16906$\pm$1509 \\ 
26.5 & 18398$\pm$1636 & 30594$\pm$2670 & 36589$\pm$3176 & 36117$\pm$3136 & 28939$\pm$2531 & 21662$\pm$1914 & 19592$\pm$1738 & 13287$\pm$1199 & 11125$\pm$1013 & 9485$\pm$871 \\ 
27.0 & 21876$\pm$1932 & 32558$\pm$2836 & 27218$\pm$2385 & 24624$\pm$2165 & 17667$\pm$1574 & 12604$\pm$1140 & 10409$\pm$951 & 6466$\pm$608 & 4966$\pm$476 & 4360$\pm$422 \\ 
27.5 & 24730$\pm$2174 & 30018$\pm$2622 & 18389$\pm$1636 & 15211$\pm$1364 & 10103$\pm$925 & 6558$\pm$616 & 4650$\pm$448 & 2395$\pm$244 & 1638$\pm$174 & 1877$\pm$196 \\ 
\end{tabular} 
 
\vspace{5mm} 
 
\begin{tabular}{l l l l l l l l l l l l l } 
Mag. & Ks & S36 & S45 & S58 & S80 & M24 & M70 & P100 & P160 & S250 & S350 & S500 \\ 
 \hline 
 \hline 
10.0 & - & - & - & - & - & - & - & 0.7$\pm$0.9 & 0.7$\pm$0.9 & 0.2$\pm$0.4 & - & - \\ 
10.5 & - & - & - & - & - & - & 0.2$\pm$0.4 & 2.5$\pm$1.8 & 6.0$\pm$2.9 & 0.5$\pm$0.7 & 0.2$\pm$0.4 & - \\ 
11.0 & - & - & - & - & - & - & 0.5$\pm$0.8 &  9.5$\pm$3.9 & 19.1$\pm$5.9 & 3.2$\pm$2.0 & 0.2$\pm$0.4 & - \\ 
11.5 & - & - & - & - & - & - & 2.4$\pm$1.7 & 24.1$\pm$6.9 & 65.6$\pm$13.5 & 16.7$\pm$5.5 & 3.7$\pm$2.2 & - \\ 
12.0 & - & - & - & - & - & - & 8.7$\pm$3.7 & 61.6$\pm$12.9 & 261$\pm$37 & 87.7$\pm$16.5 & 37.3$\pm$9.2 & 5.7$\pm$2.9 \\ 
12.5 & - & - & - & - & 0.2$\pm$0.4 & 0.2$\pm$0.4 & 21.1$\pm$6.3 & 179$\pm$28 & 661$\pm$80 & 319$\pm$44 & 177$\pm$28 & 48.9$\pm$11.0 \\ 
13.0 & - & - & - & - & 0.2$\pm$0.4 & 0.5$\pm$0.8 & 42.5$\pm$10.0 & 508$\pm$64 & 921$\pm$105 & 801$\pm$94 & 521$\pm$65 & 208$\pm$31 \\ 
13.5 & - & - & - & - & 0.8$\pm$1.0 & 1.4$\pm$1.3 & 119$\pm$21 & 808$\pm$94 & 1026$\pm$116 & 1371$\pm$149 & 979$\pm$111 & 492$\pm$62 \\ 
14.0 & 0.3$\pm$0.6 & - & 0.2$\pm$0.4 & 0.2$\pm$0.4 & 2.0$\pm$1.6 & 2.7$\pm$1.9 & 330$\pm$45 & 871$\pm$101 & 1004$\pm$114 & 1708$\pm$181 & 1265$\pm$139 & 743$\pm$88 \\ 
14.5 & 2.2$\pm$1.7 & 0.3$\pm$0.6 & - & 0.3$\pm$0.6 & 6.8$\pm$3.2 & 10.0$\pm$4.0 & 782$\pm$92 & 912$\pm$105 & 729$\pm$86 & 1533$\pm$164 & 1183$\pm$131 & 818$\pm$95 \\ 
15.0 & 7.9$\pm$3.5 & 0.5$\pm$0.7 & 0.5$\pm$0.7 & 1.2$\pm$1.2 & 15.2$\pm$5.1 & 20.6$\pm$6.2 & 1084$\pm$122 & 819$\pm$95 & 472$\pm$60 & 1157$\pm$128 & 899$\pm$103 & 673$\pm$81 \\ 
15.5 & 19.8$\pm$6.1 & 5.0$\pm$2.7 & 1.3$\pm$1.3 & 2.9$\pm$1.9 & 33.6$\pm$8.5 & 41.8$\pm$ 9.9 & 1071$\pm$120 & 581$\pm$71 & 290$\pm$41 & 747$\pm$88 & 577$\pm$71 & 485$\pm$62 \\ 
16.0 & 39.0$\pm$9.4 & 14.1$\pm$4.9 & 7.7$\pm$3.4 & 8.8$\pm$3.7 & 54.8$\pm$11.9 & 98.6$\pm$18.0 & 767$\pm$90 & 367$\pm$49 & 160$\pm$26 & 450$\pm$58 & 399$\pm$53 & 334$\pm$45 \\ 
16.5 & 89.7$\pm$16.8 & 32.6$\pm$8.4 & 21.4$\pm$6.4 & 19.1$\pm$5.9 & 88.4$\pm$16.6 & 239$\pm$35 & 450$\pm$58 & 224$\pm$33 & 105$\pm$19 & 308$\pm$43 & 277$\pm$39 & 262$\pm$38 \\ 
17.0 & 163$\pm$26 & 62.8$\pm$13.0 & 48.7$\pm$11.0 & 49.2$\pm$11.0 & 149$\pm$24 & 573$\pm$71 & 318$\pm$44 & 135$\pm$23 & 66.5$\pm$13.6 & 195$\pm$30 & 207$\pm$31 & 188$\pm$29 \\ 
17.5 & 288$\pm$40 & 139$\pm$23 & 106$\pm$19 & 90.1$\pm$16.8 & 221$\pm$33 & 1355$\pm$148 & 227$\pm$34 & 80.5$\pm$15.5 & 46.0$\pm$10.5 & 125$\pm$21 & 140$\pm$23 & 140$\pm$23 \\ 
18.0 & 505$\pm$64 & 310$\pm$43 & 224$\pm$33 & 175$\pm$28 & 333$\pm$45 & 1769$\pm$187 & 160$\pm$26 & 45.2$\pm$10.4 & 25.6$\pm$7.2 & 78.5$\pm$15.3 & 84.2$\pm$16.0 & 90.9$\pm$17.0 \\ 
18.5 & 885$\pm$102 & 702$\pm$84 & 474$\pm$60 & 382$\pm$51 & 519$\pm$65 & 1970$\pm$205 & 103$\pm$19 & 28.1$\pm$7.6 & 15.9$\pm$5.3 & 39.0$\pm$9.4 & 49.2$\pm$11.0 & 53.1$\pm$11.6 \\ 
19.0 & 1544$\pm$165 & 1578$\pm$168 & 1084$\pm$121 & 872$\pm$101 & 812$\pm$95 & 1704$\pm$180 & 64.3$\pm$13.3 & 12.6$\pm$4.6 & 8.0$\pm$3.5 & 28.5$\pm$7.7 & 35.7$\pm$8.9 & 37.0$\pm$9.1 \\ 
19.5 & 2442$\pm$249 & 2937$\pm$294 & 2399$\pm$245 & 1857$\pm$195 & 1475$\pm$159 & 978$\pm$111 & 34.2$\pm$8.6 & 12.4$\pm$4.5 & 5.9$\pm$2.9 & 17.9$\pm$5.7 & 22.1$\pm$6.5 & 20.8$\pm$6.2 \\ 
20.0 & 3792$\pm$371 & 4594$\pm$443 & 4333$\pm$419 & 3670$\pm$360 & 2841$\pm$285 & 632$\pm$77 & 23.0$\pm$6.7 & 7.0$\pm$3.2 & 2.7$\pm$1.9 & 12.9$\pm$4.6 & 14.7$\pm$5.0 & 14.1$\pm$4.9 \\ 
20.5 & 5435$\pm$518 & 6415$\pm$603 & 6479$\pm$609 & 6549$\pm$615 & 5539$\pm$526 & 409$\pm$54 & 12.8$\pm$4.6 & 5.0$\pm$2.7 & 2.3$\pm$1.7 & 9.2$\pm$3.8 & 10.0$\pm$4.0 & 13.2$\pm$4.7 \\ 
21.0 & 7471$\pm$697 & 8449$\pm$781 & 8654$\pm$799 & 10931$\pm$996 & 9961$\pm$912 & 244$\pm$36 &  9.7$\pm$3.9 & 1.7$\pm$1.4 & 0.7$\pm$0.9 &  9.5$\pm$3.9 & 8.0$\pm$3.5 & 10.5$\pm$4.1 \\ 
21.5 & 10410$\pm$953 & 11104$\pm$1011 & 11900$\pm$1080 & 16357$\pm$1462 & 15866$\pm$1420 & 156$\pm$25 & 7.0$\pm$3.2 & 1.5$\pm$1.4 & 1.3$\pm$1.3 & 8.7$\pm$3.7 &  9.7$\pm$3.9 & 10.0$\pm$4.0 \\ 
22.0 & 14741$\pm$1326 & 15157$\pm$1359 & 17134$\pm$1528 & 21138$\pm$1870 & 20990$\pm$1857 & 97.1$\pm$17.8 & 9.2$\pm$3.8 & 1.7$\pm$1.4 & 0.7$\pm$0.9 & 10.0$\pm$4.0 & 16.2$\pm$5.4 & 10.0$\pm$4.0 \\ 
22.5 & 20908$\pm$1853 & 21886$\pm$1933 & 23843$\pm$2099 & 23849$\pm$2100 & 23591$\pm$2078 & 57.1$\pm$12.2 & 10.5$\pm$4.1 & 1.2$\pm$1.2 & 1.0$\pm$1.1 & 12.1$\pm$4.5 & 17.6$\pm$5.6 & 13.2$\pm$4.7 \\ 
23.0 & 28899$\pm$2531 & 29775$\pm$2601 & 29204$\pm$2553 & 23801$\pm$2096 & 23398$\pm$2061 & 29.9$\pm$7.9 & 9.0$\pm$3.7 & 0.7$\pm$0.9 & 1.3$\pm$1.3 & 9.4$\pm$3.8 & 18.8$\pm$5.9 & 19.1$\pm$5.9 \\ 
23.5 & 37206$\pm$3233 & 34960$\pm$3039 & 31488$\pm$2746 & 21688$\pm$1916 & 21586$\pm$1908 & 21.8$\pm$6.4 & 9.4$\pm$3.8 & 0.3$\pm$0.6 & 0.5$\pm$0.7 & 8.2$\pm$3.5 & 17.4$\pm$5.6 & 19.6$\pm$6.0 \\ 
24.0 & 41654$\pm$3608 & 35913$\pm$3119 & 31040$\pm$2708 & 18234$\pm$1622 & 18162$\pm$1616 & 12.4$\pm$4.5 & 8.0$\pm$3.5 & 0.3$\pm$0.6 & 0.7$\pm$0.9 &  9.9$\pm$3.9 & 13.2$\pm$4.7 & 24.3$\pm$6.9 \\ 
24.5 & 38743$\pm$3362 & 32089$\pm$2797 & 27997$\pm$2451 & 14033$\pm$1263 & 13840$\pm$1247 & 7.5$\pm$3.3 & 12.4$\pm$4.5 & - & 0.5$\pm$0.7 & 10.2$\pm$4.0 & 10.7$\pm$4.1 & 20.3$\pm$6.2 \\ 
25.0 & 30352$\pm$2654 & 25586$\pm$2247 & 23053$\pm$2032 & 9904$\pm$907 & 9915$\pm$908 & 4.9$\pm$2.6 & 10.4$\pm$4.1 & - & 0.8$\pm$1.0 & 7.7$\pm$3.4 & 14.7$\pm$5.0 & 23.1$\pm$6.7 \\ 
25.5 & 21584$\pm$1911 & 18657$\pm$1658 & 17589$\pm$1567 & 6704$\pm$629 & 6675$\pm$626 & 1.0$\pm$1.1 & 10.9$\pm$4.2 & - & 0.7$\pm$0.9 & 8.0$\pm$3.5 & 15.7$\pm$5.3 & 15.4$\pm$5.2 \\ 
26.0 & 13944$\pm$1257 & 12613$\pm$1141 & 12319$\pm$1116 & 4359$\pm$422 & 4380$\pm$423 & 2.6$\pm$1.8 & 10.7$\pm$4.1 & 0.2$\pm$0.4 & 1.2$\pm$1.2 & 6.7$\pm$3.1 & 11.2$\pm$4.3 & 20.6$\pm$6.2 \\ 
26.5 & 7989$\pm$742 & 8253$\pm$764 & 8058$\pm$747 & 2813$\pm$282 & 2762$\pm$278 & 0.9$\pm$1.0 & 12.4$\pm$4.5 & - & 0.2$\pm$0.4 & 6.9$\pm$3.2 & 12.7$\pm$4.6 & 17.7$\pm$5.7 \\ 
27.0 & 4208$\pm$409 & 5398$\pm$514 & 5139$\pm$491 & 1774$\pm$187 & 1763$\pm$186 & 0.9$\pm$1.0 & 11.9$\pm$4.4 & - & 0.3$\pm$0.6 & 6.5$\pm$3.1 & 12.9$\pm$4.6 & 17.6$\pm$5.6 \\ 
27.5 & 2219$\pm$228 & 3350$\pm$331 & 3201$\pm$318 & 1129$\pm$126 & 1112$\pm$124 & 0.9$\pm$1.0 & 8.7$\pm$3.7 & - & 1.2$\pm$1.2 & 7.5$\pm$3.4 & 12.2$\pm$4.5 & 13.9$\pm$4.9 \\ 
\end{tabular} 
\normalsize 
\end{center} 
\end{table*}

\begin{table*} 
 \begin{center} 
  \caption{DEVILS photometric catalogue columns available in \texttt{DEVILS\_PhotomCat\_15\_10\_2020\_v0.5}. The *B* values are available for \textit{FUV NUV ugrizYJHK$_{s}$ S36 S45 S58 S80 MIPS24 MIPS70 P100 P160 S250 S350 S500}$\mu$m bands. } 
  \label{Tab:cat} 
  \footnotesize
 \setlength\tabcolsep{3pt} 
 \begin{tabular}{l l l l }  
Column Name & Unit & UCD & Description   \\ 
 \hline 
 \hline 

UID  &  none  &  meta.id  &  UID based on field location, and source RAmax and DECmax   \\ 
NewSegID  &  none  &  meta.id  &  Segment ID derived from segmentation maps - unique over full field  \\ 
field  &  none  &  meta.id  &  DEVILS Field identifier (D02/D03/D10)  \\ 
RAcen  &  deg  &  pos.eq.ra  &  Right ascension of the ProFound segment centre in the VISTA Y-band image  \\ 
DECcen  &  deg  &  pos.eq.dec  &  Declination of the ProFound segment centre in the VISTA Y-band image  \\ 
RAmax  &  deg  &  pos.eq.ra  &  Right ascension of the brightest pixel in ProFound segment in the VISTA Y-band image  \\ 
DECmax  &  deg  &  pos.eq.dec  &  Declinations of the brightest pixel in ProFound segment in the VISTA Y-band image  \\ 
mask  &  none  &  meta.id  &  Mask flag to identify is segment falls in a masked region. 0=not masked, 1=masked  \\ 
artefactFlag  &  none  &  meta.id  &  Artefact flag 0=Not artefact, >0=artefact  \\ 
starFlag  &  none  &  meta.id  &  Star/galaxy classifier. 0=galaxy, 1=star  \\ 
R50\_Y  &  $^{\prime\prime}$   &  phys.angSize;em.IR.Y  &  The segment radius contain 50\% of the VISTA Y-band flux  \\ 
N100\_Y  &  none  &  phys.area
  &  Number of segment pixels in the VISTA Y-band  \\ 
subFrame  &  none  &  meta.id  &  ProFound processing subFrame within the field that the segment is taken from  \\ 
segIDFrame  &  none  &  meta.id  &  Segment ID within subFrame  \\ 
xcenFrame  &  pixel  &  instr.pixel  &  x position of the segment centre within the subFrame in the VISTA Y-band   \\ 
ycenFrame  &  pixel  &  instr.pixel  &  y position of the segment centre within the subFrame in the VISTA Y-band   \\ 
xmaxFrame  &  pixel  &  instr.pixel  &  x position of the segment brightest pixel within the subFrame in the VISTA Y-band   \\ 
ymaxFrame  &  pixel  &  instr.pixel  &  y position of the segment brightest pixel within the subFrame in the VISTA Y-band   \\ 
EBV  &  none  &  phys.excitParam  &  Galactic Extinction coefficient applied - derived from Planck maps  \\ 
flux\_*B*  &  Jy  &  phot.flux.density;em.*B*  &  Total flux measured in *B*  \\ 
flux\_err\_*B*  &  Jy  &  stat.error;phot.flux.density;em.*B*  &  Error on total flux measured in *B*   \\ 
flux\_*B*\_c  &  Jy  &  phot.flux.density;em.*B*  &  Colour-optimised flux measured in *B*    \\ 
flux\_err\_*B*\_c  &  Jy  &  stat.error;phot.flux.density;em.*B*  &  Error on colour-optimised flux measured in *B* \\ 

\end{tabular} 
\normalsize 
\end{center} 
\end{table*}

\section{Summary}

We have derived a new photometric catalogue for the DEVILS regions for use in multiple scientific projects within the DEVILS team, which will then be provided to the wider community. Photometry is derived from the up-to-date imaging in these regions using the new \textsc{ProFound} source detection and photometry code, which is designed to overcome a number of known issues present in previous photometry. We perform a number of detailed checks and manual fixing of the \textsc{ProFound} segmentation maps to improve the validity of our photometric measurements and then drive new source photometry measurements from the UV-FIR. We apply a masking, artefact flagging and star/galaxy separation process to identify true extragalactic sources. We then undertake a series of comparisons to existing photometry in the D10 (COSMOS) region to highlight the validity of our catalogues and their use for scientific projects within DEVILS and the wider community. We also use these tests to suggest that our photometry is comparable or better than existing photometry in these regions. We note that these updates to the photometry can have a strong impact on derived galaxy evolution properties (such as the evolution of the sSFR-M* relation). Finally, we use our catalogues to derive deep number counts in all photometric bands analysed in this work, which are measured consistently from the FUV to 500\,$\mu$m, show consistency between individual fields and provide state-of-the-art galaxy number counts to faint magnitudes for the astronomical community.

\section*{Acknowledgements}

 LJMD, ASGR and LC  acknowledge support from the \textit{Australian Research Councils} Future Fellowship scheme (FT200100055, FT200100375 and FT180100066, respectively). JET is supported by the Australian Government Research Training Program (RTP) Scholarship. SB and SPD acknowledge support from the \textit{Australian Research Councils} Discovery Project scheme (DP180103740). MS has been supported by the European Union's Horizon 2020 Research and Innovation programme under the Maria Sklodowska-Curie grant agreement (No. 754510), the Polish National Science Centre (UMO-2016/23/N/ST9/02963), and the Spanish Ministry of Science and Innovation through the Juan de la Cierva-formacion programme (FJC2018-038792-I). MJJ acknowledges  support from the UK Science and Technology Facilities Council [ST/S000488/1 and ST/N000919/1], the Oxford Hintze Centre for Astrophysical Surveys which is funded through generous support from the Hintze Family Charitable Foundation and the South African Radio Astronomy Observatory (SARAO), and ICRAR for financial support during a sabbatical visit. MBi is supported by the Polish National Science Center through grants no. 2020/38/E/ST9/00395, 2018/30/E/ST9/00698 and 2018/31/G/ST9/03388, and by the Polish Ministry of Science and Higher Education through grant DIR/WK/2018/12. MV acknowledges support from the Italian Ministry of Foreign Affairs and International Cooperation (MAECI Grant Number ZA18GR02) and the South African Department of Science and Innovation's National Research Foundation (DSI-NRF Grant Number 113121) as part of the ISARP RADIOSKY2020 Joint Research Scheme. NM acknowledges support from the Bundesministerium f{\"u}r Bildung und Forschung (BMBF) award 05A20WM4. MBr acknowledges the support of the University of Western Australia through a Scholarship for International Research Fees and Ad Hoc Postgraduate Scholarship. DEVILS is an Australian project based around a spectroscopic campaign using the Anglo-Australian Telescope. DEVILS is part funded via Discovery Programs by the \textit{Australian Research Council} and the participating institutions. The DEVILS website is https://devilsurvey.org. The DEVILS data is hosted and provided by AAO Data Central (\url{datacentral.org.au}).

\section{Data Availability}

The photometry catalogues and associated meta data derived in this work are currently made available to the DEVILS team for internal use via a Data Management Unit (DMU) describing the data origin, and providing all photometric measurements and other useful parameters provided by \textsc{ProFound}. Each DMU is both version and date stamped for ease of use and to keep track of updates. The catalogue used in this paper is \texttt{DEVILS\_PhotomCat\_15\_10\_2020\_v0.5} with columns described in Table \ref{Tab:cat}. The data are currently available for proprietary data projects within DEVILS. However, once these data have been used for core DEVILS science projects, it will be made available to the wider community via AAO data central\footnote{\url{https://datacentral.org.au/}}.

\bsp	
\label{lastpage}
\end{document}